\documentclass{aa}
\usepackage{graphicx}
\usepackage[varg]{txfonts}
\usepackage{aalongtable}
\usepackage{natbib}
\usepackage{subfig}
\begin{document}
   \title{A mid-IR interferometric survey with MIDI/VLTI: resolving the second-generation protoplanetary disks around post-AGB binaries. 
\thanks{Based on observations made with ESO Telescopes at the La Silla Paranal Observatory under programmes ID 073.A-9002, 073.A-9014, 
   073.D-0610, 075.D-0605, 077.D-0071, 078.D-0113, 079.D-0013, 080.D-0059, 081.D-0089, 082.D-0066, 083.D-0011, 083.D-0013, 084.D-0009, 093.D-0914,
   and 094.D-0778. Some observations were obtained in the framework of the Belgian Guaranteed Time allocation on VISA.}}
   \titlerunning{MIDI on post-AGB binaries}
   \author{M. Hillen \inst{1} 
           \and 
           H. Van Winckel \inst{1}
           \and
           J. Menu \inst{1} 
           \and
           R. Manick \inst{1}
           \and
           J. Debosscher \inst{1}
           \and
           M. Min \inst{2,3}
           \and
           W.-J. de Wit \inst{4}
           \and
           T. Verhoelst \inst{5}
           \and
           D. Kamath \inst{1}
           \and
           L.B.F.M. Waters \inst{2}
         }
   \institute{Instituut voor Sterrenkunde (IvS), KU Leuven,
              Celestijnenlaan 200D, B-3001 Leuven, Belgium\\
              \email{Michel.Hillen@ster.kuleuven.be}
            \and
            Anton Pannekoek Institute for Astronomy, University of Amsterdam, 1090 GE Amsterdam, The Netherlands \\
             \and
            SRON, Netherlands Institute for Space Research, Sorbonnelaan 2, 3584 CA Utrecht, The Netherlands \\
           \and
            European Southern Observatory, Alonso de Cordova 3107, Vitacura, Santiago, Chile \\
            \and
            Belgian Institute for Space Aeronomy, Brussels, Belgium \\
            \and
            SRON, Netherlands Institute for Space Research, Sorbonnelaan 2, 3584 CA Utrecht, The Netherlands
   }
   \date{Received ? ?, 2016; accepted ? ?, ?}
   \authorrunning{Hillen et al.}

  \abstract
  {}
  {We present a mid-IR interferometric survey of the circumstellar environment of a specific class of post-Asymptotic Giant Branch (post-AGB) binaries. For this class the presence of a compact dusty disk has been postulated on the basis of various spatially unresolved measurements. The aim is to
  determine the angular extent of the N-band emission directly and to
  resolve the compact circumstellar structures.}
  {Our interferometric survey was performed with the MIDI instrument on the VLTI. In total 19 different systems were observed using variable baseline configurations. Combining all the visibilities at a single wavelength at 10.7~$\mu$m, we 
  fitted two parametric models to the data: a uniform disk (UD) and a ring model mimicking  a temperature gradient. We compared 
  our observables of the whole sample, with synthetic data computed from a grid of radiative transfer models of passively irradiated disks in hydrostatic equilibrium. These models are computed with 
  a Monte Carlo code that has been widely applied to describe the structure of protoplanetary disks around young stellar objects (YSO).
  }
  {The spatially resolved observations show that the majority of our targets cluster closely together in the distance-independent size-colour diagram, 
  and have extremely compact N-band emission regions. The typical uniform disk diameter of the N-band emission region is $\sim40$\,mas, which 
  corresponds to a typical brightness temperature of 400-600~K. The resolved objects display very similar characteristics 
  in the interferometric observables and in the spectral energy distributions. Therefore, the physical properties
  of the disks around our targets must be similar. Our results are discussed in the light of recently published sample studies of 
  young stellar objects (YSOs) to compare quantitatively the secondary discs around post-AGB stars to the ones around YSOs. 
  }
  { Our high-angular-resolution survey further confirms the disk nature of the circumstellar structures present around wide post-AGB binaries.
  The grid of protoplanetary disk models covers very well the observed objects. 
  Much like for young stars, the spatially resolved N-band emission region is determined by the hot inner rim of the disk. 
  Continued comparisons between post-AGB and protoplanetary disks will help to understand grain growth and disk evolution processes, 
  and to constrain planet formation theories. 
  These second-generation disks are an important missing ingredient in binary evolution theory of intermediate-mass stars. 
  }  

   \maketitle
%

\section{Introduction}
The impact of binarity  on the evolution of low- to intermediate-mass
stars is an important, yet poorly understood, domain of stellar
astrophysics. Interactions between the loosely bound envelope of a
(super)giant and the gravitational pull of its companion result in a
diverse zoo of peculiar objects. Binarity is not only responsible for
spectacular phenomena like supernovae type Ia, sub-luminous
supernovae, gravitational wave sources and novae explosions, but also
for less energetic systems like sub-dwarf B stars, barium stars and
cataclysmic variables. Furthermore, binary interaction processes
influence the shaping of planetary nebulae (PNe). Therefore, binary
interaction physics plays a fundamental role in understanding stellar
evolution.

In this work we focus on the circumstellar environment (CE) of a
distinct class of evolved objects. All sample objects show a
characteristic strong near-IR excess
which can be explained as due to thermal emission of warm dust in the
close environment around the central binary.  The energy source in these
systems is an evolved low-mass object in its post-Asymptotic Giant
Branch (post-AGB) phase of evolution \citep{2003ARAAVanWinckel,
  2006AAdeRuyter}, with a typical spectral type between G and late
A. The characteristic spectral energy distribution (SED) in
these objects indicates the presence of a stable, compact, dusty reservoir in the form of a circumbinary disk
\citep[e.g.][]{2006AAdeRuyter, 2007AADeroo, 2014AAHillen, 2015AAHillen}.
  The Keplerian rotation of the circumstellar
structure was already resolved in two objects \citet{2005AABujarrabal,
2015AABujarrabal} and recently much more detailed maps were made
with the Atacama Large Millimetre Array (ALMA) by
\citet{2013AABujarrabalC}.  The single-dish CO line survey of
\citet{2013AABujarrabalB} confirmed, by the detection of very narrow CO
line profiles, that rotation is likely widespread, which
is a strong observational indicator of long-term stability. Other indications of longevity are 
the observational evidence of strong dust grain processing in the form of a high degree of crystallinity 
\citep[e.g.][]{2008AAGielen,2011AAGielen} and the presence of large grains \citep[e.g.][]{2005AAdeRuyter}.

These specific SEDs can be intimately linked to the
binary nature of the central object. The luminous evolved component
has a likely unevolved companion with a very minor contribution to the
energy budget \citep[e.g.][]{2009AAVanWinckel, 2016AAHillen}. Based on the SED, new
such objects can be efficiently identified.  In recent searches for
post-AGB stars in the Large and Small Magellanic Clouds
\citep{2011AAVanAarle,2014MNRASKamath, 2015MNRASKamath}, disk sources
represent about half of the population of optically bright post-AGB
stars. Disks also appear at lower luminosities, indicating them to be
post-Red Giant Branch (post-RGB) stars, rather than post-AGB stars
\citep{kamath16}.
Evolved sources with these specific SED characteristics are very
common, and the formation, structure, and evolution of these
disks represents an important gap in our understanding of binary
interaction physics.

\begin{table*}   
  \caption{List of binary post-AGB stars in our interferometric survey. We give the coordinates together with the spectral type retrieved
    from Simbad. Most objects are confirmed binaries and some orbital parameters (orbital period, semi-major axis $a\sin(i)$, and mass
    function $f(M)$) are provided. We note for which targets the photometric 'RVb' phenomenon has been reported.}\label{tab:sample}
\begin{footnotesize}
\begin{tabular}{rlccccccccl} \hline \hline
\multicolumn{1}{c}{Nr.} &
\multicolumn{1}{c}{IRAS} &
\multicolumn{1}{c}{name} &
\multicolumn{1}{c}{$\alpha_{2000}$} &
\multicolumn{1}{c}{$\delta_{2000}$} &
\multicolumn{1}{c}{Spectral} &
\multicolumn{1}{c}{P$_{\rm{binary}}$} &
\multicolumn{1}{c}{a$_1$ sin($i$)} &
\multicolumn{1}{c}{$f(M)$} &
\multicolumn{1}{c}{RVb} &
\multicolumn{1}{c}{Ref.}\\
\multicolumn{1}{c}{} &
\multicolumn{1}{c}{} &
\multicolumn{1}{c}{} &
\multicolumn{1}{c}{[h m s]} &
\multicolumn{1}{c}{[\,\,$^o$\,\,'\,\,'']} &
\multicolumn{1}{c}{Type} &
\multicolumn{1}{c}{[days]} &
\multicolumn{1}{c}{[AU]} &
\multicolumn{1}{c}{[$M_{\odot}$]} &
\multicolumn{1}{c}{} &
\multicolumn{1}{c}{} \\
\hline
1 & 04440+2605 & RV\,Tau    & 04 47 06.7 &   +26 10 45.6 & K3pv    & $1180\pm15$   &       &       & y & \cite{1996AASZsoldos}  \\  
2 & 07008+1050 & HD\,52961  & 07 03 39.6 &   +10 46 13.1 & F6I     & $1297\pm7$    & 1.54  & 0.29  & y & \cite{1999AAVanWinckel} \\
3 & 07284-0940 & U\,Mon     & 07 30 47.5 & $-$09 46 36.8 & K0IIb   & $\sim2600$    &       &       & y & \cite{2006MmSAIPollard}\\
4 & 08011-3627 & AR\,Pup    & 08 03 01.6 & $-$36 35 47.9 & F0I     & $1250\pm300$  &       &       & y & \cite{2007MNRASKiss}\\
5 & 08544-4431 &            & 08 56 14.2 & $-$44 43 10.7 & F3      & $508\pm2$     & 0.39  & 0.03  & - & \cite{2003AAMaas} \\
6 & 09256-6324 & IW\,Car    & 09 26 53.3 & $-$63 37 48.9 & F7I     & $\sim1440$    &       &       & y & \cite{1996MNRASPollard} \\
7 & 10158-2844 & HR\,4049   & 10 18 07.6 & $-$28 59 31.2 & A4Ib/II & $434\pm1$     & 0.61  & 0.16  & y & \cite{1991AAWaelkensB}\\
8 & 10174-5704 &            & 10 19 16.9 & $-$57 19 26.0 & -       & $323\pm50$    & 0.17  & 0.007 & -  & \cite{2003PhDTMaas} \\             
9 & 10456-5712 & HD\,93662  & 10 47 38.4 & $-$57 28 02.7 & K5      & $572\pm6$     & 0.015 & $1.4\times10^{-6}$ & - & \cite{2003PhDTMaas}\\ 
10& 11385-5517 & HD101584   & 11 40 58.8 & $-$55 34 25.8 & A6Ia    & $218\pm1$\tablefootmark{a} &       &       & y & \cite{1996AABakker} \\
11& 12185-4856 & SX\,Cen    & 12 21 12.6 & $-$49 12 41.1 & G3V     & $592\pm13$    & 1.23  & 0.70  & y & \cite{2006AADeroo} \\
12& 12222-4652 & HD\,108015 & 12 24 53.5 & $-$47 09 07.5v& F4Ib/II & $914\pm4$     & 0.29  & 0.004 & -  & \cite{2009AAVanWinckel} \\
13& 15469-5311 &            & 15 50 43.8 & $-$53 20 43.3 & F3      & $390\pm1$     & 0.42  & 0.07  & -  & \cite{2009AAVanWinckel}\\
14& 17038-4815 &            & 17 07 36.6 & $-$48 19 08.6 & G2p     & $1386\pm12$   & 1.5   & 0.22  & y  & Manick et al., subm. \\
15& 17243-4348 & LR\,Sco    & 17 27 53.6 & $-$43 50 46.3 & G2      & $\sim475$     &       &       & -  & \cite{2003PhDTMaas}\\
16& 17534+2603 & 89\,Her    & 17 55 25.2 &   +26 03 00.0 & F3      & $288\pm1$     & 0.08  & $8\times10^{-4}\,$ & - & \cite{1993AAWaters} \\
17& 18281+2149 & AC\,Her    & 18 30 16.2 &   +21 52 00.6 & F2Iep   & $1194\pm 6$   & 1.39  & 0.25  & -  & \cite{1998AAvanWinckel} \\
18& 19125+0343 &            & 19 15 01.2 &   +03 48 42.7 & F2      & $520\pm2$     & 0.58  & 0.097 & -  & \cite{2009AAVanWinckel}\\
19& 22327-1731 & HD\,213985 & 22 35 27.5 & $-$17 15 26.9 & A0III   & $258.6\pm0.3$ & 0.79  & 0.97  & y & \cite{1995ApSSWaelkens} \\
\hline
\end{tabular}
\tablefoot{\tablefoottext{a}{\citet{2007IAUSDiaz} reported a period of 144~d on an IAU symposium.}} 
\end{footnotesize}
\end{table*}

The inner rim is an important structural component of a post-AGB disk, just like in young stellar objects (YSO).
The rim captures the radiation of the central source and gas pressure makes that the scale height becomes large. 
A significant fraction of the stellar luminosity is re-radiated by the disk in the infrared \citep[e.g.][]{2010ARAADullemond}.
As the disks are compact, interferometric tools are needed to resolve them.
This is nicely illustrated by our recent study of one of the evolved binaries, IRAS08544-4431 \citep{2016AAHillen}.
Using an interferometric image reconstruction technique,
\citet{2016AAHillen} resolved the inner circumstellar environment of IRAS08544-4431. The image, with a spatial resolution of 1.3 mas,
shows that the inner rim of the disk is indeed at the sublimation radius. Moreover, a contribution of the companion star
is spatially resolved and likely originating from a small gas accretion disk \citep{2016AAHillen}.

Here we aim at constraining the disk properties for a significant number of stars.
We report on the first N-band interferometric survey of these
binary post-AGB stars, providing direct insight into the spatial
structure of the circumstellar environment. The observations were
performed with the MID-infrared Interferometric instrument \citep[MIDI,][]{2003SPIELeinert} on
the Very Large Telescope Interferometer \citep{2012SPIEHaguenauer}. MIDI
operates in the N-band, which is where the circumstellar environment emits the peak of its
energy. Several case studies were presented in previous papers
\citep[e.g.][and references therein]{2014AAHillen, 2015AAHillen}.
These results inspired our survey of 19 objects. The intent of this paper is to
gain insight into the overall sample characteristics and study the
typical physical scales of the disks. The challenge of this work is that many objects 
are observed, but each with a limited baseline coverage. 
Recently, similar surveys with MIDI were 
published about the torii in active galactic nuclei \citep{2013AABurtscher}, the structures around massive 
YSOs \citep{2013AABoley}, and the circumstellar disks around Herbig Ae/Be stars \citep{2015AAMenu}.
Our methodology is heavily inspired by the last, which allows us to make a direct comparison between the two samples.

The paper is organized as follows: first, we describe the different
programme stars (Sect.~\ref{section:sampledescription}) and their data (Sect.~\ref{section:datadescription}). 
We report on our construction of detailed SEDs for all objects in Sect.~\ref{section:SEDfitting} and on the fitting of 
the interferometric observables in Sect.~\ref{section:modelfitting}. The results on some individual 
objects are highlighted in Sect.~\ref{section:individualtargets} and
we analyse the sample as a whole in Sect.~\ref{section:analysis}. We confront the interferometric observables 
with the results of a grid of radiative transfer models in Sec.~\ref{section:RTgrid} and focus on comparing the circumbinary 
structures around our evolved binaries with the circumstellar structures in young stellar objects.
Finally, we discuss the implications of our results and conclude the paper in Sect.~\ref{section:discussion}.

\section{Description of the sample}\label{section:sampledescription}
The sample of Galactic post-AGB stars for which we obtained N-band interferometric data is listed in 
Table~\ref{tab:sample}. The sample of 19 stars was selected from the catalogue of
\citet{2006AAdeRuyter} of post-AGB objects with a suspected circumstellar disk, with the addition of HD101584 \citep{1996AABakker}. 
The primary criteria to select the science targets to be observed with MIDI were the estimated correlated N-band flux-level as well as the accessibility 
from the Paranal observatory. We have used both the Unit Telescopes (UT, 8~m diameter) and the Auxiliary Telescopes (AT, 1.8~m diameter), 
allowing the inclusion of targets with both moderate and high N-band fluxes.
About half of our sources were observed within the Belgian Guaranteed Time programme (VISA) on the ATs.

There is good observational evidence that all 19 objects are indeed binaries. The orbital elements themselves have been determined for 
13 of them. For U\,Mon, LR\,Sco and IRAS\,10174-5704 the radial velocity variability is linked to binary motion but the
full orbital elements have not yet been quantified.

For some sources, a photometric time series can constrain the orbital period as well. In these objects the total line-of-sight
extinction is varying with the orbital motion of the primary and this induces a variable line-of-sight reddening, in phase with the orbital
period. This method has been widely applied to this type of source
\citep[e.g.][]{1985MNRASEvans,1987AAWaelkens,1999AAVanWinckel,2003AAMaas,2007MNRASKiss}
and is employed here to establish the likely orbital period of AR\,Pup, RV\,Tau and IW\,Car. Since this photometric phenomenon was
discovered for RV\,Tauri stars and labeled the `RVb phenomenon', we use the same terminology, irrespective of whether the object is really an
RV\,Tauri pulsator. We list whether this phenomenon was detected in a lightcurve in Table~\ref{tab:sample}. 
It is an important qualifier for a system, because it strongly constrains the inclination of the
circumstellar structure: when the 'RVb phenomenon' is observed, the
system must be seen under an inclination such that the line-of-sight
grazes the circumbinary disk and the line-of-sight reddening changes
during the orbital motion.

\section{The data}\label{section:datadescription}
\subsection{Interferometric data}\label{subsection:MIDI}

We do not list the detailed log of every observation, but graphically present 
the UV coverage for every source in Sect.~\ref{section:modelfitting} along with the results of our model fitting.

For a detailed account of the data reduction and calibration strategy, we refer to \citet{2015AAMenu}.
In short, the data processing is done in two steps: 1) an extraction of raw correlated and total fluxes from
the recorded data frames, and 2) the calibration of these raw fluxes into the science-worthy observables.
The first data reduction step is performed with the 2.0 version of the Expert Work Station (EWS) software \citep{2004SPIEJaffe},
using a Python wrapper to extract observation by observation of both science and calibrator targets.
The second step consists of determining the wavelength-dependent interferometric transfer function to calibrate the 
science observations from the interferometer's response to a point source. Science targets are observed in between calibrators to 
estimate the transfer function, which is subsequently interpolated to the times of the science observations. 

In this paper we make use of the absolute correlated fluxes instead of the classically used visibilities. 
The visibility is equal to the correlated flux normalized with the zero-baseline (i.e. total) flux.
Operating in the mid-IR, the zero-baseline flux measurements with MIDI are limited by the background subtraction, 
unlike the correlated flux measurements because of the way MIDI is designed to 
record the mid-IR interferogram \citep[for more details, see ][]{2015AAMenu}. Hence, if the calibration is able 
to capture the effective transfer function that corresponds to the science measurement, then correlated fluxes 
are preferred over visibilities. In contrast, the calibration of visibilities is principally more robust because
effects that affect correlated and zero-baseline fluxes equally are divided away. 
Generally, we believe this is the case for our sample. For a few measurements 
we have doubts about the reliability of the transfer function estimation, and these are not included in our further analysis.
For the estimation of the zero-baseline flux we combine the available MIDI fluxes (if reliable) with external 
measurements already published in the literature.

Although MIDI is a spectro-interferometric
instrument, originally designed for studies of the amorphous and crystalline N-band dust features, 
in this paper we focus on a single wavelength in-between the peaks of these features, 10.7~$\mu$m.
The calibration step makes that the data are correlated in wavelength (e.g., due to seeing, etc.). 
Taking these correlations into account would complicate our analysis significantly, with limited gain. 
Moreover, at the long wavelength end of the N band the measured fluxes become increasingly unreliable due to the increase in 
background noise. 
We choose the same reference wavelength of 10.7~$\mu$m as in the interferometric surveys already mentioned previously
\citep[][]{2015AAMenu,2013AABoley,2009ApJMonnier}. This allows us to compare our results to these young stellar object 
disk samples. Our correlated fluxes are the weighted average of the measured fluxes between 10.55 and 10.85~$\mu$m, including 
only the error term due to photon noise in the weights. For the total error, the resulting photon noise term is added in quadrature with the 
error due to the calibration.

\begin{figure*}
   \centering
   \includegraphics[width=18cm]{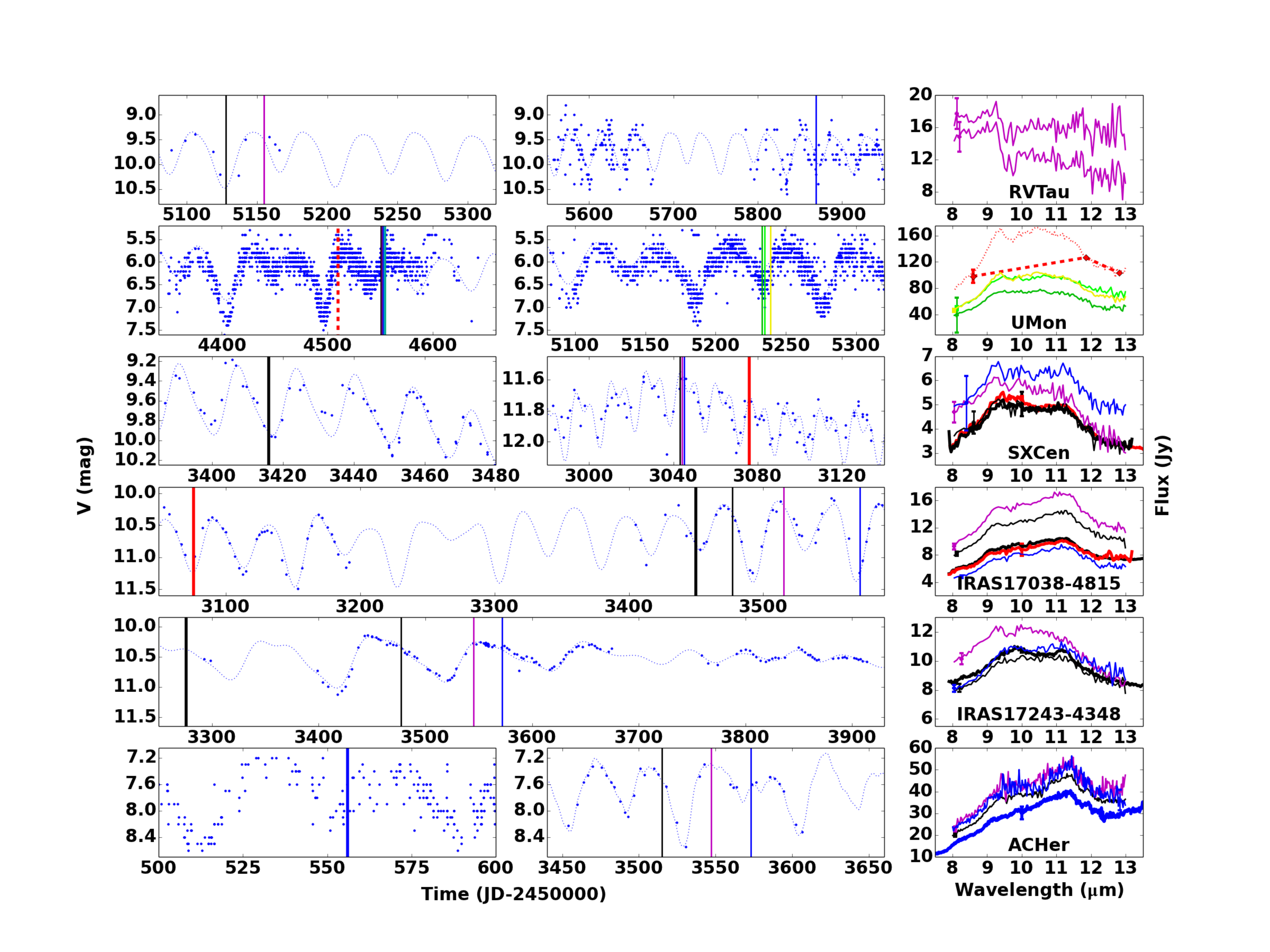}
   \caption{Left panels: The AAVSO and ASAS V band light curves of the six stars with pulsation-related flux variations larger 
   than 0.5 mag. The blue dotted line is a fit to the full light curve as described in Sect.~\ref{section:variability}, only shown to guide the eye.
   The coloured vertical lines indicate the time of observation of the mid-IR spectra shown in the right panels. Thick black, thick blue and 
   thick red lines represent data from Spitzer, ISO and Timmi2, respectively. The thick red dashed line indicates the three-filter VISIR photometry
   of U Mon \citep[published by][]{2011MNRASLagadec}. The thin coloured lines represent our MIDI measurements at various 
   epochs. Right panels: The measured mid-IR fluxes for the indicated epochs in the left panels. The minimum error over the shown wavelength 
   range is indicated for each spectrum with a bar. For RV Tau and U Mon, a few very noisy MIDI spectra are not included in the 
   right panel. The red dotted line in the right panel of U Mon is the orange spectrum scaled to the VISIR fluxes.}
   \label{figure:lightcurves}
\end{figure*} 

\subsection{Additional mid-IR spectra}\label{subsection:spectra}
Infrared spectra of most sources were discussed with the focus
on the solid state band profiles as probes for the dust grain
processing \citep{2007AAGielen, 2009AAGielen, 2011AAGielen}. We re-use
these data here to define the slope of the continuum in the N-band
outside the silicate feature. For these spectra an error of 15\% on the absolute flux calibrations is assumed.
Two of our sources, RV\,Tau and U\,Mon, were not included in the sample of \citet{2011AAGielen}. For U\,Mon, we therefore 
also include the mid-IR fluxes measured by \citet{2011MNRASLagadec} in our analysis (see Sect.~\ref{section:variability}).
These fluxes were extracted from unresolved, but diffraction-limited, images in three narrowband filters (centred at 
8.59, 11.85, and 12.81~$\mu$m) with the VISIR instrument \citep{2004MsngrLagage} on the Very Large Telescope.

\subsection{Mid-IR variability} \label{section:variability}
Many of our targets show pulsational variability, often with very high amplitudes. In this section we 
consider whether some of the temporal variation in the MIDI spectra may be related to the stellar pulsation,
and hence affect the interpretation of the interferometric observables.
Not much is known about variability in the mid-IR of post-AGB disk sources, except the claims for a sine-like variation in the 
10~$\mu$m fluxes of AC Her by \citet{1971PhDTGehrz} and \citet{1992AAShenton}. Table~\ref{tab:pulsationprop} lists the pulsation
properties of all our sample stars. We list whether the star is identified as a pulsator, and whether the pulsation is of RV Tau type. 
Objects for which no pulsation period is known in the literature are listed as pAGB. We also give typical peak-to-peak amplitudes of
V band brightness variations that are attributed to pulsations. Brightness variations of the RVb type are not taken into 
account here, since we attribute these to variable line-of-sight extinction during binary motion.
Intrinsic mid-IR variability is more likely the result of a coupling between the stellar luminosity variation and the inner disk emissivity.

\begin{table}   
  \caption{Pulsation properties of the sample stars. Objects in which the characteristic RV Tau features have been identified 
  in the light curve are listed as RV Tau. Otherwise, if a pulsation period has been detected it is listed as 'Puls.', and if no 
  period is known in the literature the source is denoted as 'pAGB'. The (maximum) amplitude of pulsational variability in the V band is 
  given, as well as the main pulsation period wherever relevant. }\label{tab:pulsationprop}
\begin{footnotesize}
\begin{tabular}{rlcccc} \hline \hline
\multicolumn{1}{c}{Nr.} &
\multicolumn{1}{c}{IRAS} &
\multicolumn{1}{c}{type} &
\multicolumn{1}{c}{$\Delta$V$_{\rm{puls}}$\tablefootmark{a}} &
\multicolumn{1}{c}{P$_{\rm{pulsation}}$\tablefootmark{b}} &
\multicolumn{1}{c}{Ref.} \\
\hline
1 &04440+2605 & RVTau & 1.2   & 78.7 & 1,2 \\  
2 &07008+1050 & Puls. & 0.15  & 72$\pm$3 & 3 \\
3 &07284-0940 & RVTau & 1.1   & 92$\pm$3 & 4,5 \\
4 &08011-3627 & RVTau & 0.5   & 76$\pm$4 & 4 \\
5 &08544-4431 & Puls. & 0.15  & $\sim$72 & 4,6 \\
6 &09256-6324 & Puls. & 0.45  & 72$\pm$1& 4 \\
7 &10158-2844 & pAGB  & $<0.25$ & - & 7 \\
8 &10174-5704 & pAGB  & $<0.10$ & - & 4 \\ 
9 &10456-5712 & pAGB  & $<0.25$ & - & 4 \\
10&11385-5517 & pAGB  & $<0.10$ & - & 8 \\
11&12185-4856 & RVTau & 0.6  & 32.8642 & 9 \\
12&12222-4652 & Puls. & 0.15  & $\sim$60 & 4,10 \\
13&15469-5311 & Puls. & 0.15  & $\sim$50 & 4 \\
14&17038-4815 & RVTau & 1.50  & 37.9$\pm$1.5 & 11 \\
15&17243-4348 & RVTau & 0.8   & 100$\pm$4 & 4,12 \\
16&17534+2603 & Puls. & $<0.25$ & 65$\pm$3 & 13,14 \\
17&18281+2149 & RVTau & 1.8   & 75$\pm$2 & 10 \\
18&19125+0343 & pAGB  & $<0.15$ & -        & 4 \\
19&22327-1731 & pAGB  & $<0.05$ & -        & 4 \\
\hline
\end{tabular}
\tablefoot{\tablefoottext{a}{Peak-to-peak amplitude.}
           \tablefoottext{b}{The main pulsation period.}
           \tablefoottext{1}{\citet{2009AstLTaranova}}
           \tablefoottext{2}{\citet{1996AASZsoldos}}
           \tablefoottext{3}{\citet{1991AAWaelkens}}
           \tablefoottext{4}{\citet{2007MNRASKiss}}
           \tablefoottext{5}{\citet{1998JAVSOPercy}}
           \tablefoottext{6}{\citet{2003AAMaas}}
           \tablefoottext{7}{\citet{1991AAWaelkensB}}
           \tablefoottext{8}{\citet{1996AABakker}}
           \tablefoottext{9}{\citet{1994AAShenton}}
           \tablefoottext{10}{\citet{2009yCatSamus}}
           \tablefoottext{11}{Manick et al, subm.}
           \tablefoottext{12}{\cite{2003PhDTMaas}}
           \tablefoottext{13}{\citet{2011MNRASDubath}}
           \tablefoottext{14}{\citet{2000PASPPercy}}
}
\end{footnotesize}
\end{table}

We identify six targets in our sample with amplitudes in the visual light curve larger than 0.5 mag ($>35\%$ in flux variation). 
These are the candidates for which variable mid-IR emission  may be detectable in the MIDI data, in response to the photospheric variations of the central luminous star. 
To check this hypothesis, we downloaded the light curves for these sources, from the ASAS catalogue 
\citep{2003AcAPojmanski,2004AcAPojmanski,2005AcAPojmanski} if available, and otherwise 
from the AAVSO. AAVSO data are only used for the final MIDI epoch of RV\,Tau, all epochs of U\,Mon and the ISO epoch of AC\,Her.
In all other cases we prefer the ASAS data due to their superior quality. 

Fig.~\ref{figure:lightcurves} shows the resulting light curves, with the epochs of our mid-IR observations indicated by coloured vertical lines. To guide the eye, we include a fit to each of the visual light curves with frequencies determined from a Fourier analysis. The mid-IR spectra corresponding to these epochs are shown in the right panels, using the same colour and line-style coding. For each spectrum, the error is indicated at the wavelength with the smallest error in the 8-13~$\mu m$ wavelength range.

A close inspection of Fig.~\ref{figure:lightcurves} shows that the epochs of observations are not in extrema of the lightcurve for most stars, except for U\,Mon and IRAS17038-4815. For U\,Mon, the VISIR fluxes were taken near a pulsation maximum,
and are significantly larger than the high-quality MIDI spectra obtained at a secondary minimum. That affects the analysis of the MIDI correlated fluxes (see Sect.~\ref{section:modelfitting}).
The clearest case is IRAS17038-4815, however. For this source the quality of the MIDI spectra is very good and the temporal sampling is 
favourable (three near-minimum phases with significantly lower mid-IR fluxes than the two near-maximum phases).

In summary, for the remainder of this paper we consider the mid-IR fluxes as intrinsically constant, except for U\,Mon and IRAS17038-4815, which warrant a separate analysis at the different pulsation epochs.

\section{SED fitting} \label{section:SEDfitting}
Our interferometric model fitting in the next section requires two input parameters, the stellar effective temperature T$_{\rm{eff}}$ and 
the stellar angular diameter $\theta_\star$ (Eq.~\ref{eq:subradius}). Here we explain how we estimate both parameters. In addition, the 
results in this section allow us to estimate the stellar contribution to the N band emission.

\begin{figure*}
   \centering
   \includegraphics[width=17cm]{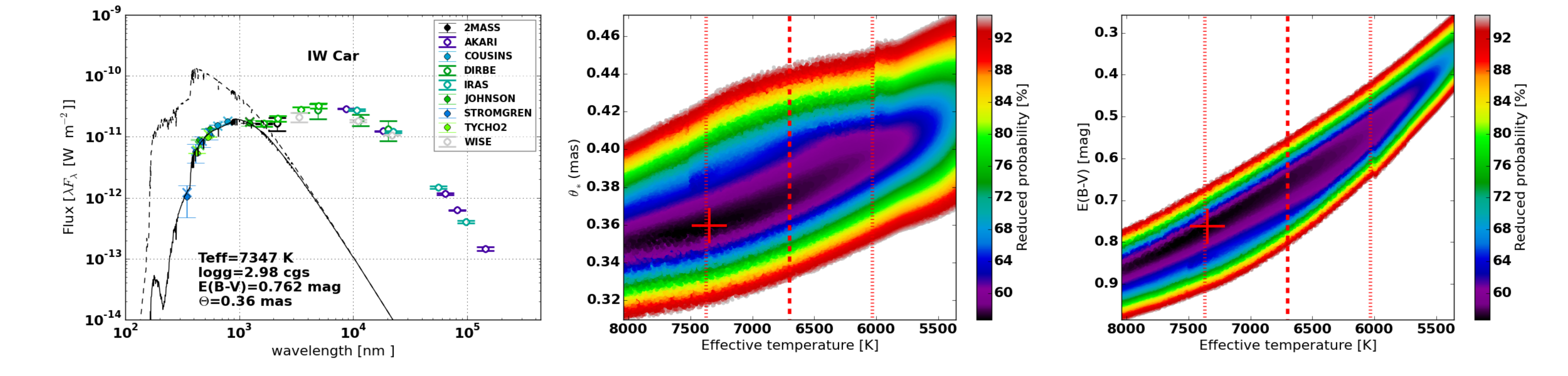}
   \caption{An example (IW Car) of an SED of a post-AGB star with negligible pulsational variability.
   The left panel shows the measured photometry (circles), which are filled if the data were used in the fitting process. The colour indicates 
   the photometric system, as listed in the legend. The full black line is the best-fit SED model that abides the spectroscopic 
   temperature constraint (see the red cross in the middle and right panels). The exact parameter values of this model are given in the figure. 
   The dashed black line is the same model, but unreddened. A consistent vertical scale, and 
   colour coding for the photometric systems, is used for all stars in the sample (see the online appendix).
   Middle panel: the confidence interval of the angular diameter $\theta_\star$ vs. the effective temperature T$_{\rm{eff}}$. 
   The dashed red line corresponds to the spectroscopically measured effective temperature. The dotted red lines indicate 
   the 3$\sigma$ lower and upper limit on this temperature.
   Right panel: the confidence interval of the reddening E(B-V) vs. the effective temperature T$_{\rm{eff}}$. 
   }
     \label{figure:SEDexample1}
\end{figure*} 

\begin{figure*}
   \centering
   \includegraphics[width=17cm]{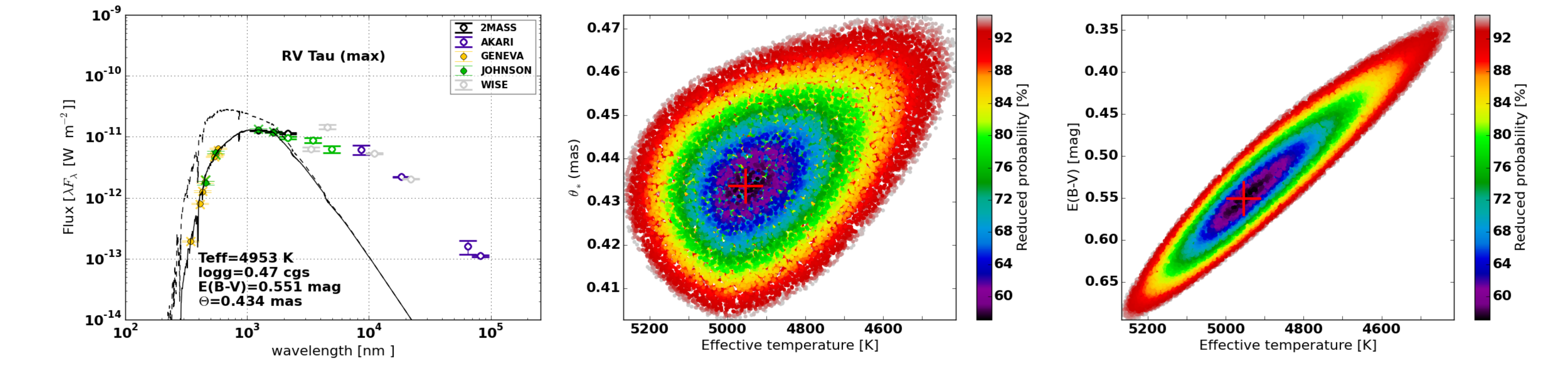}
   \includegraphics[width=17cm]{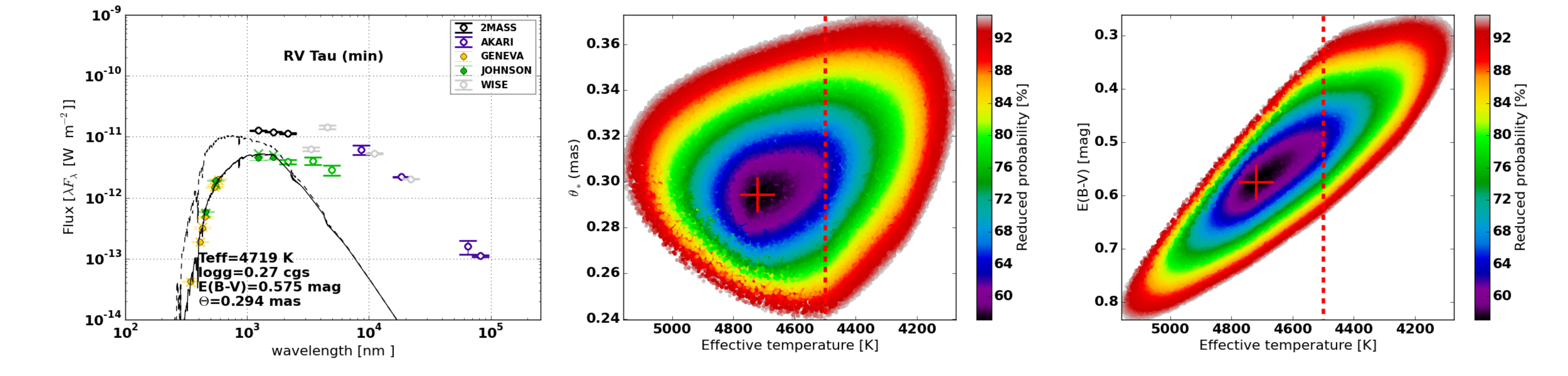}
   \caption{Same as Fig.~\ref{figure:SEDexample1} but for a post-AGB star with strong pulsational variability (RV Tau itself). The upper and 
   lower panels show optical and near-IR photometry at visual maximum and minimum pulsation phases, respectively. 
   For visual reference we also include photometry for which no proper phase attribution can be 
   made (mainly Tycho-2, 2MASS, and all mid- and far-IR data). In some cases these data may correspond to a different RVb phase.}
     \label{figure:SEDexample2}
\end{figure*} 

\onlfig{4}{
\begin{figure*}
   \centering
   \includegraphics[width=17cm]{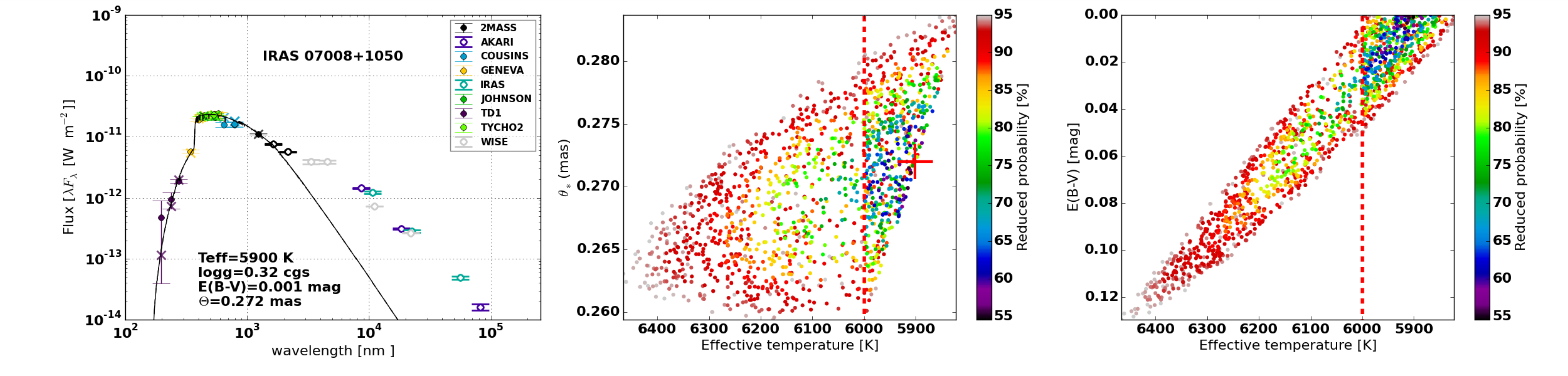}
   \includegraphics[width=17cm]{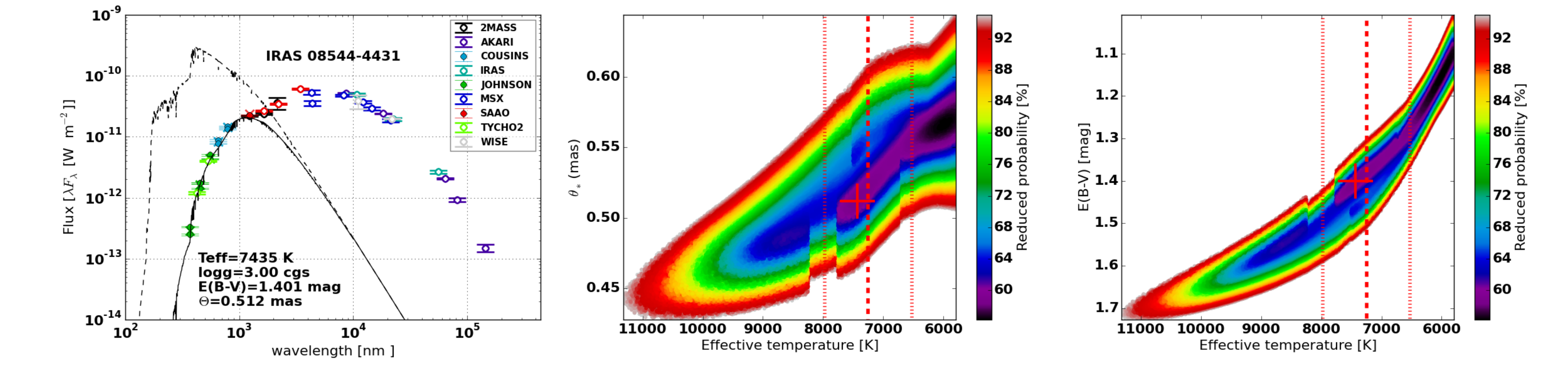}
   \includegraphics[width=17cm]{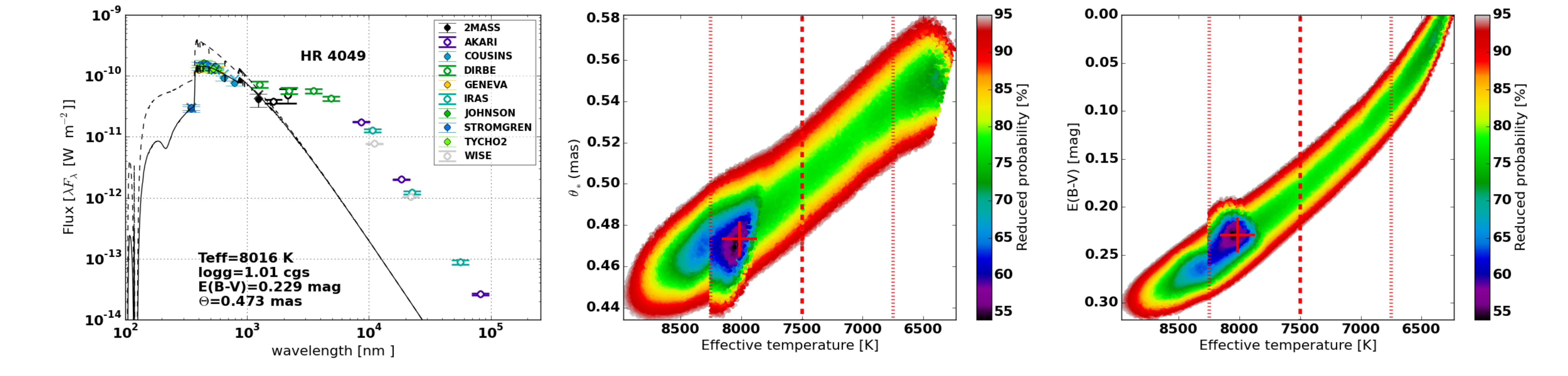}
   \includegraphics[width=17cm]{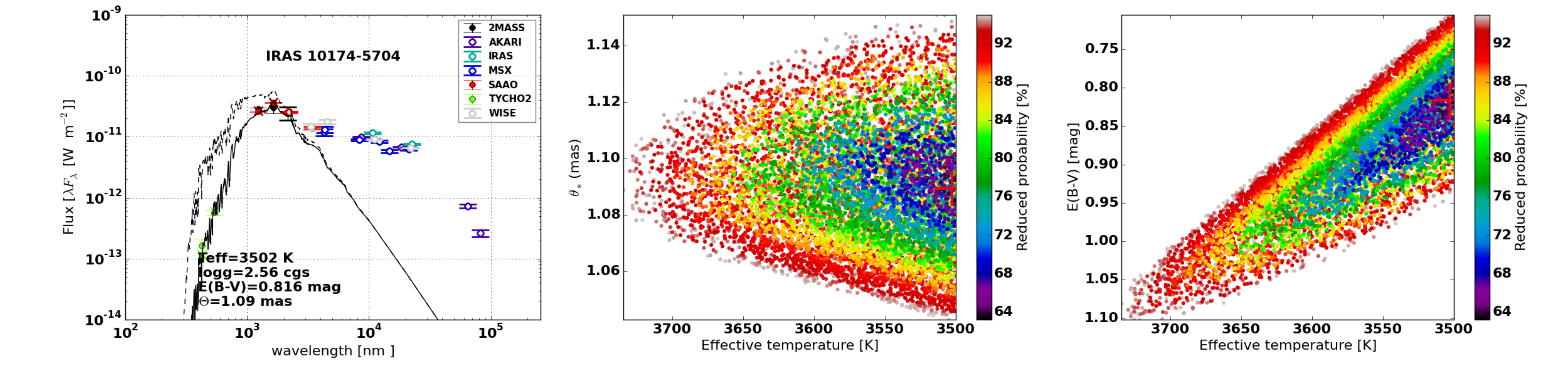}
   \includegraphics[width=17cm]{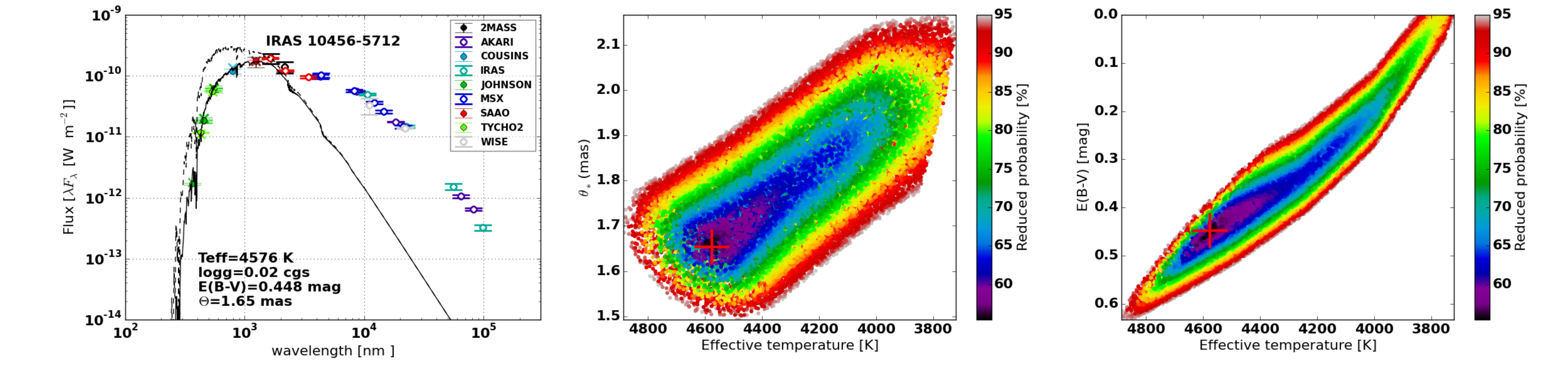}
   \includegraphics[width=17cm]{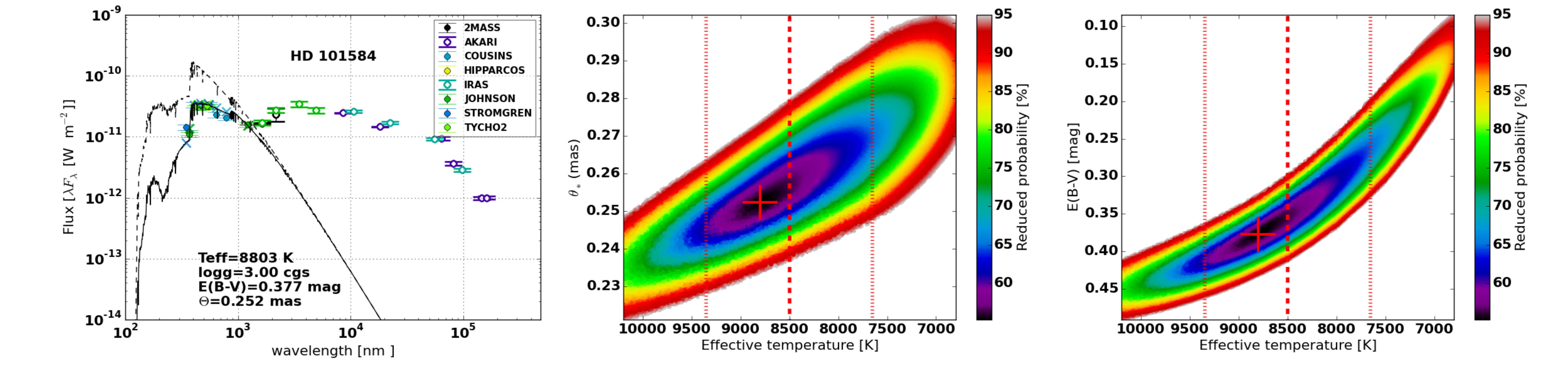}
   \caption{Same as Fig.~\ref{figure:SEDexample1} of the main text, for the other non-variable sources in our sample.}
   \label{figure:SEDexample3}
\end{figure*} 
}

\onlfig{5}{
\begin{figure*}
   \centering
   \includegraphics[width=17cm]{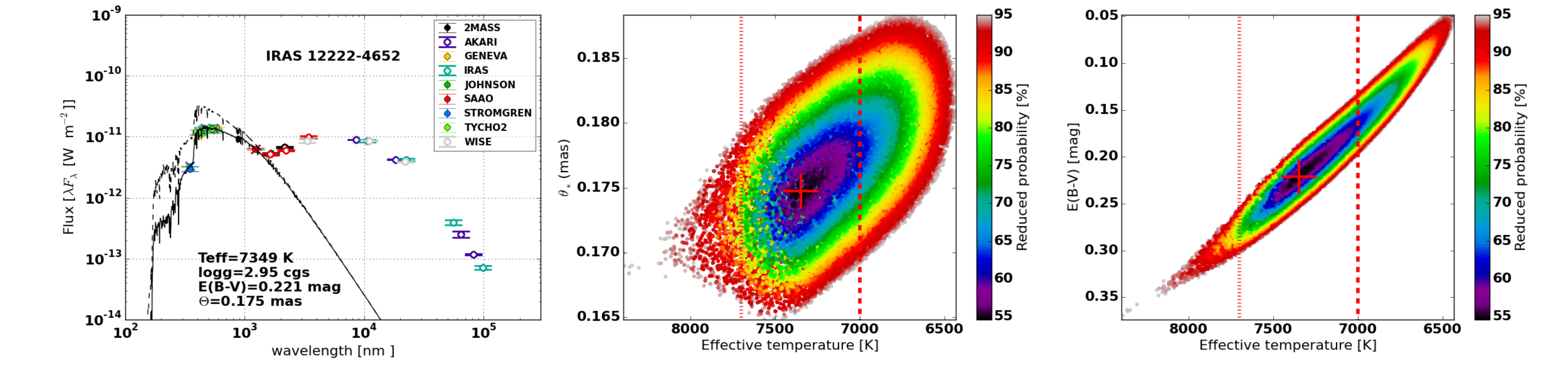}
   \includegraphics[width=17cm]{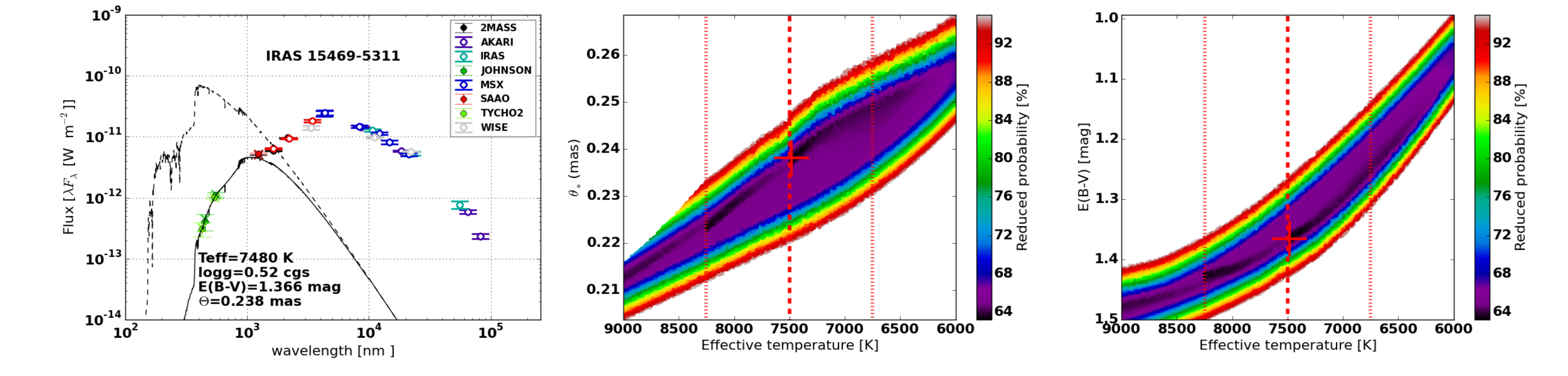}
   \includegraphics[width=17cm]{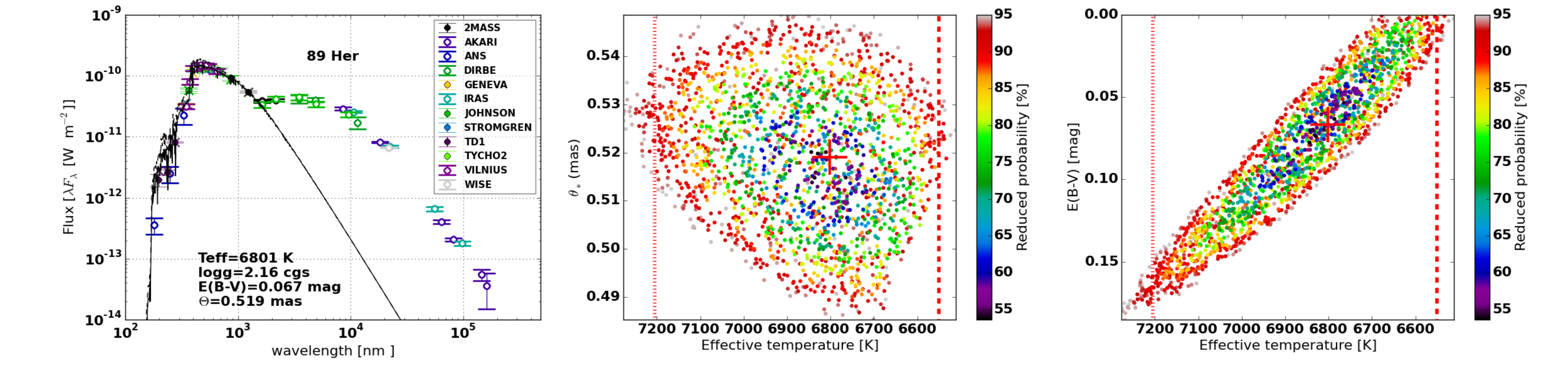}
   \includegraphics[width=17cm]{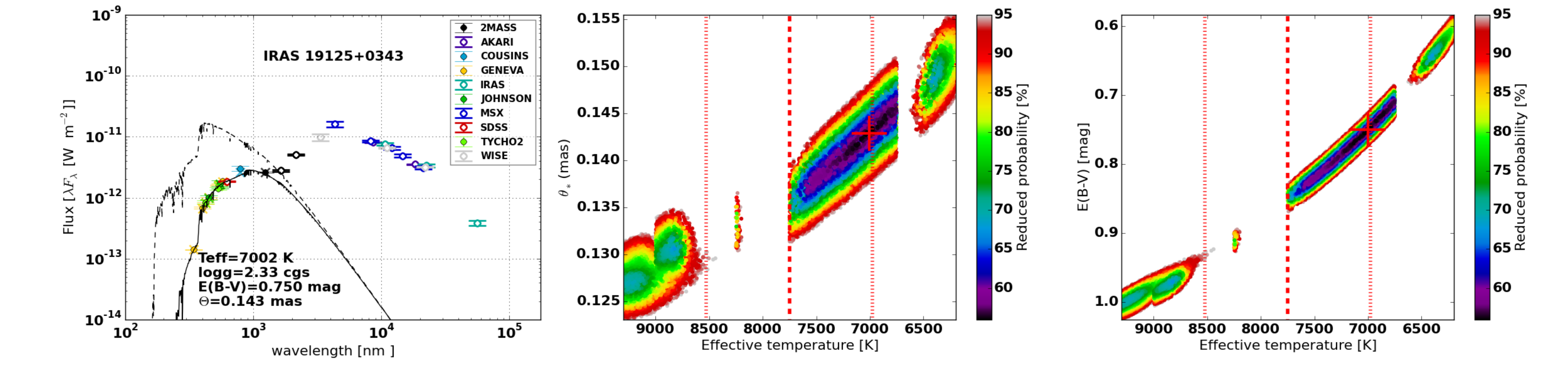}
   \includegraphics[width=17cm]{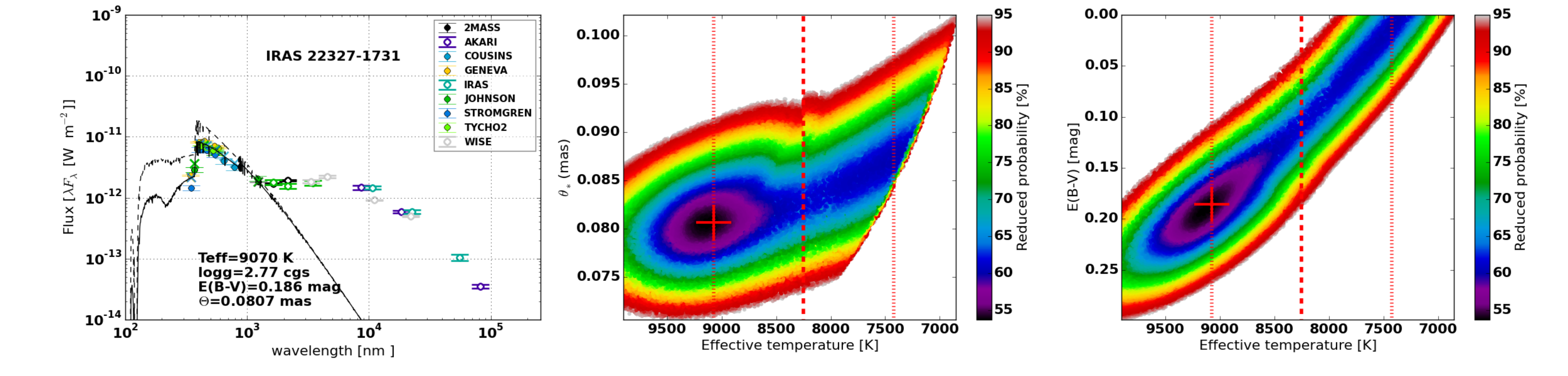}
   \caption{Same as Fig.~\ref{figure:SEDexample1} of the main text, for the other non-variable sources in our sample.}
   \label{figure:SEDexample4}
\end{figure*} 
}

\onlfig{6}{
\begin{figure*}
   \centering
   \includegraphics[width=17cm]{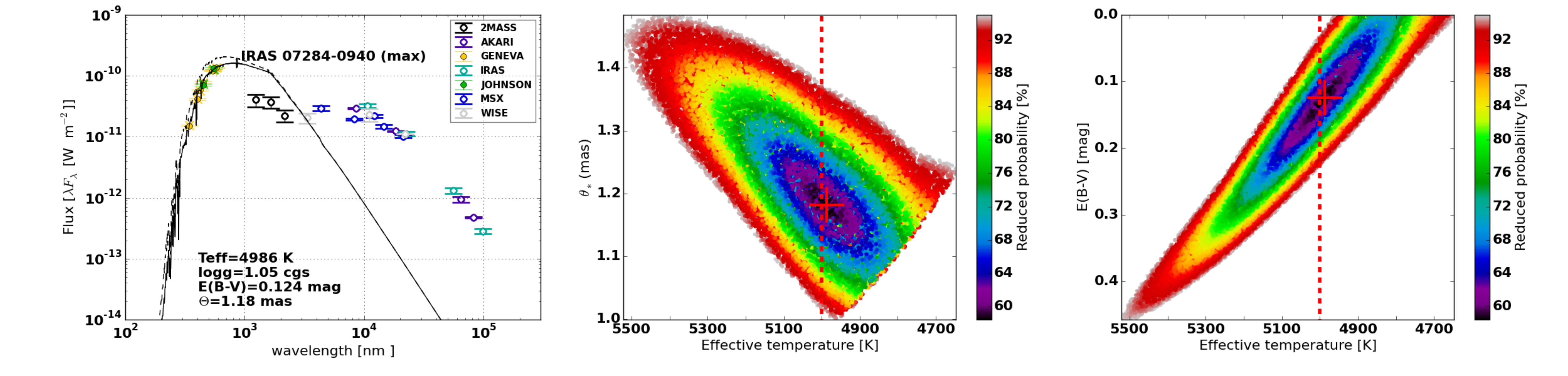}
   \includegraphics[width=17cm]{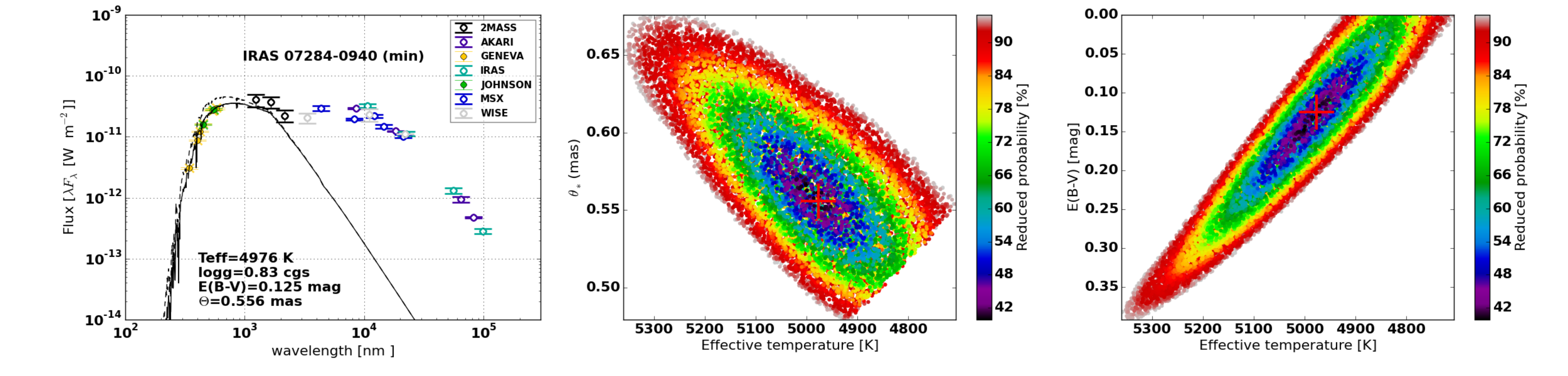}
   \includegraphics[width=17cm]{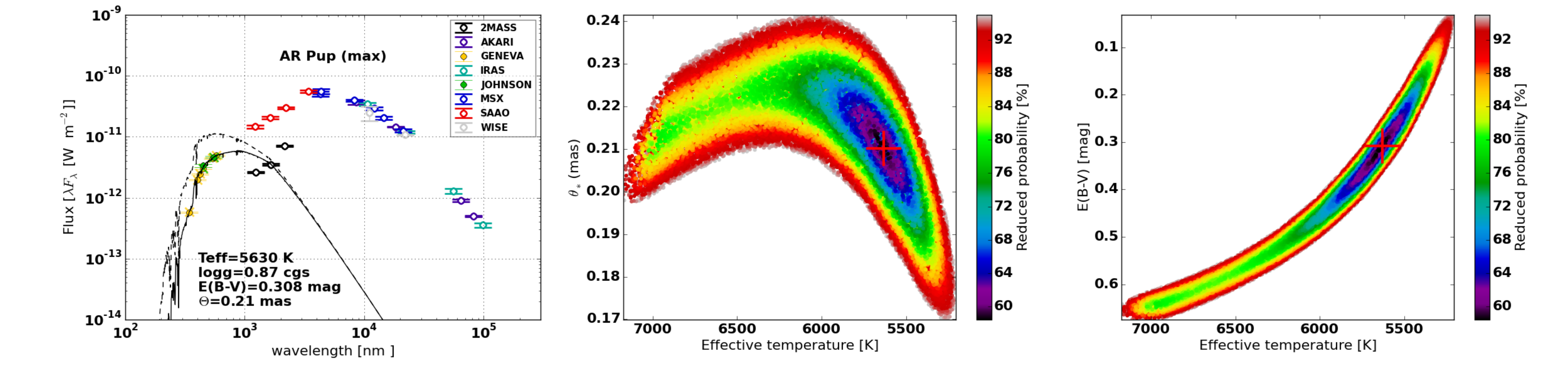}
   \includegraphics[width=17cm]{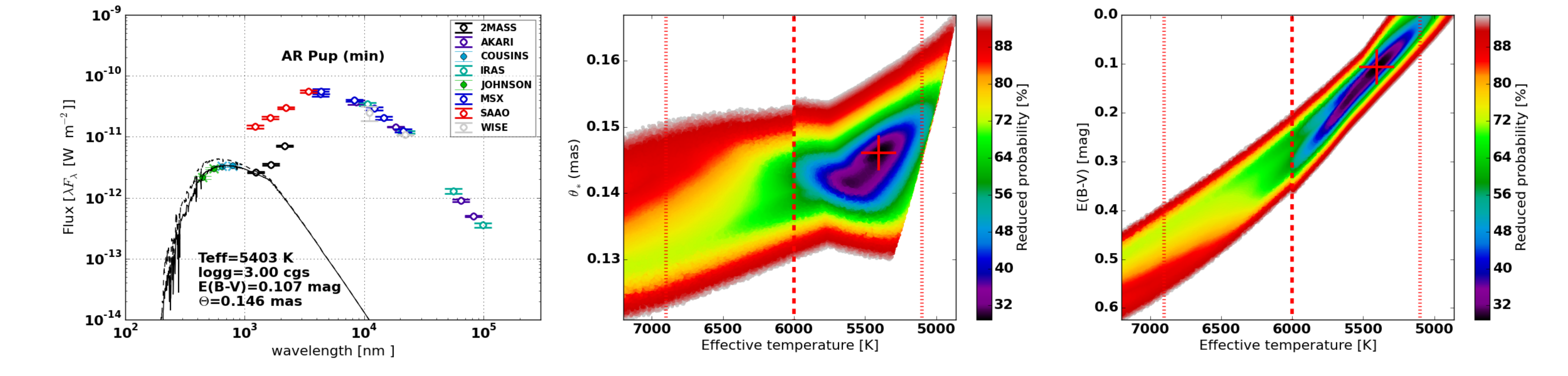}
   \includegraphics[width=17cm]{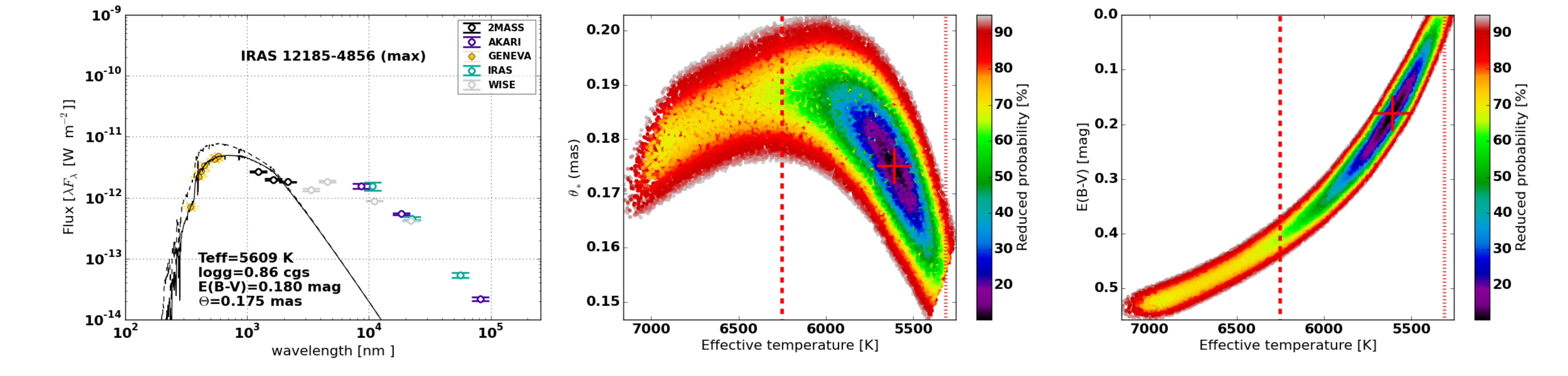}
   \includegraphics[width=17cm]{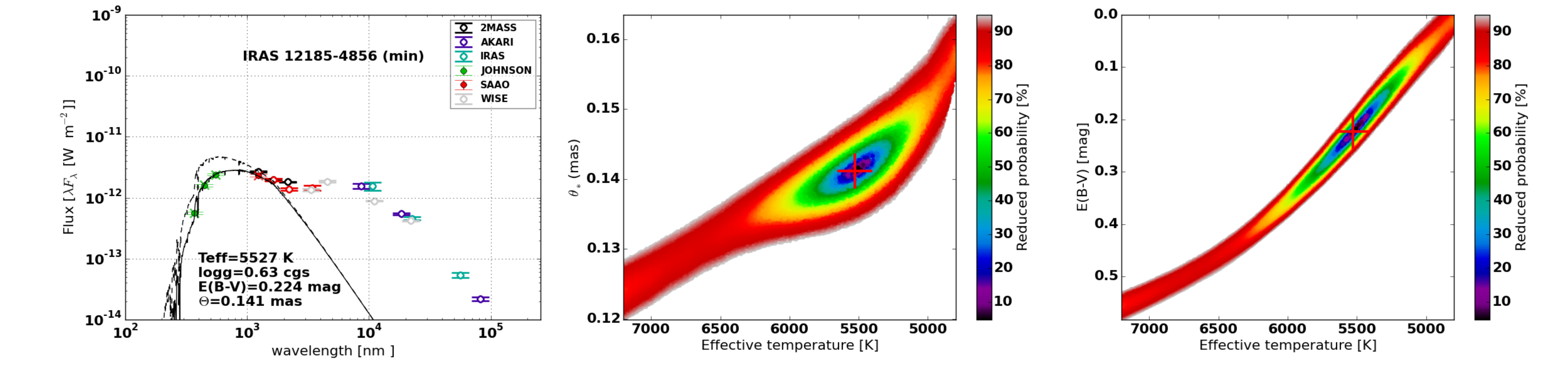}
   \caption{Same as Fig.~\ref{figure:SEDexample2} of the main text, for the variable sources in our sample.}
   \label{figure:SEDexample5}
\end{figure*} 
}

\onlfig{7}{
\begin{figure*}
   \centering
   \includegraphics[width=17cm]{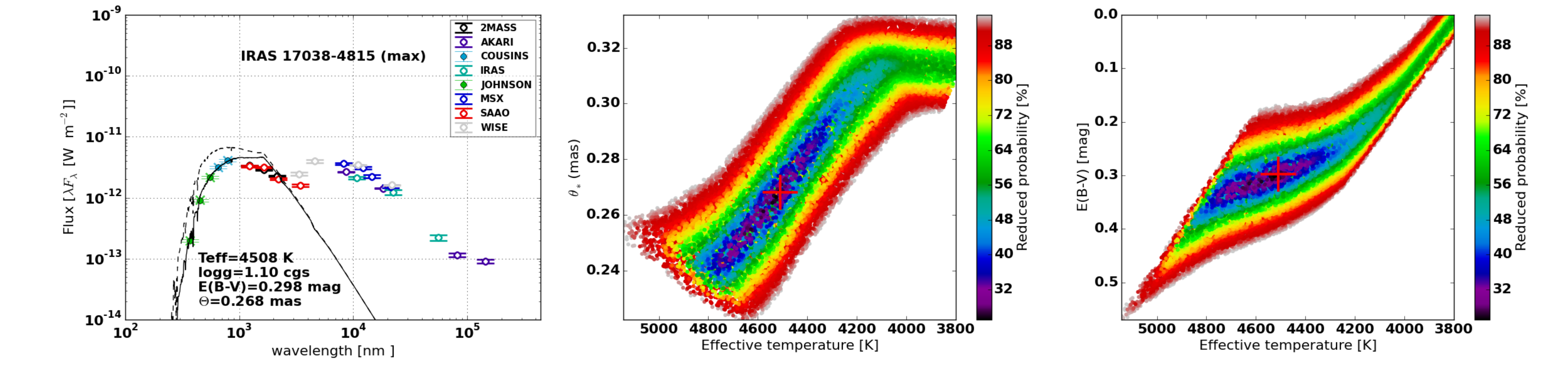}
   \includegraphics[width=17cm]{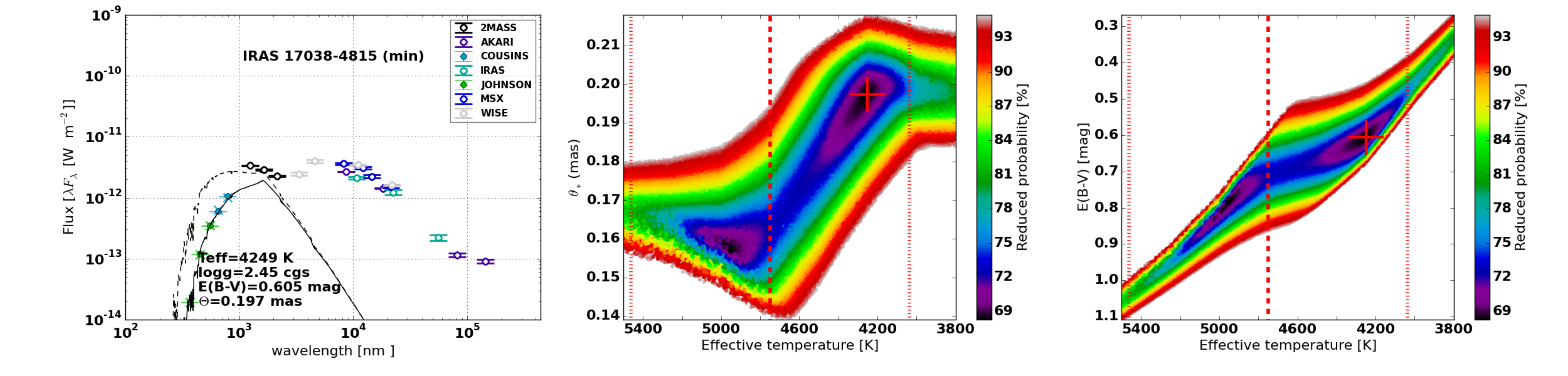}
   \includegraphics[width=17cm]{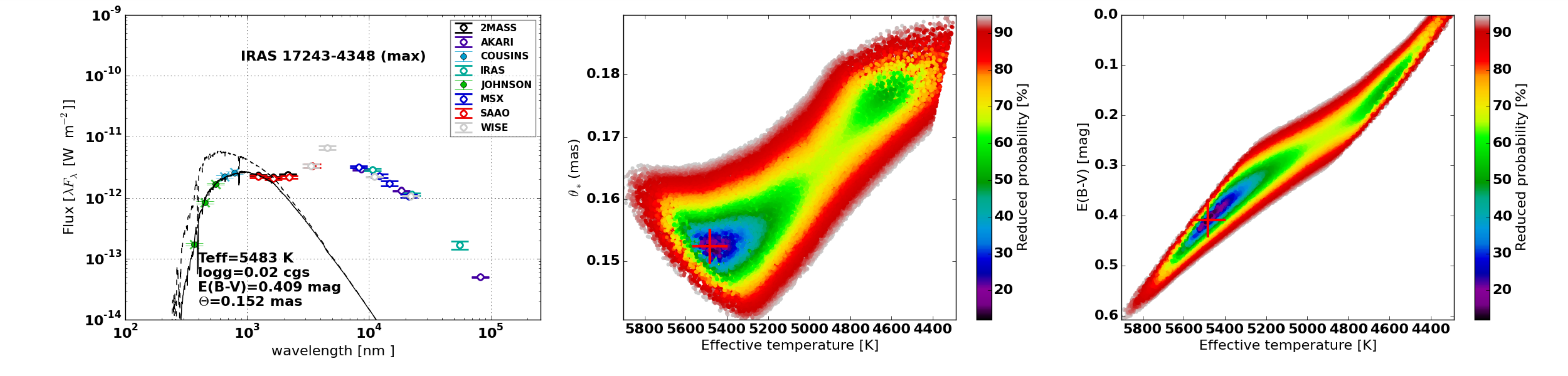}
   \includegraphics[width=17cm]{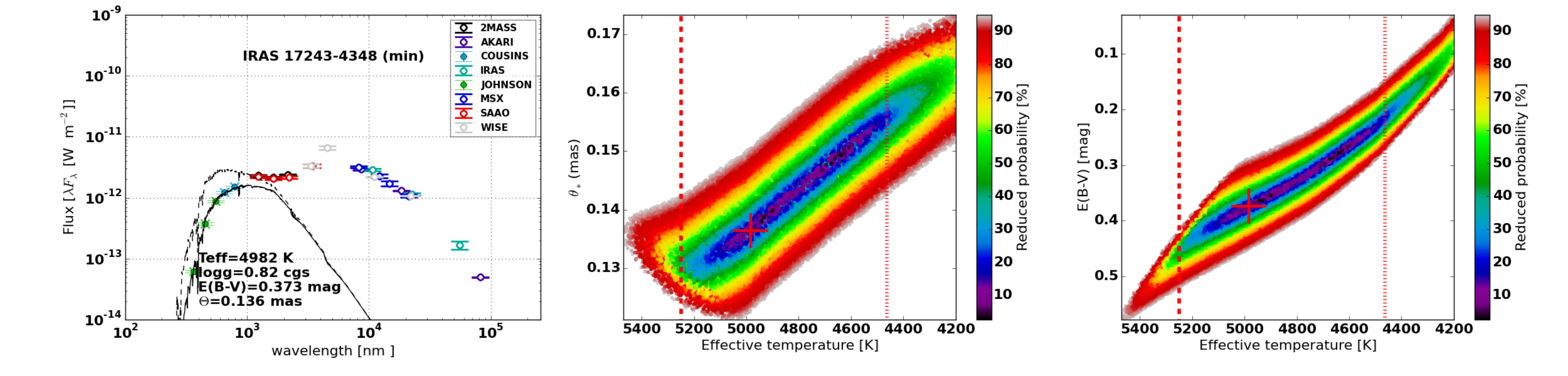}
   \includegraphics[width=17cm]{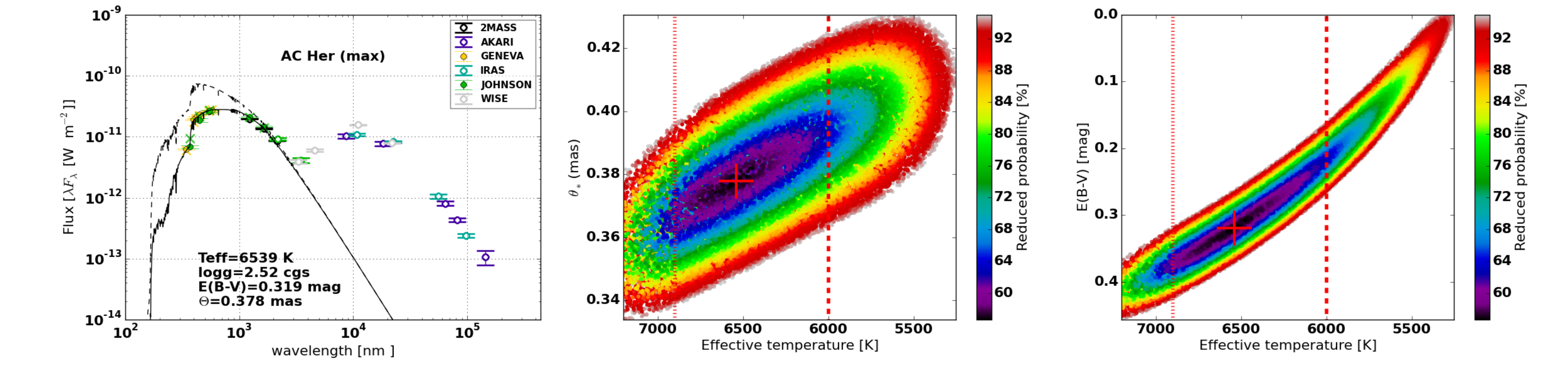}
   \includegraphics[width=17cm]{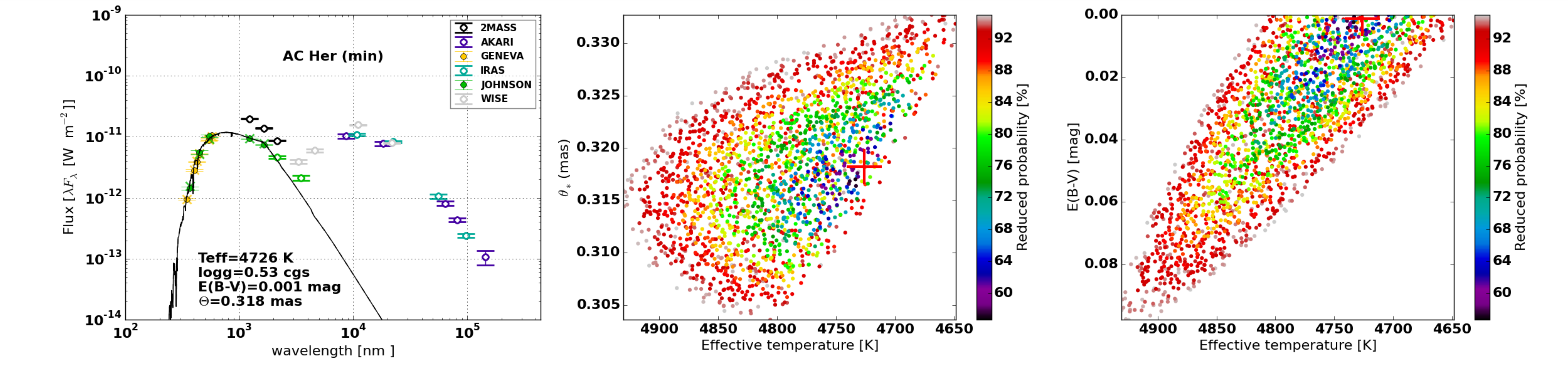}
   \caption{Same as Fig.~\ref{figure:SEDexample2} of the main text, for the variable sources in our sample.}
   \label{figure:SEDexample6}
\end{figure*} 
}

Since there are no published angular diameter measurements available for the sources in our sample, we resort to a uniform fitting 
of the stellar part of the SED. The fit parameters that determine a fully scaled stellar SED model at a single time are
the effective temperature T$_{\rm{eff}}$, the gravity $\log g$, the angular diameter $\theta_\star$, the reddening E(B-V), 
and the metallicity [Fe/H]. We use Kurucz model atmospheres since the effective temperatures of our stars are generally
above 3500~K. For one star, IRAS~10174-5704, this lower bound is reached in the fitting, but we do not attempt a more detailed analysis
as there is not sufficient photometry available for this source.


We keep the total reddening as a free parameter and assume that a
single reddening law \citep[the one of][]{2004ASPCFitzpatrick} can
represent the interstellar as well as the circumstellar contribution,
the latter is expected to be particularly important for the RVb sources. We fix
the metallicities based on the values listed in \citet{2006AAdeRuyter}
and  take the closest value available in our standard grid of Kurucz
model atmospheres, or assume a solar metallicity if no spectroscopic
determination is available.

The T$_{\rm{eff}}$, $\log g$, $\theta_\star$ and E(B-V) are
fitted with a grid-based method, as described in
\citet{2011AADegroote}. A reddened model is integrated over the
observed photometric passbands before scaling it to the measured
fluxes by adapting the angular diameter. Confidence intervals
are determined using a $\chi^2$ statistic with four degrees of
freedom (95\% probability or 2$\sigma$).

Our strategy differs depending on the significance of the
pulsation amplitude, relative to our
ability to make a proper pulsation phase attribution. For the
sources with an amplitude $V < 0.5$~mag,
we include all reliable photometry in a single fit. 

For the strong pulsators we collect photometry for a maximum and 
a minimum phase, and fit each set independently. For the RVb sources
we try to collect both photometric sets during a long-term maximum phase, 
because this long-term variability is not caused by variations in 
the stellar properties.

We collect various photometric measurements from the literature
(only for the pulsating sources) and from assorted on-line data
catalogues. The sources of the photometry that is included in our
analysis, are listed in Sect.~\ref{section:photorigin}. We carefully
select the photometric passbands to be included in our fitting
process, the details of which are given in the Appendix
(Sect.~\ref{section:photorigin}).

Only measurements with effective wavelengths up to the J (and in
specific cases H) band were included in the fitting process.  
We avoid these passbands if there is an indication 
that there is already an excess at the respective 
wavelengths, but prefer to include them whenever possible.

Figs.~\ref{figure:SEDexample1} and~\ref{figure:SEDexample2} show the
SEDs of a single illustrative static and variable source,
respectively. Also included are the confidence intervals of the T$_{\rm{eff}}$ versus
the $\theta_\star$ and E(B-V), respectively. In the online appendices
similar figures (Fig.~\ref{figure:SEDexample3} to \ref{figure:SEDexample6}) are included for all the other sources.  
The confidence intervals for $\log g$ generally remain unconstrained to the range (3.0;0.0).  For large-amplitude pulsators, there
are two caveats. First, due to the scarcity of the used photometry,
any systematic error on a subset has a direct impact on the
results. In particular, the lack of reliable near-IR data is
troublesome in certain cases (e.g. IRAS~17038-4815).  Second, our method
assumes that a sequence of hydrostatic Kurucz models can represent
the highly dynamic atmosphere of these stars.

All sources except IRAS~10174-5704 and IRAS~10456-5712 have a
spectroscopic temperature determination \citep{2006AAdeRuyter}.
Taking the pulsation phase into account, we use these spectroscopic
temperatures as an a posteriori constraint (included as the vertical
red lines in the middle and right panels of
Figs~\ref{figure:SEDexample1} and~\ref{figure:SEDexample2}). The
2$\sigma$ errors on the spectroscopic temperatures are uniformly
assumed to be 10\% for the static sources and 15\% for the strong
pulsators. The final parameter values, as listed in Table~\ref{tab:stellarPars},
are determined by taking the median value of the confidence intervals that 
remain after removing the points that do not agree with the spectroscopic 
temperature constraints. The final 2$\sigma$ errors are determined
similarly, by taking the minimum and maximum value instead of
the median.

In this SED analysis we neglect any scattered light contribution. For at least one
source, 89 Her, this condition is severely violated, as was found by
\citet{2013AAHillen} and \citet{2014AAHillen}. The very small
reddening that we find for a few sources (e.g. for IRAS07008+1050
it is smaller than the interstellar reddening in the
Galactic extinction maps of \citet{1998ApJSchlegel} and
\citet{2001ApJDrimmel}), might also be an indication for a
non-negligible scattered-light flux fraction. Without resolved
observations at near-IR or optical wavelengths, there is no general
way of estimating this scattered light contribution. Our angular
diameters are thus likely overestimated.  Fortunately, the sublimation
radius only depends linearly on the angular diameter (see
Eq.~\ref{eq:subradius}), but
quadratically on the effective temperature (which is spectroscopically 
constrained, hence less affected by scattering).

\begin{table}   
  \caption{Stellar parameters resulting from the SED fitting. The pulsating sources have two entries, one for a minimum and 
  a maximum pulsation phase. Metallicities are taken from \citet{2006AAdeRuyter} unless otherwise noted.}\label{tab:stellarPars}
\begin{footnotesize}
\begin{tabular}{lllcll}\hline
\multicolumn{1}{l}{Nr.} &
\multicolumn{1}{c}{IRAS} &
\multicolumn{1}{l}{E(B-V) $^{+2\sigma}_{-2\sigma}$} &
\multicolumn{1}{c}{T$_{\rm{eff}}$ $^{+2\sigma}_{-2\sigma}$} &
\multicolumn{1}{c}{$\theta_\star$ $^{+2\sigma}_{-2\sigma}$} &
\multicolumn{1}{l}{[Fe/H]} \\ 
\multicolumn{1}{l}{} &
\multicolumn{1}{c}{} &
\multicolumn{1}{c}{(mag)} &
\multicolumn{1}{c}{(K)} &
\multicolumn{1}{c}{(mas)} &
\multicolumn{1}{l}{(dex)} \\ 
\hline
1 &04440+2605 & 0.5$^{+0.2}_{-0.2}$    & 4875$^{+400}_{-450}$ & 0.44$^{+0.03}_{-0.04}$ & -0.5 \\ 
  &           & 0.5$^{+0.3}_{-0.3}$    & 4550$^{+600}_{-475}$ & 0.31$^{+0.06}_{-0.07}$ & -0.5 \\ 
2 &07008+1050 & 0.04$^{+0.09}_{-0.04}$ & 6050$^{+400}_{-225}$ & 0.270$^{+0.014}_{-0.010}$ & -4.0 \\ 
3 &07284-0940 & 0.18$^{+0.3}_{-0.18}$  & 5050$^{+450}_{-400}$ & 1.2$^{+0.2}_{-0.2}$    & -1.0 \\ 
  &           & 0.16$^{+0.2}_{-0.16}$  & 5025$^{+350}_{-300}$ & 0.58$^{+0.10}_{-0.10}$ & -1.0 \\ 
4 &08011-3627 & 0.4$^{+0.2}_{-0.4}$    & 5925$^{+1250}_{-725}$ & 0.22$^{+0.03}_{-0.05}$ & -1.0 \\ 
  &           & 0.2$^{+0.3}_{-0.2}$    & 5875$^{+1025}_{-775}$ & 0.14$^{+0.02}_{-0.02}$ & -1.0 \\ 
5 &08544-4431 & 1.40$^{+0.19}_{-0.2}$ & 7225$^{+750}_{-700}$ & 0.54$^{+0.09}_{-0.08}$ & -0.5 \\ 
6 &09256-6324 & 0.7$^{+0.2}_{-0.3}$    & 6700$^{+675}_{-650}$ & 0.38$^{+0.07}_{-0.07}$ & -1.0 \\ 
7 &10158-2844 & 0.21$^{+0.09}_{-0.16}$ & 7750$^{+525}_{-975}$ & 0.49$^{+0.07}_{-0.06}$ & -4.0 \\ 
8 &10174-5704 & 0.88$^{+0.2}_{-0.18}$ & 3500$^{+175}_{-175}$ & 1.09$^{+0.06}_{-0.05}$ &  0.0\tablefootmark{b} \\ 
9 &10456-5712 & 0.34$^{+0.3}_{-0.34}$ & 4275$^{+600}_{-550}$ & 1.8$^{+0.4}_{-0.3}$    &  0.0\tablefootmark{b} \\ 
10&11385-5517 & 0.35$^{+0.11}_{-0.15}$ & 8350$^{+1000}_{-700}$ & 0.26$^{+0.03}_{-0.03}$ &  0.0\tablefootmark{c} \\
11&12185-4856 & 0.3$^{+0.2}_{-0.3}$    & 5925$^{+1225}_{-625}$ & 0.18$^{+0.02}_{-0.03}$ & -1.0 \\ 
  &           & 0.3$^{+0.3}_{-0.3}$    & 5850$^{+1350}_{-1050}$ & 0.138$^{+0.03}_{-0.018}$ & -1.0 \\ 
12&12222-4652 & 0.17$^{+0.13}_{-0.12}$ & 7000$^{+700}_{-575}$ & 0.178$^{+0.010}_{-0.013}$ &  0.0 \\ 
13&15469-5311 & 1.34$^{+0.16}_{-0.2}$  & 7425$^{+825}_{-675}$ & 0.23$^{+0.03}_{-0.02}$ &  0.0 \\ 
14&17038-4815 & 0.3$^{+0.2}_{-0.3}$    & 4375$^{+775}_{-575}$ & 0.29$^{+0.04}_{-0.07}$ & -1.5 \\ 
  &           & 0.7$^{+0.4}_{-0.3}$    & 4575$^{+900}_{-525}$ & 0.18$^{+0.04}_{-0.04}$ & -1.5 \\ 
15&17243-4348 & 0.3$^{+0.3}_{-0.3}$    & 5075$^{+825}_{-800}$ & 0.16$^{+0.03}_{-0.02}$ &  0.0 \\ 
  &           & 0.3$^{+0.2}_{-0.2}$    & 4850$^{+625}_{-400}$ & 0.14$^{+0.03}_{-0.02}$ &  0.0 \\ 
16&17534+2603 & 0.07$^{+0.10}_{-0.07}$ & 6850$^{+350}_{-325}$ & 0.52$^{+0.03}_{-0.03}$ &  0.0 \\ 
17&18281+2149 & 0.26$^{+0.17}_{-0.26}$ & 6200$^{+700}_{-950}$ & 0.39$^{+0.04}_{-0.05}$ & -1.5 \\ 
  &           & 0.03$^{+0.07}_{-0.03}$ & 4800$^{+150}_{-150}$ & 0.318$^{+0.014}_{-0.015}$ & -1.5 \\ 
18&19125+0343 & 0.79$^{+0.15}_{-0.07}$ & 7275$^{+1250}_{-300}$ & 0.140$^{+0.009}_{-0.011}$ & -0.5 \\ 
19&22327-1731 & 0.12$^{+0.14}_{-0.12}$ & 8275$^{+800}_{-850}$ & 0.085$^{+0.013}_{-0.013}$ & -1.0 \\ 
\hline
\end{tabular}
\tablefoot{\tablefoottext{a}{H band fluxes are also included in the fit for sources 1, 8 and 17},
           \tablefoottext{b}{This metallicity is assumed},
           \tablefoottext{c}{The spectroscopic constraints for this source come from \citet{2005BaltAKipper}},
           \tablefoottext{f}{We are cautious about IRAS10174-5704, given the limited data and the 
           limiting T$_{\rm{eff}}=3500$~K in the atmosphere grid.}}
\end{footnotesize}
\end{table}

Finally, we note that our SEDs confirm the presence of variability in the mid-IR for U~Mon and IRAS~17038-4815, as attested by the 
large scatter in the mid-IR fluxes as compared to the static sources. Similarly, it seems likely that RV~Tauri and 
AC~Her are intrinsically variable in the mid-IR as well, in contrast to the other stars in our category of 'pulsators'. This is consistent 
with the difference in visual pulsation amplitude between these objects.

The SEDs of our sources show a diversity of total infrared excesses, but near 10~$\mu$m the dust always contributes an order of 
magnitude more flux than the photosphere.

\begin{figure}
   \centering
   \includegraphics[width=9cm]{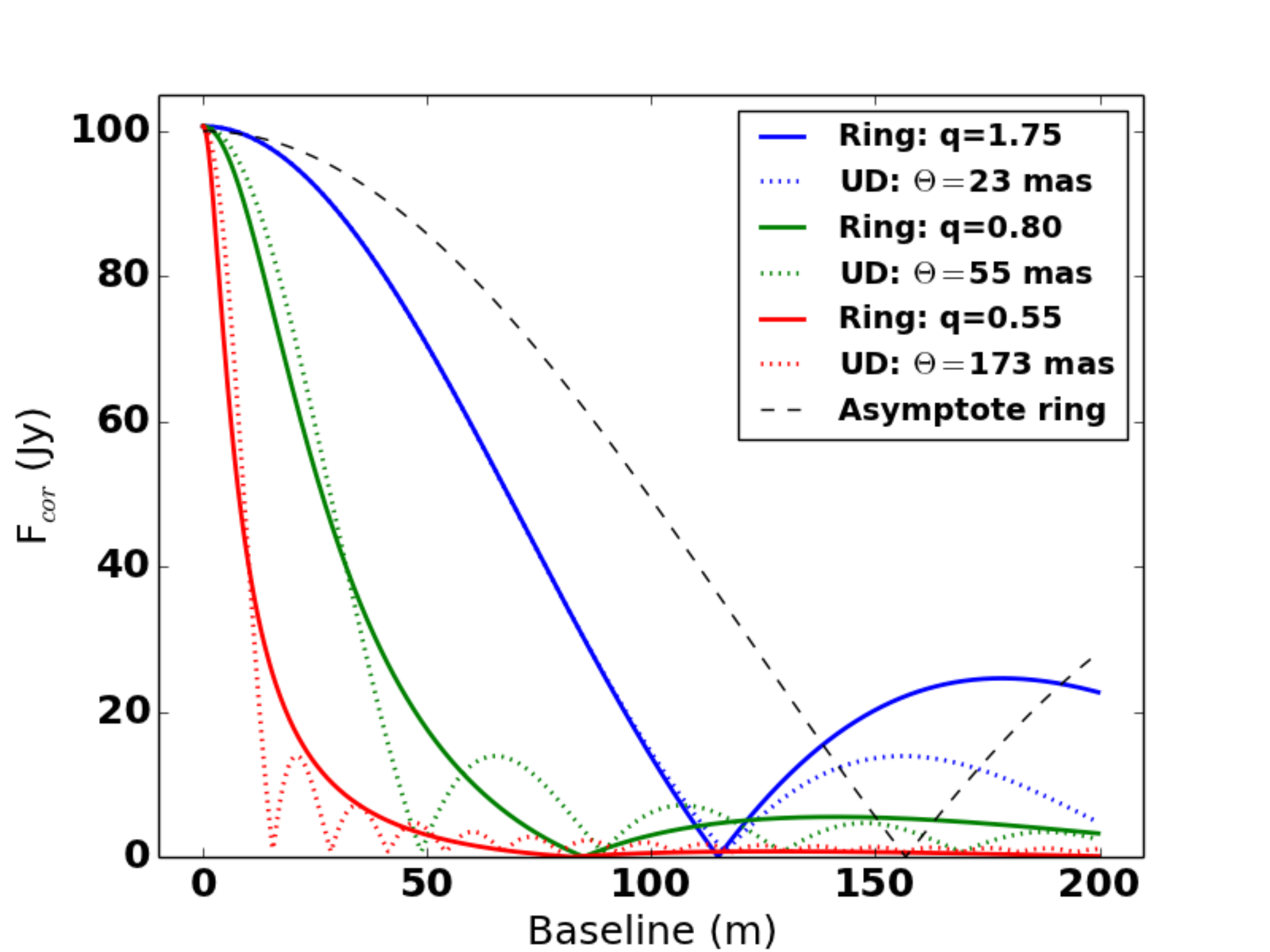}
   \caption{Simulated model visibility curves. Three ring models with different power law values are shown as full lines (see the legend for the 
   colour coding). The inner and outer ring radii are fixed to the values adopted for the 89 Her system. The correlated flux at zero baseline (i.e., the 
   total flux) is fixed at 100 Jy for comparison purposes. For each ring model, the uniform disk model that crosses the correlated flux curve 
   at 40 Jy is shown for comparison. The black dashed line represents the asymptotic limit of the ring model: an infinitesimally thin 
   ring at the sublimation radius. }
   \label{figure:modelviscomparison}
\end{figure}

\begin{figure}
   \centering
   \includegraphics[width=9cm]{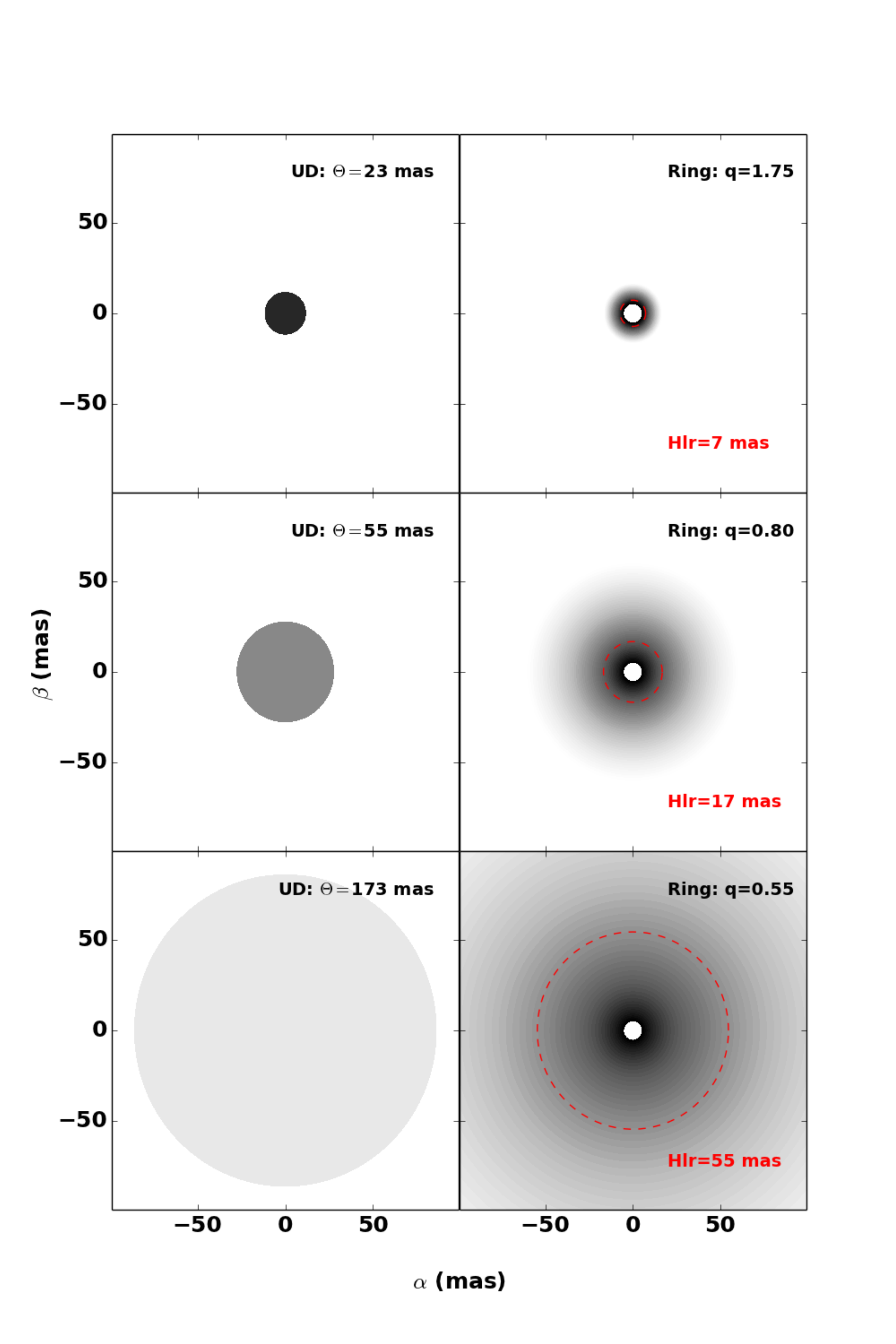}
   \caption{Simulated model images corresponding to the visibility curves shown in Fig.~\ref{figure:modelviscomparison}. Left panels: uniform 
   disks; right panels: the semi-physical ring models. A common intensity normalization is used across the panels. Also the same logarithmic greyscale
   is applied for all panels. The half-light radius for each of the ring models (see Sect.~\ref{subsection:halflightradius}) is 
   indicated by the red dashed circle. The field of view of 
   the image is limited for clarity of presentation.}
   \label{figure:imagecomparison}
\end{figure} 

\section{Interferometric model fitting}\label{section:modelfitting}
\subsection{The models} \label{subsection:models}

Although our sample is significant in number of stars, the uv-coverage per source is
limited, in particular in terms of baseline position angle. For the current
global analysis, we opt for a parametric approach. We
make use of two parametric models to fit the mid-IR intensity distribution on the
sky in 1D. For a large majority of our sources the baseline position angle
coverage does not allow the detection of variations in the intensity distribution caused by 
inclination effects \citep[for a more extensive discussion on this topic, see][]{2015AAMenu}. 
The effect of the inclination will be included in our sample-wise comparison with radiative transfer models (Sect.~\ref{section:RTgrid}).

First we fit our data with the simplest possible model: a uniform
disk. This has only the outer angular diameter $\Theta$ and the total
integrated flux as free parameters.  The advantage of this model is
that it is widely used and allows to make a simple angular size
estimate for each object in the sample.

Second, we use the semi-physical disk model applied by \citet{2015AAMenu} to a large sample of protoplanetary disks observed with MIDI. 
The specific intensity of this model is parameterized as a function of the angular radius $\rho$ and optical depth $\tau$ as follows: 
\begin{equation}
 I_\nu(\rho)  =  (1-\exp(-\tau))\, B_\nu(T(\rho)) \qquad \rm{ for } \qquad \rho_{\rm{sub}} < \rho < \rho_{\rm{out}} , \\
\end{equation}
and zero elsewhere, with 
\begin{equation}
 T(\rho) = T_{\rm{sub}} \left(\frac{\rho}{\rho_{\rm{sub}}}\right)^{-q}.
\end{equation}

Here, T$_{\rm{sub}}$ and $\rho_{\rm{sub}}$ are the sublimation temperature and angular sublimation radius of the dust. 
We keep the dust sublimation temperature fixed at $T_{\rm{sub}} = 1500$~K. We assume that the dust disk starts at sublimation radius.
The angular sublimation radius is calculated as follows:
\begin{equation}
 \rho_{\rm{sub}} = \frac{\theta_\star}{2} \left(\frac{T_{\rm{eff}}}{T_{\rm{sub}}}\right)^2
\end{equation}\label{eq:subradius}
\citep{2010ARAADullemond}, with $\theta_\star$ and $T_{\rm{eff}}$ the stellar angular diameter and effective temperature, respectively. 
We keep the outer radius fixed at $\rho_{\rm{out}}=100 \rho_{\rm{sub}}$ but our results are insensitive to 
the exact choice.

The correlated flux corresponding to a given baseline and wavelength is then computed as the 1D Hankel transform of the 
intensity distribution, integrated over the radial angular coordinate on the sky. The only free fit parameters in this model are 
the temperature gradient $q$ and the total integrated flux. The average optical depth is a derived parameter. 

Fig.~\ref{figure:modelviscomparison} compares the correlated flux
behaviour of our two models. The corresponding images for these models
are displayed in Fig.~\ref{figure:imagecomparison}. Two extreme cases of a large ($q=0.55$) and a small ($q=1.75$) ring are used, 
together with the median value in our sample ($q=0.8$; see Sect.~\ref{subsection:interferometricfitresults}). The total flux F$_{\rm{tot}} = 100$~Jy 
and sublimation radius are kept fixed in all three models (using the stellar parameters of 89 Her).
A uniform disk is shown that reaches a correlated flux of 40~Jy at the same baseline 
length as in the corresponding ring model. Fig.~\ref{figure:modelviscomparison} shows that for correlated fluxes above 40\% the two models are 
indistinguishable. Only at longer baselines the models behave significantly different. By applying both parametric models 
to our observations, we have a better diagnostic to analyse the sample properties. 

Although the value of $q$ should not be over-interpreted, classical models of flaring protoplanetary disks 
give rise to temperature profiles of $T(\rho) \propto \rho^{-1/2}$ \citep{1987ApJKenyon}. The steepest profile one would expect 
for a passively irradiated disk is obtained in the geometrically thin limit, and corresponds 
to $T(\rho) \propto \rho^{-3/4}$ \citep{2010astrophArmitage,1987ApJKenyon}. For a spherical stellar wind, 
the expected temperature profile corresponds to $T(\rho) \propto \rho^{-2/5}$ \citep{1999iswbookLamers}.
In reality, the radial optical depth profile  
is important as well, while it is assumed to be constant here ($\tau < 1$).
Our fitted value of $q$ can thus not be directly associated with the true temperature gradient in a given source, but merely acts 
as a parameter that distributes flux over subsequent radial annuli. Our model assumes the dust emission to start at the sublimation 
radius and to be continuous outwards. For sources where there is only dust further out, the fit will compensate the too small assumed 
inner radius by making $q$ smaller (i.e. a larger apparent structure, see Figs.~\ref{figure:modelviscomparison} and~\ref{figure:imagecomparison}).

One can define the apparent size of a given ring model with a single parameter, called the half-light radius (\textit{hlr}), which 
is computed from the following implicit equation (see \citet{2015AAMenu}):
\begin{equation}
\frac{F_{tot,\nu}}{2} = \int^{hlr}_{\rho_{\rm{sub}}} d\rho 2 \pi \rho I_\nu(\rho).
\end{equation}

The half-light radius represents the radius within which half of the mid-IR flux is emitted. The half-light radius of each model 
in Fig.~\ref{figure:imagecomparison} is indicated in red.

\subsection{The fit strategy} \label{subsection:interferometricfitresults}
We describe the fit strategy adopted in this work.
Both our models have two fit parameters: $q$ or $\Theta$ and the total flux F$_{\textrm{tot}}$. 

To locate best-fitting models and to estimate parameter uncertainties,
the Markov chain Monte Carlo (MCMC) method is widely used.  In this
paper we use the \textit{emcee} package implemented in Python
\citep{2013PASPForeman}. An MCMC allows to randomly draw
realizations of the model parameters according to the Bayesian
posterior probability distribution determined by the data,
$P(M|D)$. Bayes' theorem states that this posterior probability
distribution is proportional to the product of the prior probability
of the model $P(M)$ with the likelihood $P(D|M)$. We assume uniform
priors, allowing F$_{\rm{tot}} \in (0,500)$~Jy, and $\Theta \in
(0,500)$~mas (the upper limit being the field-of-view of MIDI). For
the ring model we assume $q \in (0,2.5)$. The lower value corresponds to
a uniform ring with the temperature equal to the
sublimation temperature. The upper limit can be considered as the
start of the asymptotic regime toward the infinitesimally thin ring
(see Fig.~\ref{figure:UD-hlr}). The likelihood is proportional to
$\exp(-\chi^2/2)$, which assumes Gaussian distributed measurement errors. The
$\chi^2$ includes the MIDI fluxes, but also the additional fluxes
collected for many sources (see Sect.~\ref{subsection:spectra}).  For
each parameter we take the median, and the 16th and 84th percentile of
the marginalized distribution as the final parameter value, and its
lower and upper 1~$\sigma$ error, respectively (see Table~\ref{tab:interferometryPars}).

We fit F$_{\textrm{tot}}$, instead of fixing it to the average of all
zero-baseline fluxes, because short-baseline correlated fluxes are
also sensitive to this parameter. 

Due to the limited uv-coverage per source, and the S/N-dependent super-resolution capability of an interferometer, 
we determine whether a source is \textit{partially resolved} by means of a statistical criterion.
We use the concept of Bayesian model selection, in which the evidence provided by the data is weighed against or in 
favour of a null hypothesis \citep{1995Kass}.
Our null hypothesis is defined as \textit{the object is unresolved}, i.e. $\Theta = 0$, 
versus the alternative hypothesis that \textit{the object is resolved}, or $\Theta > 0$. We 
compute the Bayes factor B$_{10}$ and reject the null hypothesis in favour of the 
alternative if B$_{10} > 10$ (see Table~\ref{tab:interferometryPars}).
We fit the uniform disk model twice (with and without $\Theta = 0$), and take the ratio of 
the Bayesian evidences (provided by \textit{emcee}) as the Bayes factor.
We do not use the ring model for deciding the resolvedness of each source because 
it incorporates a lower limit on the size (see Sect.~\ref{subsection:models}).


\subsection{The fit results}
\begin{figure*}
   \centering
   \includegraphics[width=9cm]{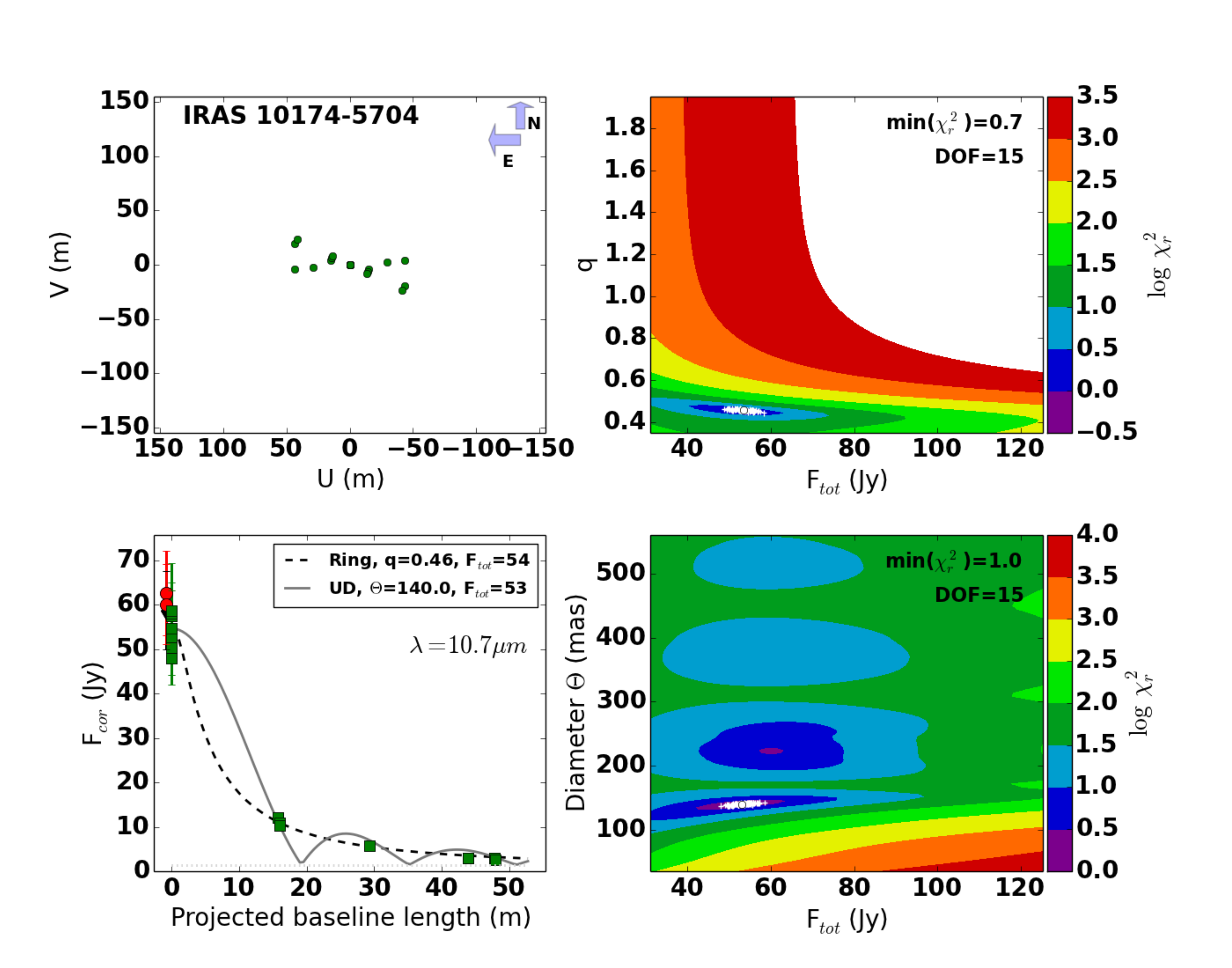}
   \includegraphics[width=9cm]{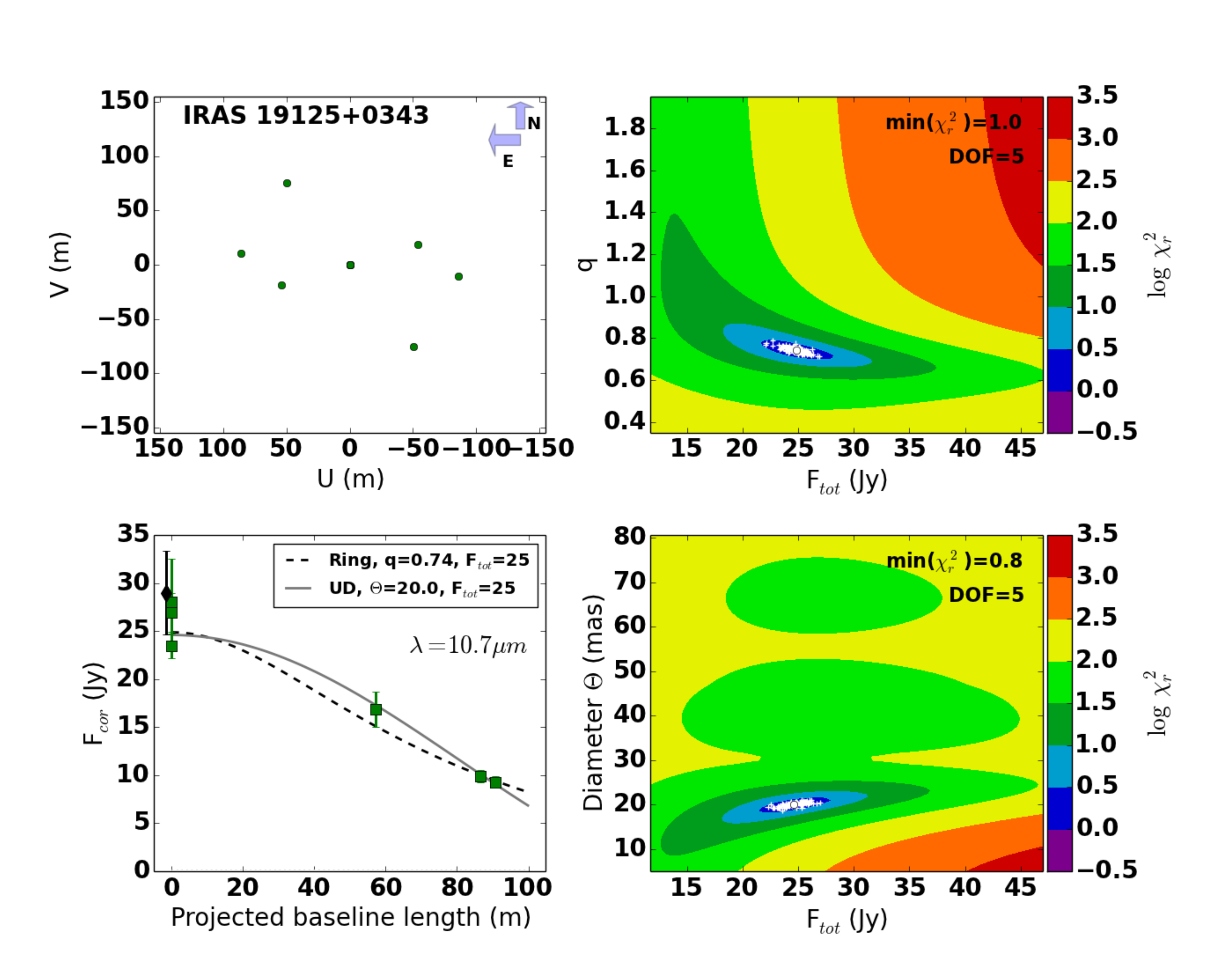}
   \caption{Results of the interferometric fit for two sources, IRAS 10174-5704 (left) and IRAS 19125+0343 (right). The former is an 
   outlier in our sample, while the latter is a typical and representative source. For each figure, the left panels show the 2D uv-coverage (upper) and 
   the corresponding correlated fluxes as a function of projected baseline length (lower), respectively.  
   The right panels show $\chi^2$-maps for the two models used in this paper, a uniform disk (upper) and 
   the semi-physical disk model (lower). The minimum reduced $\chi^2$ and the degrees-of-freedom are indicated in the upper right corner of both panels.
   The white crosses in both $\chi^2$-maps are 100 random realisations of the posterior 
   probability distribution that was sampled by the MCMC chain. The final adopted values are indicated with a white dot. 
   Visibility curves corresponding to these final parameter values are included in the lower left panel as the full grey and black dashed lines
   for the uniform disk and semi-physical disk model, respectively.}
   \label{figure:visfit1}
\end{figure*}

The interferometric fit results are graphically depicted in Fig.~\ref{figure:visfit1} for
two sources in our sample, the remainder is included in the online
appendix (Fig.~\ref{figure:visfit2} to Fig.~\ref{figure:visfit5}  ). The quantified results are given in Table~\ref{tab:interferometryPars}. 
The figures show the 2-dimensional uv-coverage, the
correlated fluxes as a function of projected baseline length and
$\chi^2$-maps for both models. One hundred realizations randomly drawn
according to the posterior probability distribution that was derived with the MCMC are shown
on top of the $\chi^2$-maps as well. The quantified results are given in table~\ref{tab:interferometryPars}.
We note that the half-light radii found for the minimum and maximum pulsation phase of U Mon 
and IRAS17038-4815 are very similar (but the optical depths differ), even though the total flux increases by $\geq$50\%.
This is not unexpected because the inner rim radius cannot change dramatically over the course of a 
pulsation cycle. However, the inner rim temperature, which we assume fixed, can respond almost instantaneously \citep{2010ApJNagel}.

All sources in our sample, except SX Cen \citep[see also][]{2006AADeroo} and IRAS22327-1731, are resolved.  
For IRAS22327-1731 we only have one correlated flux measurement.
Even though this flux is marginally smaller than the two total flux measurements, statistically we must consider 
this source to be unresolved.

In the following section we discuss some peculiar individual objects in more detail.

\onlfig{11}{
\begin{figure*}
   \centering
   \includegraphics[width=9cm]{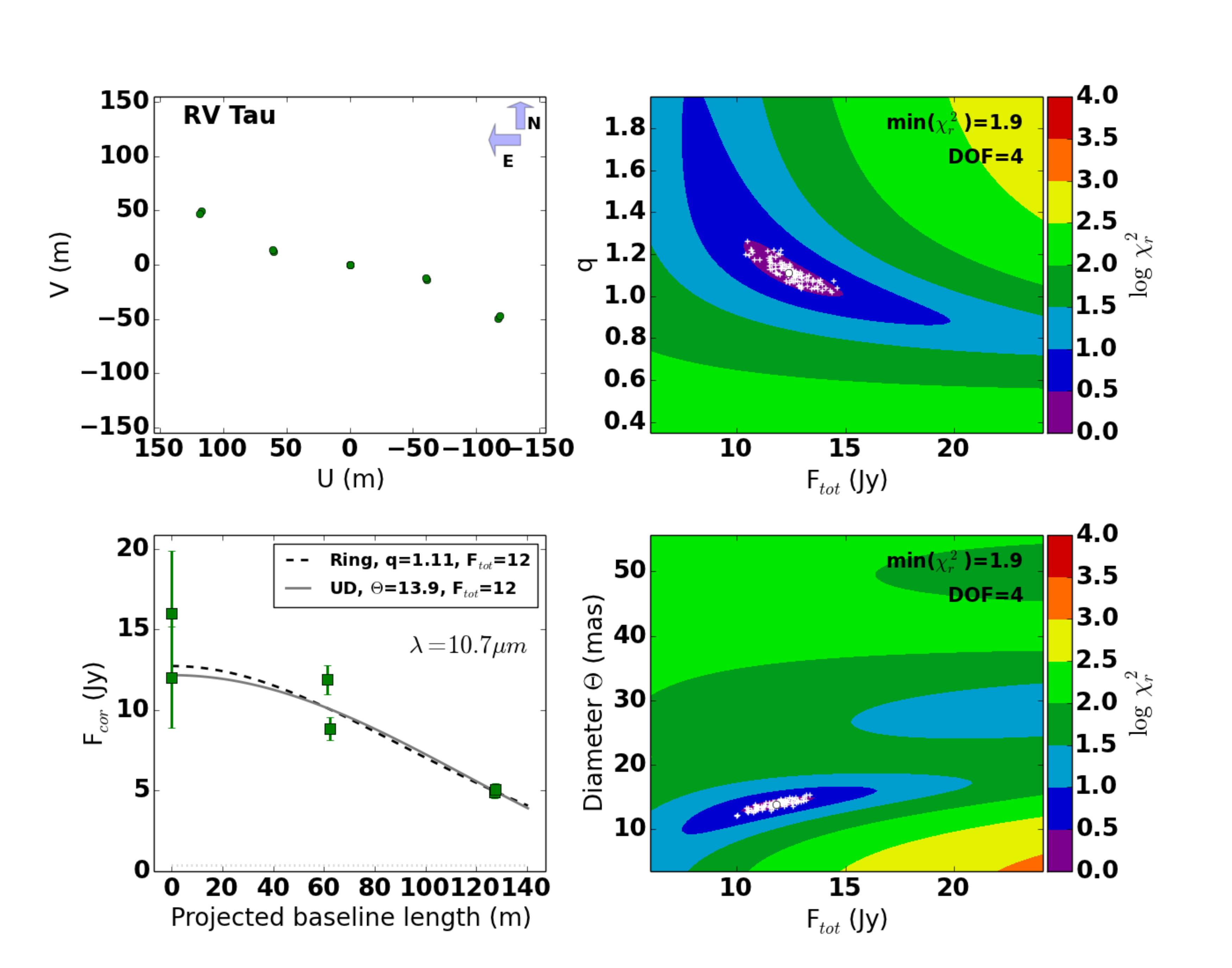}
   \includegraphics[width=9cm]{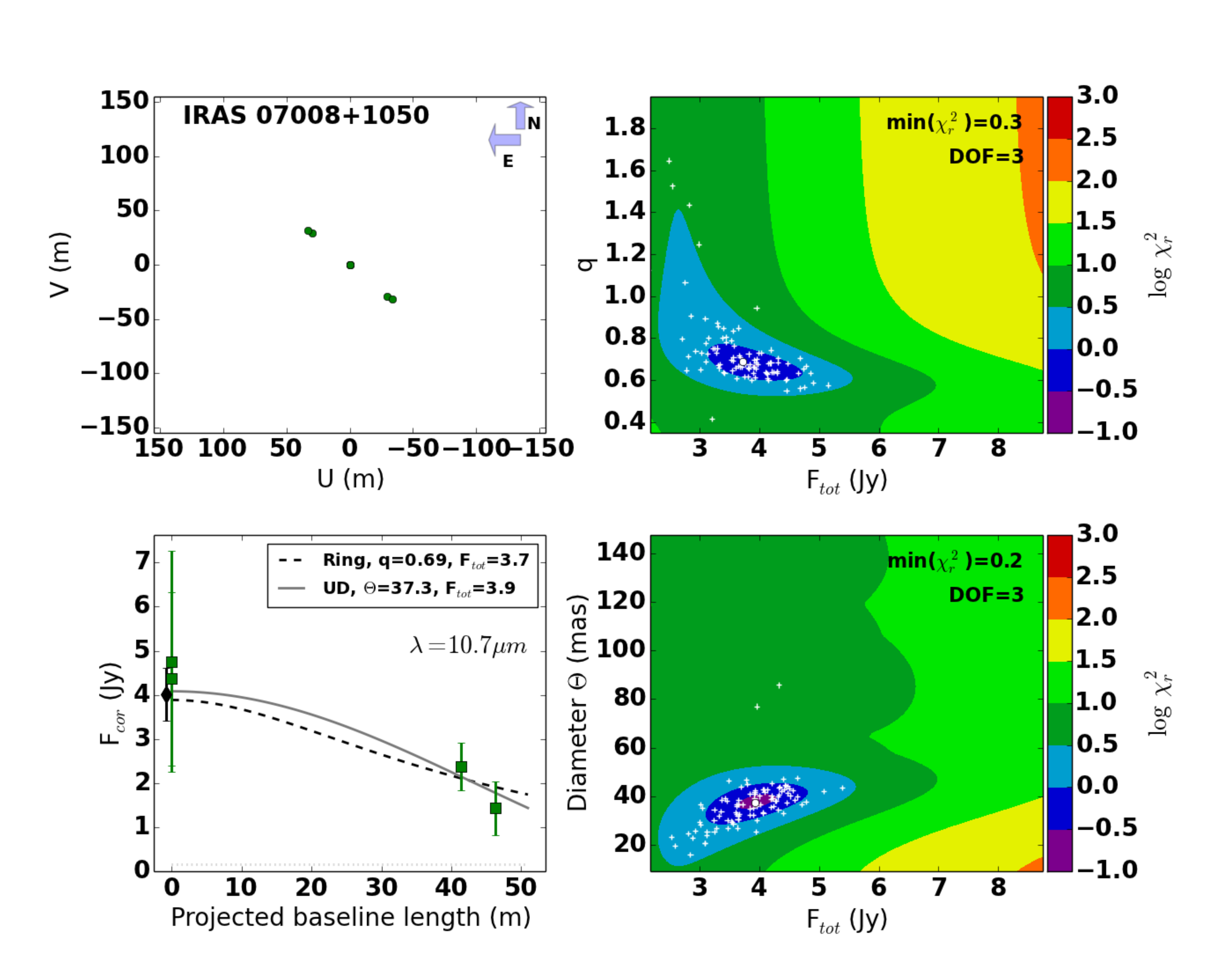}
   \includegraphics[width=9cm]{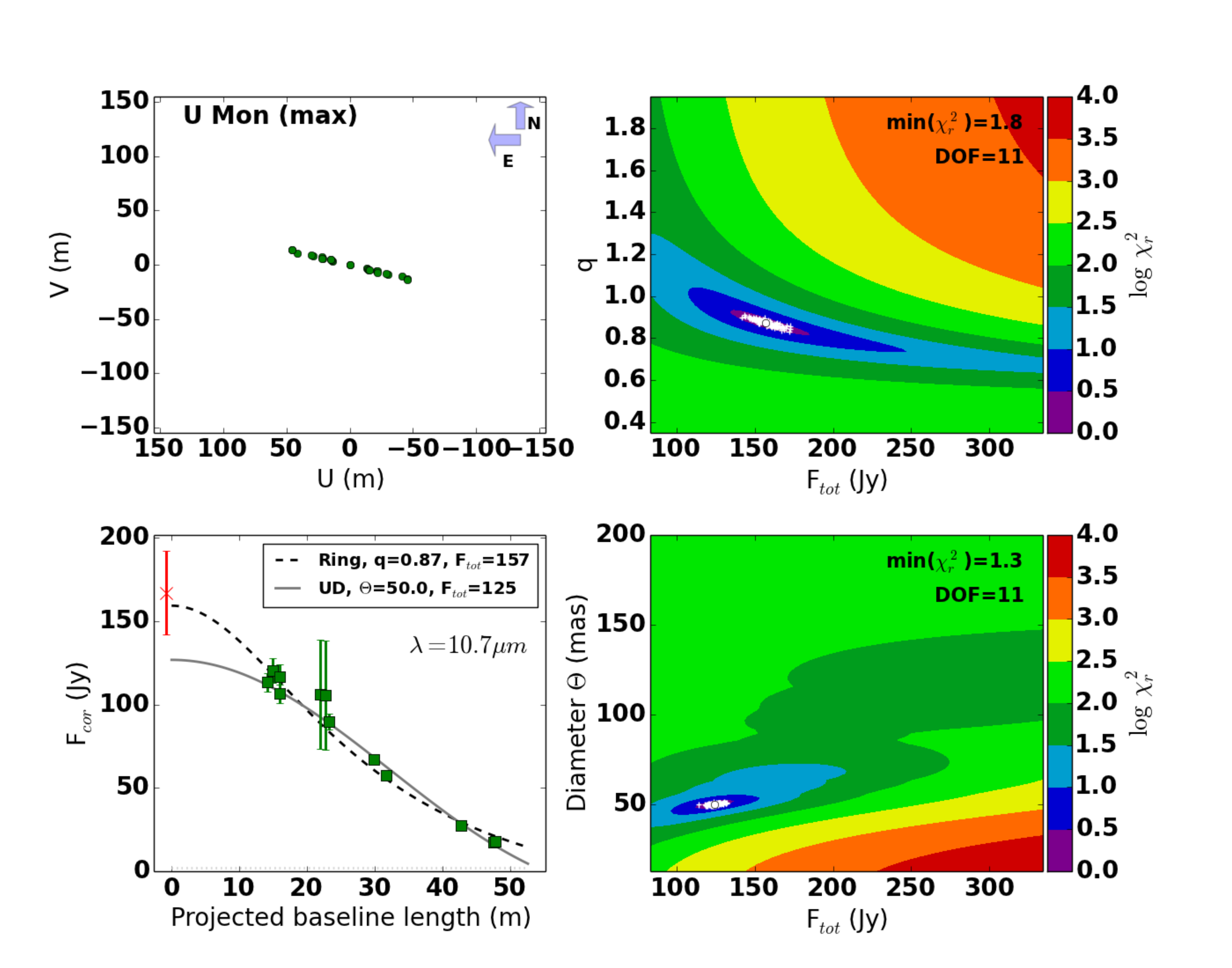}
   \includegraphics[width=9cm]{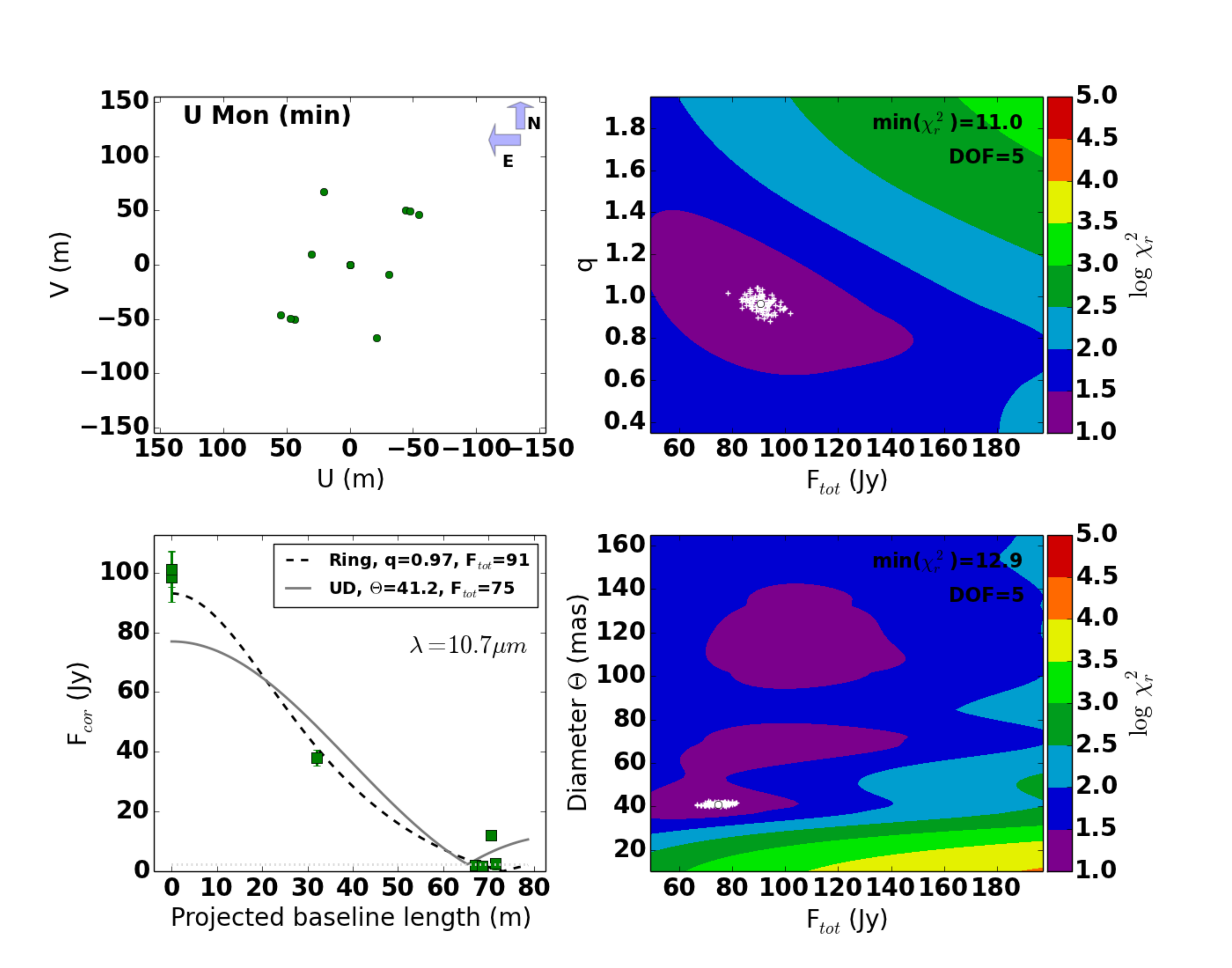}
   \includegraphics[width=9cm]{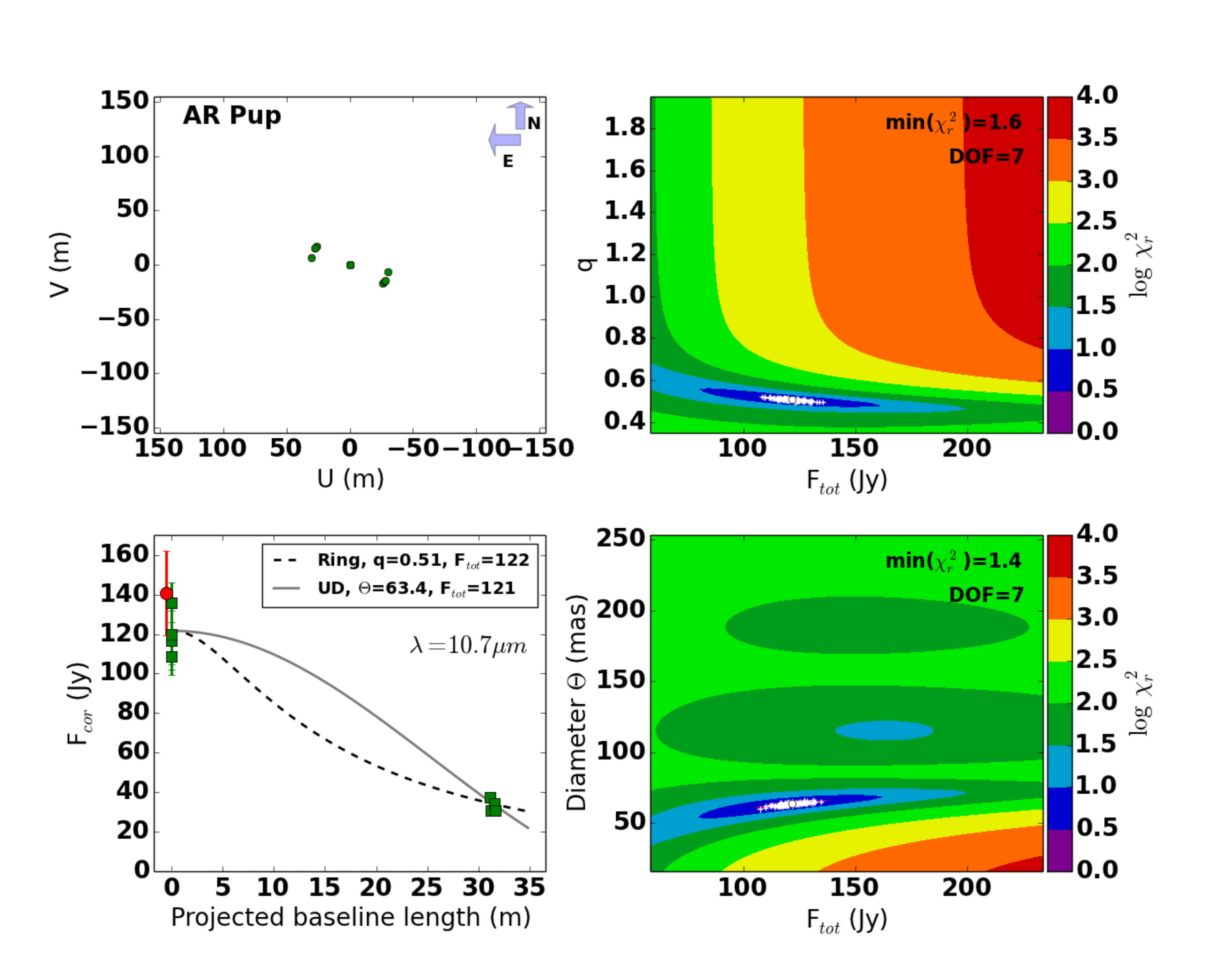}
   \includegraphics[width=9cm]{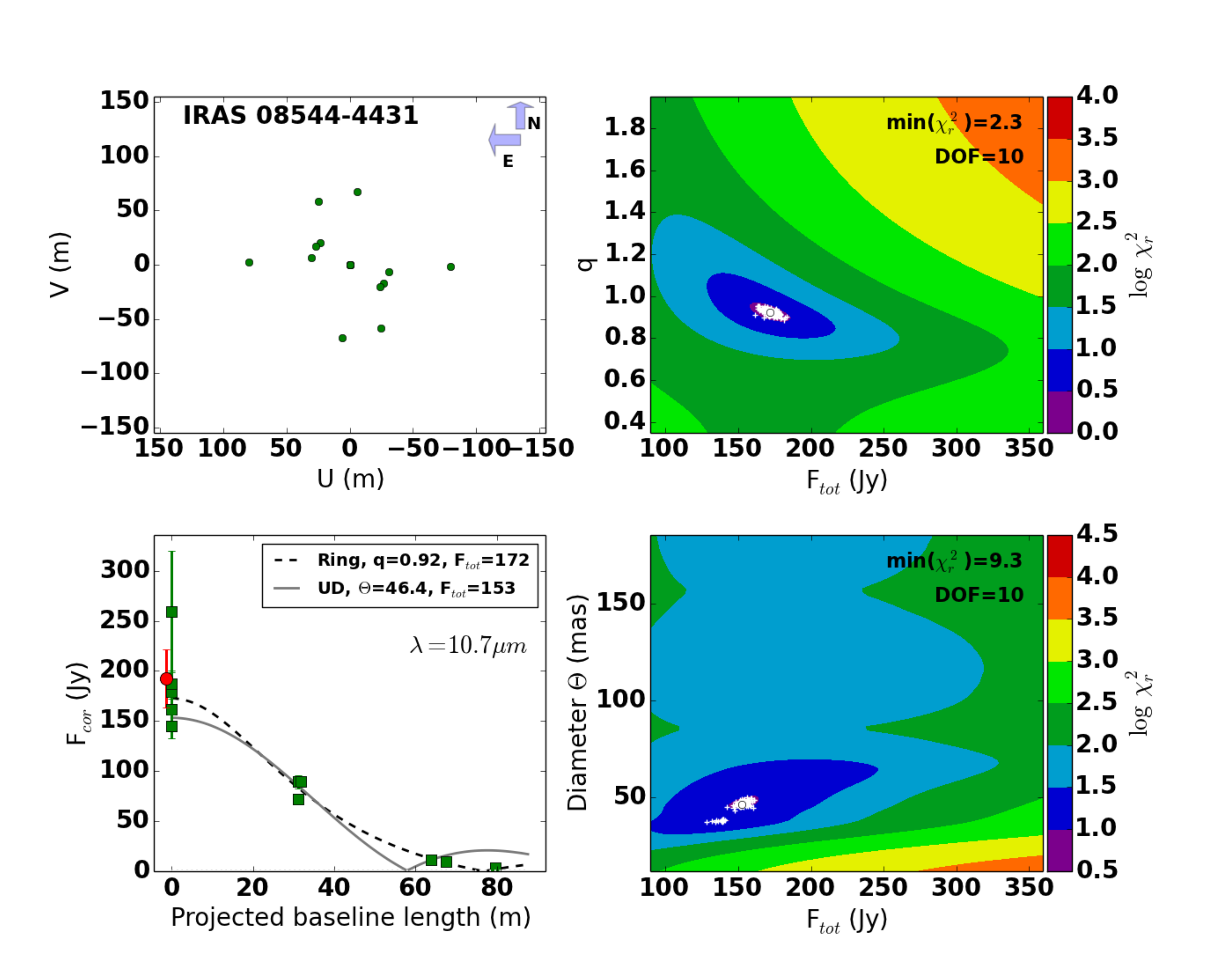}
   \caption{Same as Fig~\ref{figure:visfit1} of the main text, but for the other sources in the sample.}
   \label{figure:visfit2}
\end{figure*} 
}

\onlfig{12}{
\begin{figure*}
   \centering
   \includegraphics[width=9cm]{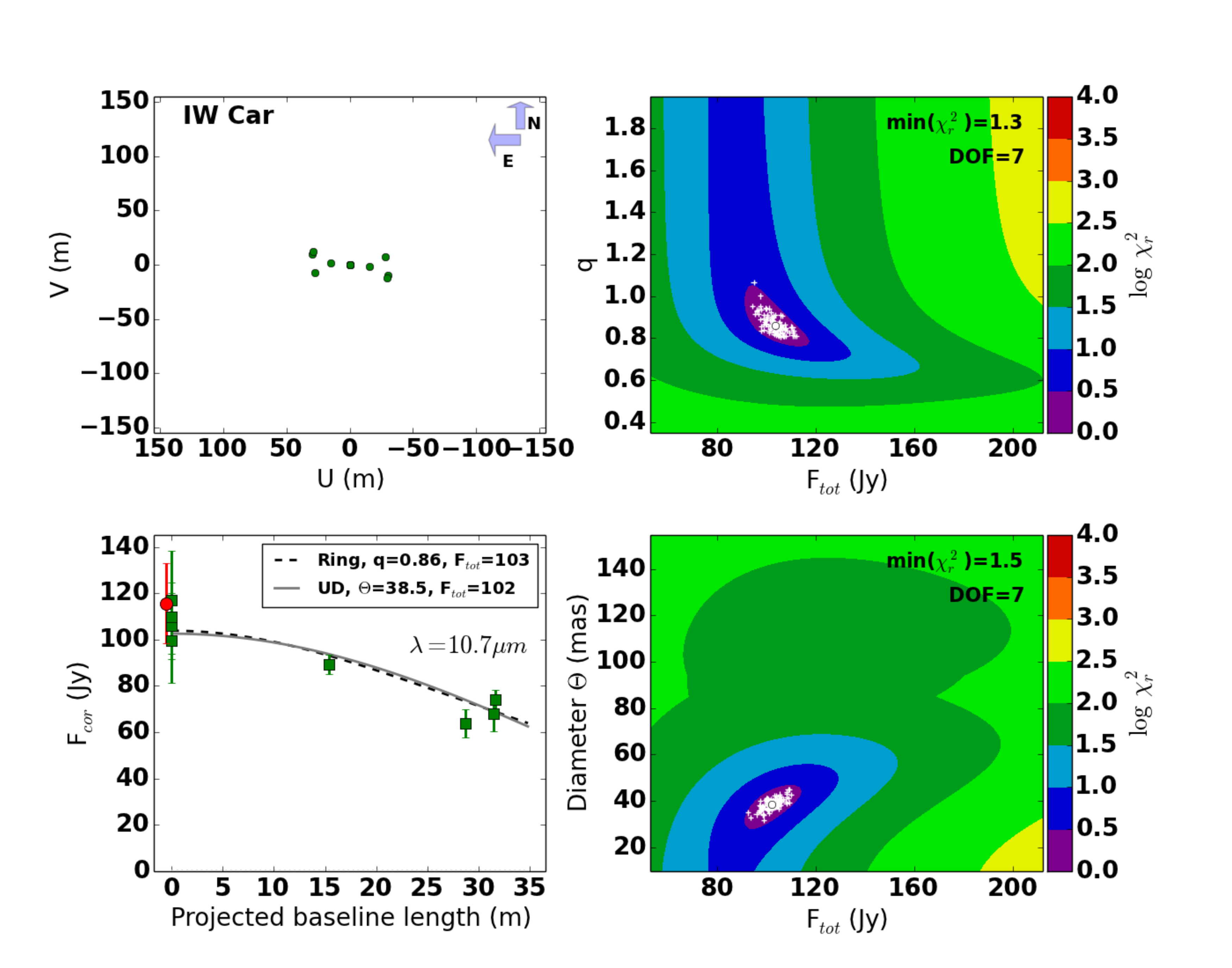}
   \includegraphics[width=9cm]{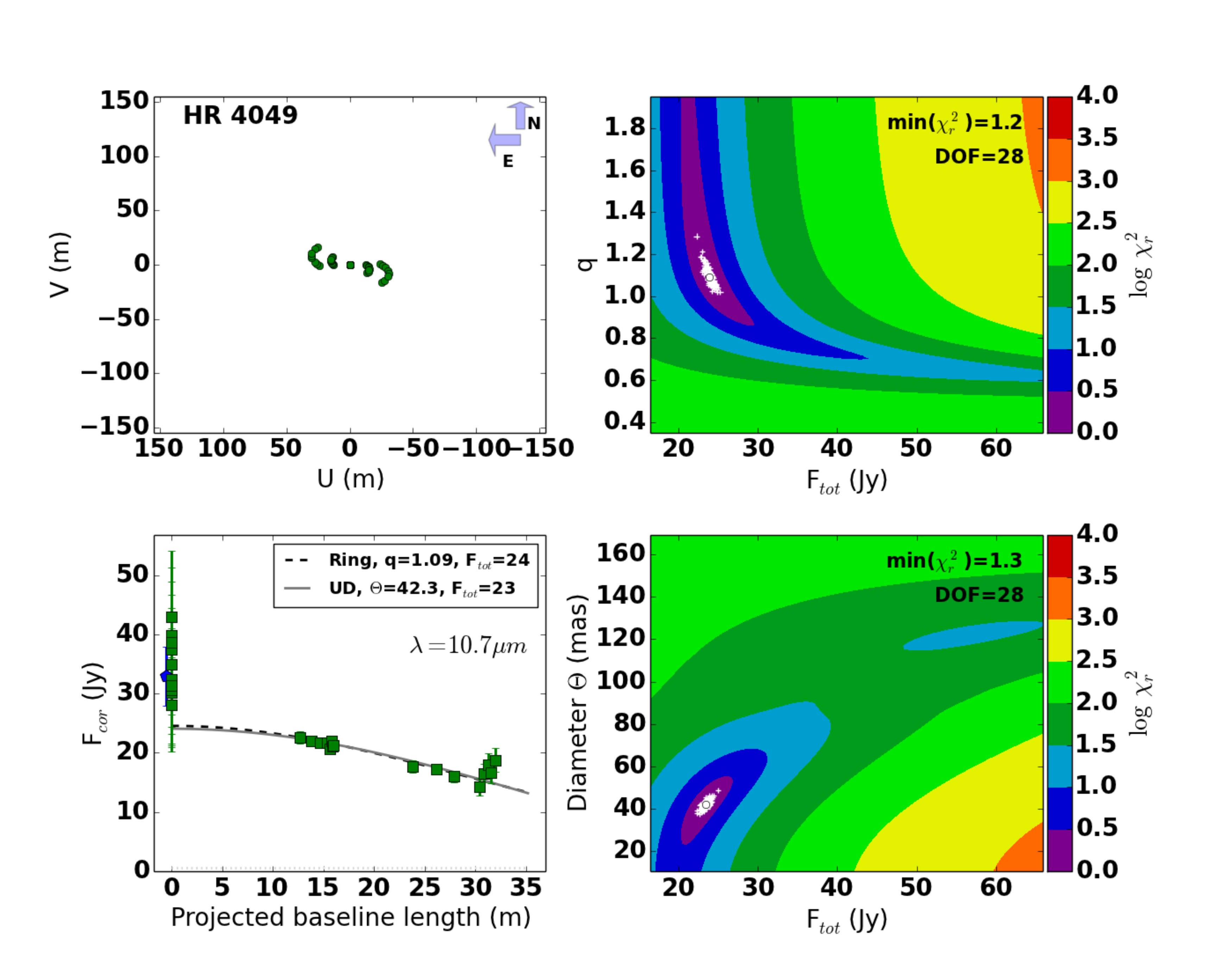}
   \includegraphics[width=9cm]{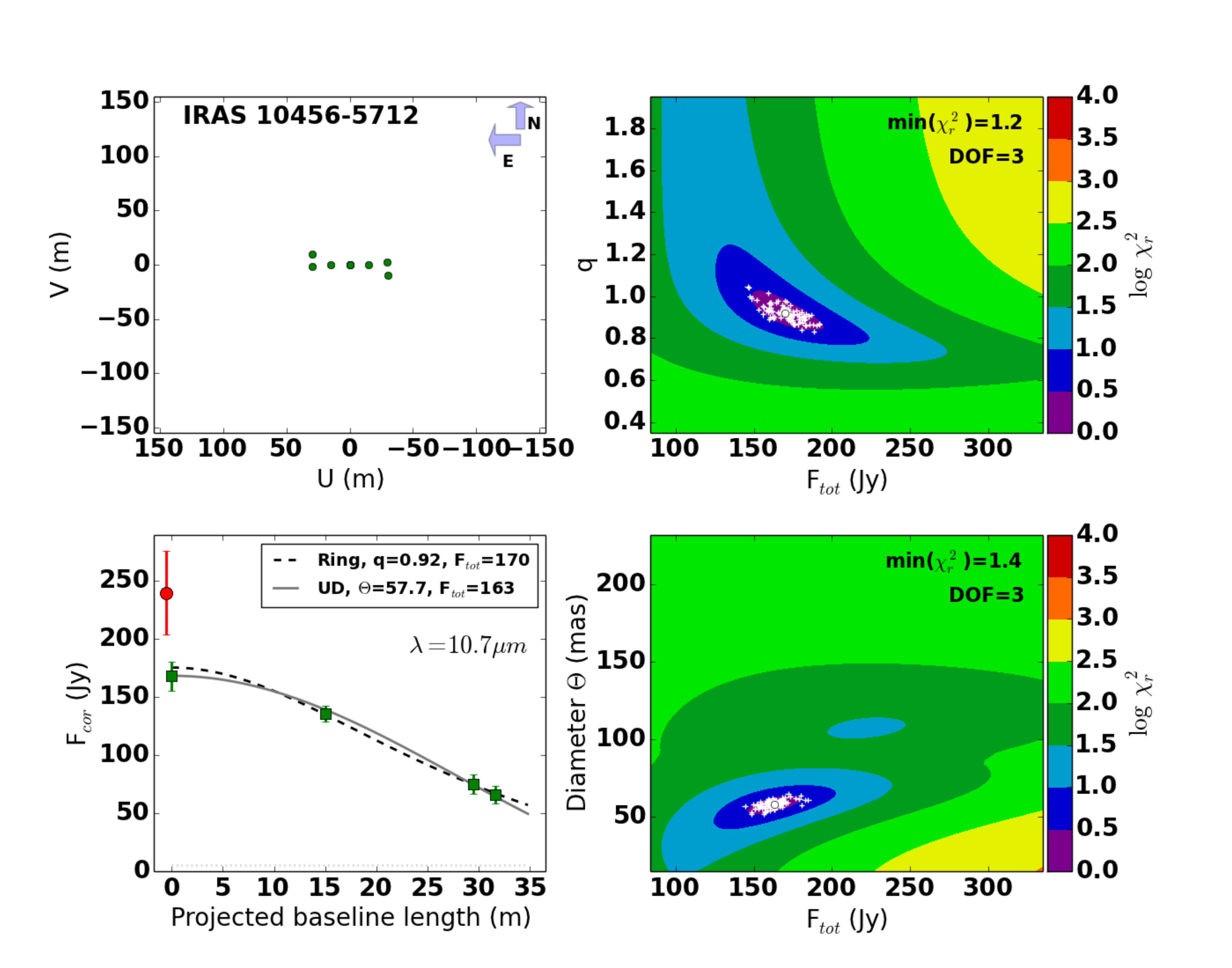}
   \includegraphics[width=9cm]{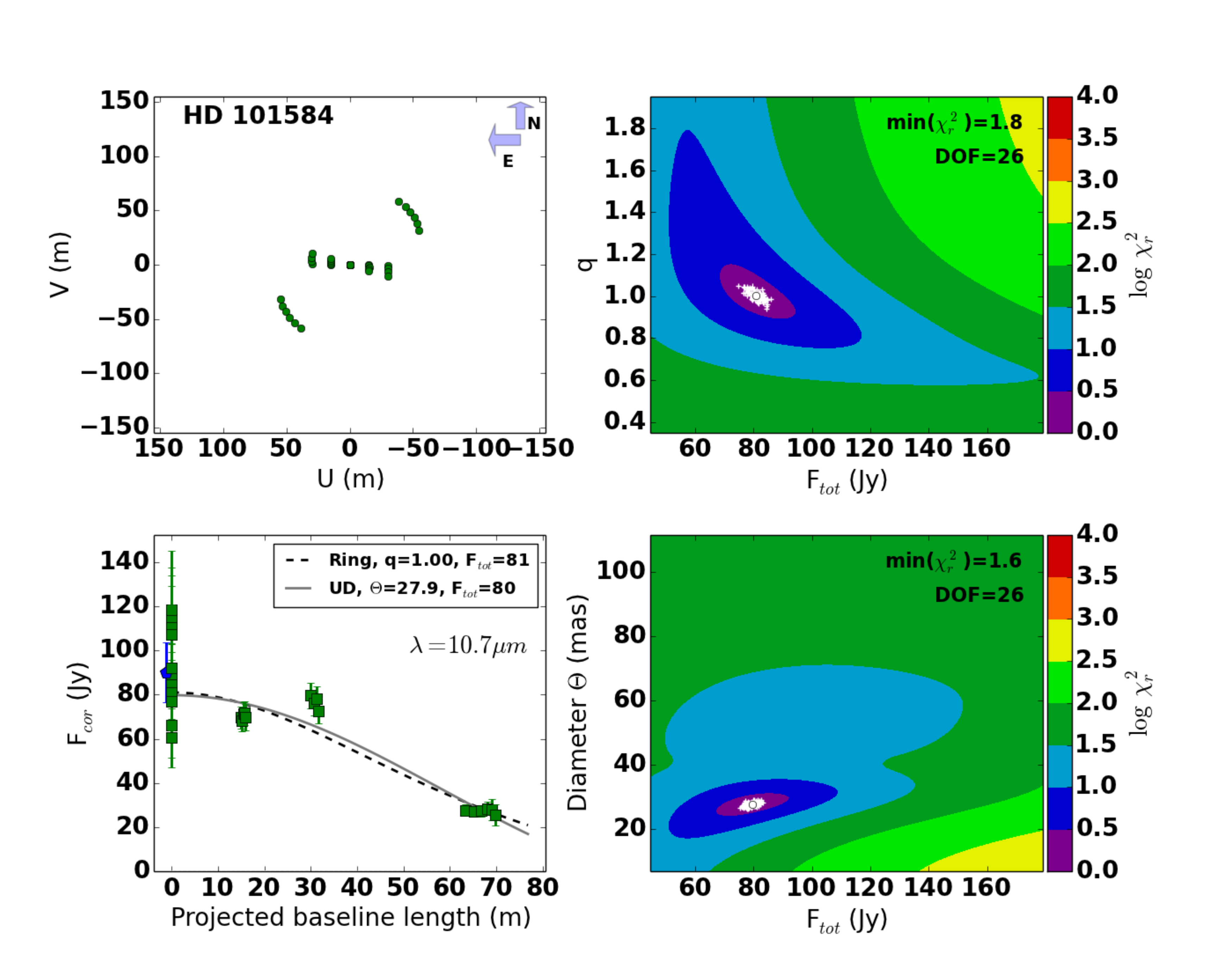}
   \includegraphics[width=9cm]{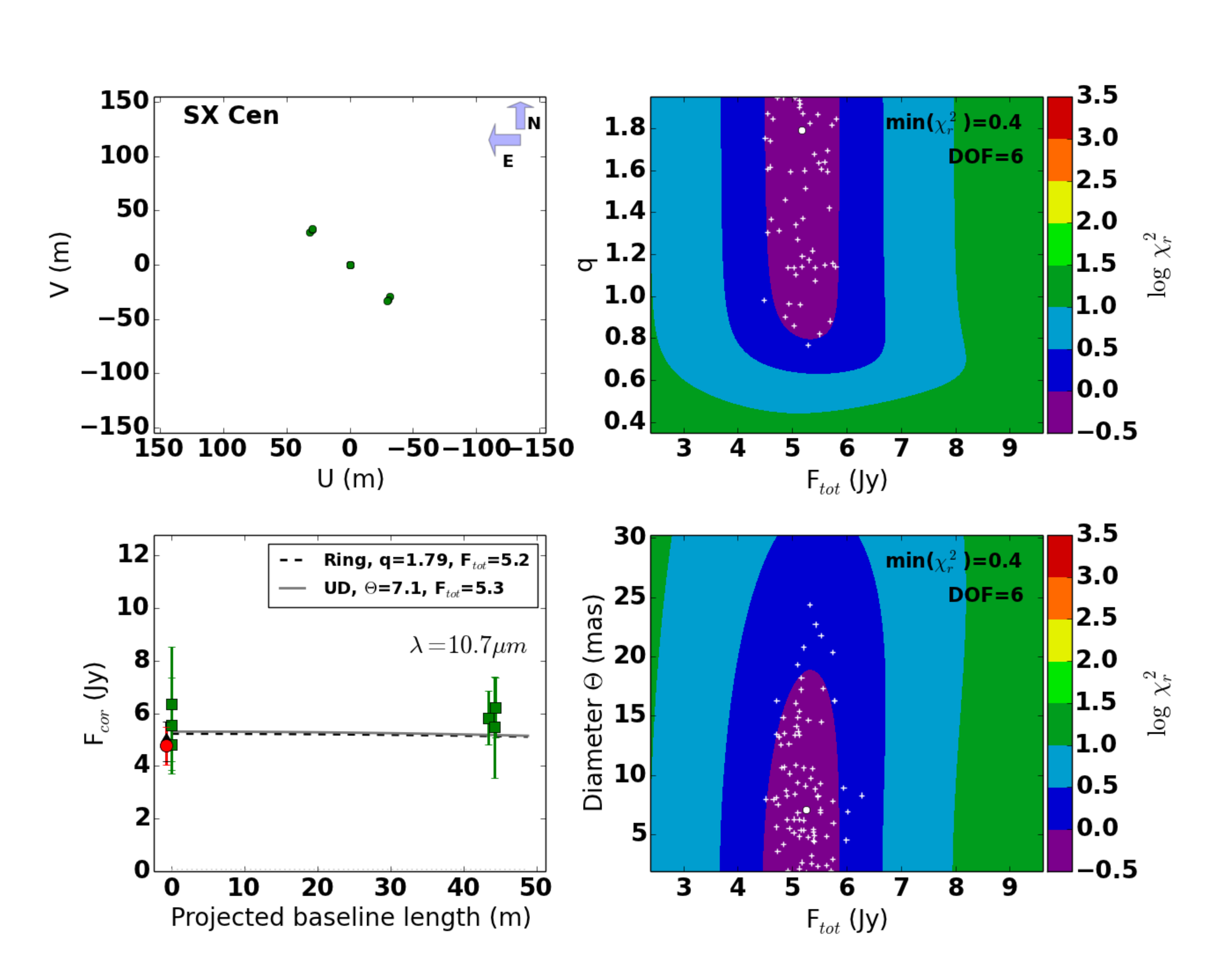}
   \includegraphics[width=9cm]{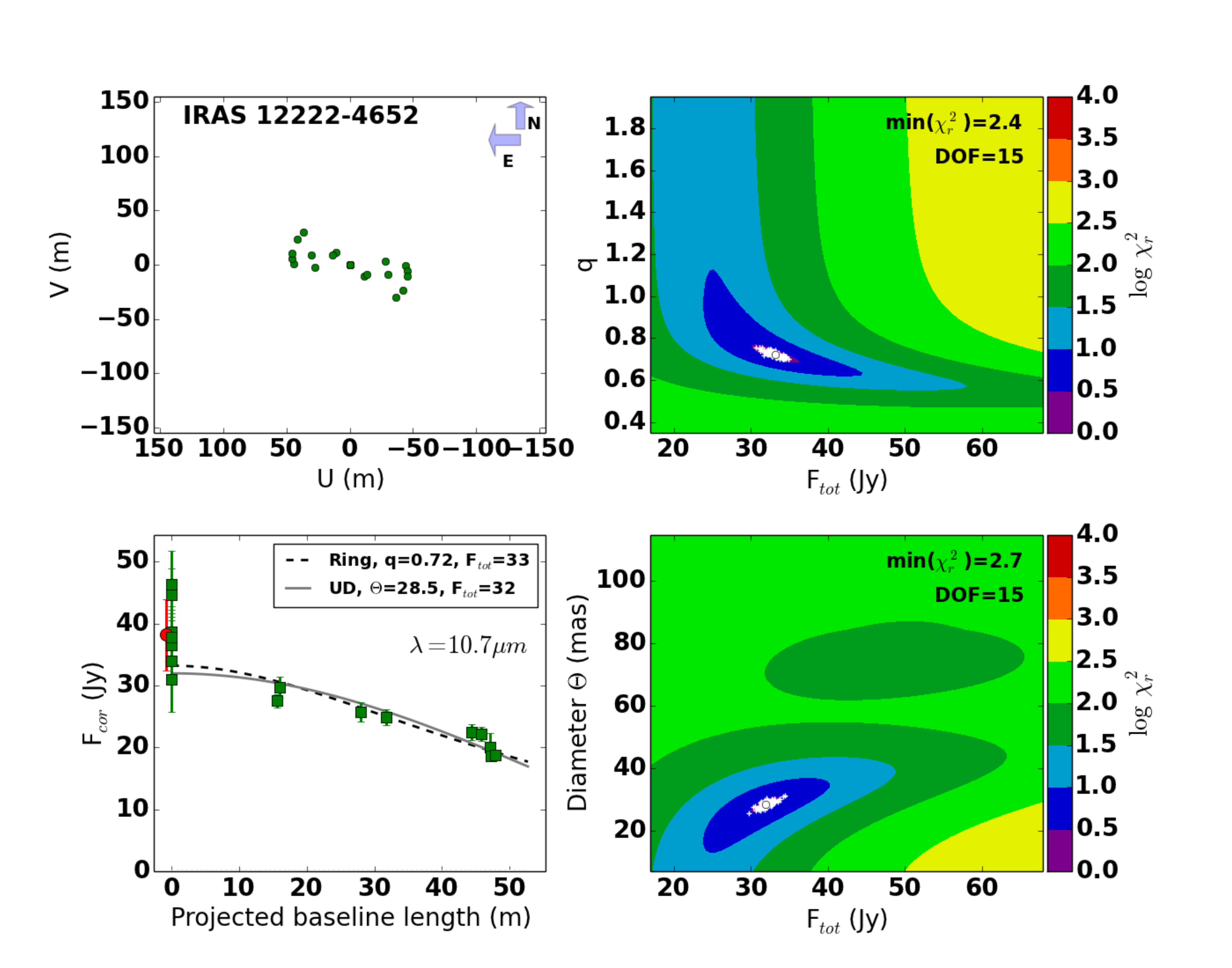}
   \caption{Continued.}
      \label{figure:visfit3}
\end{figure*} 
}

\onlfig{13}{
\begin{figure*}
   \centering
   \includegraphics[width=9cm]{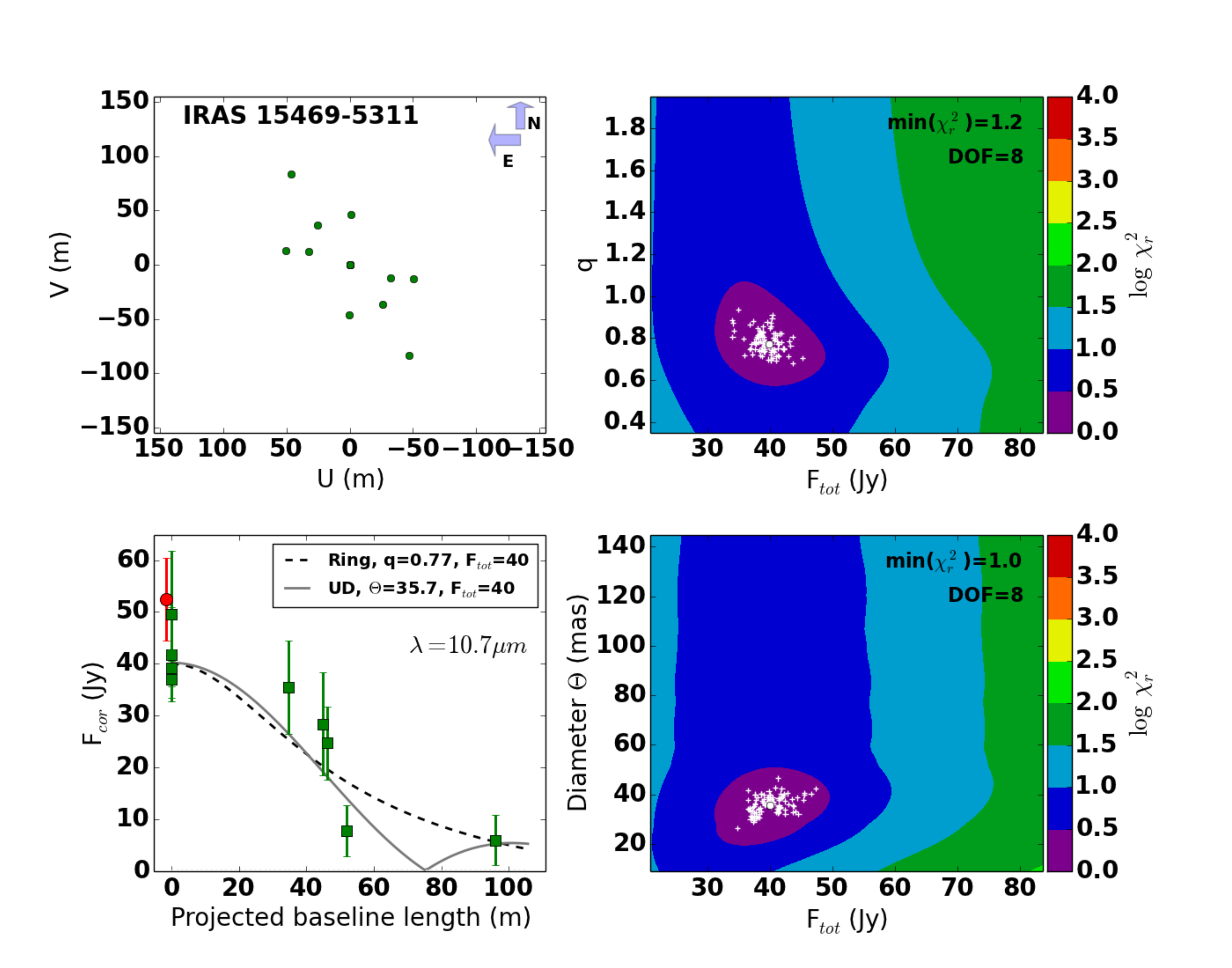}
   \includegraphics[width=9cm]{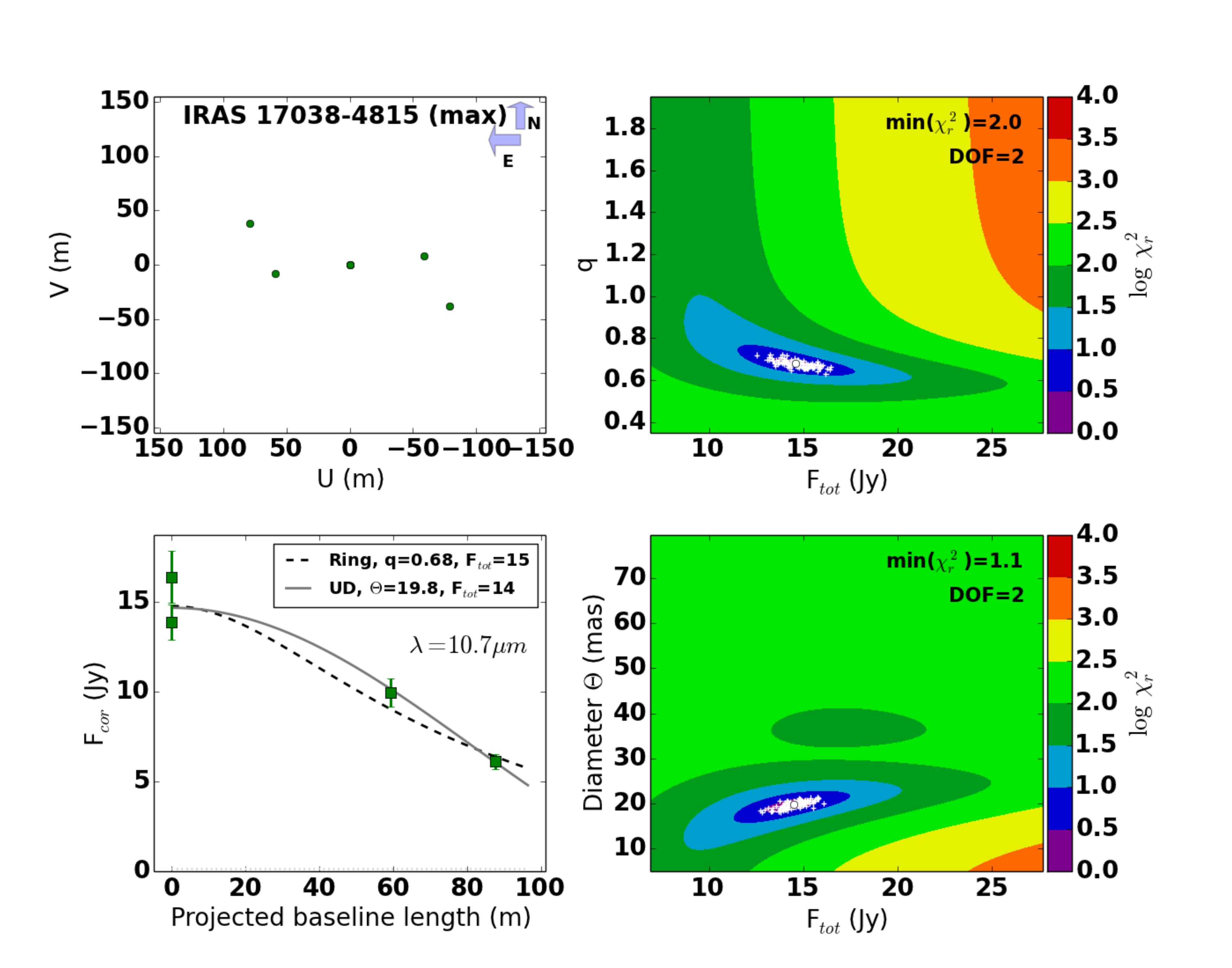}
   \includegraphics[width=9cm]{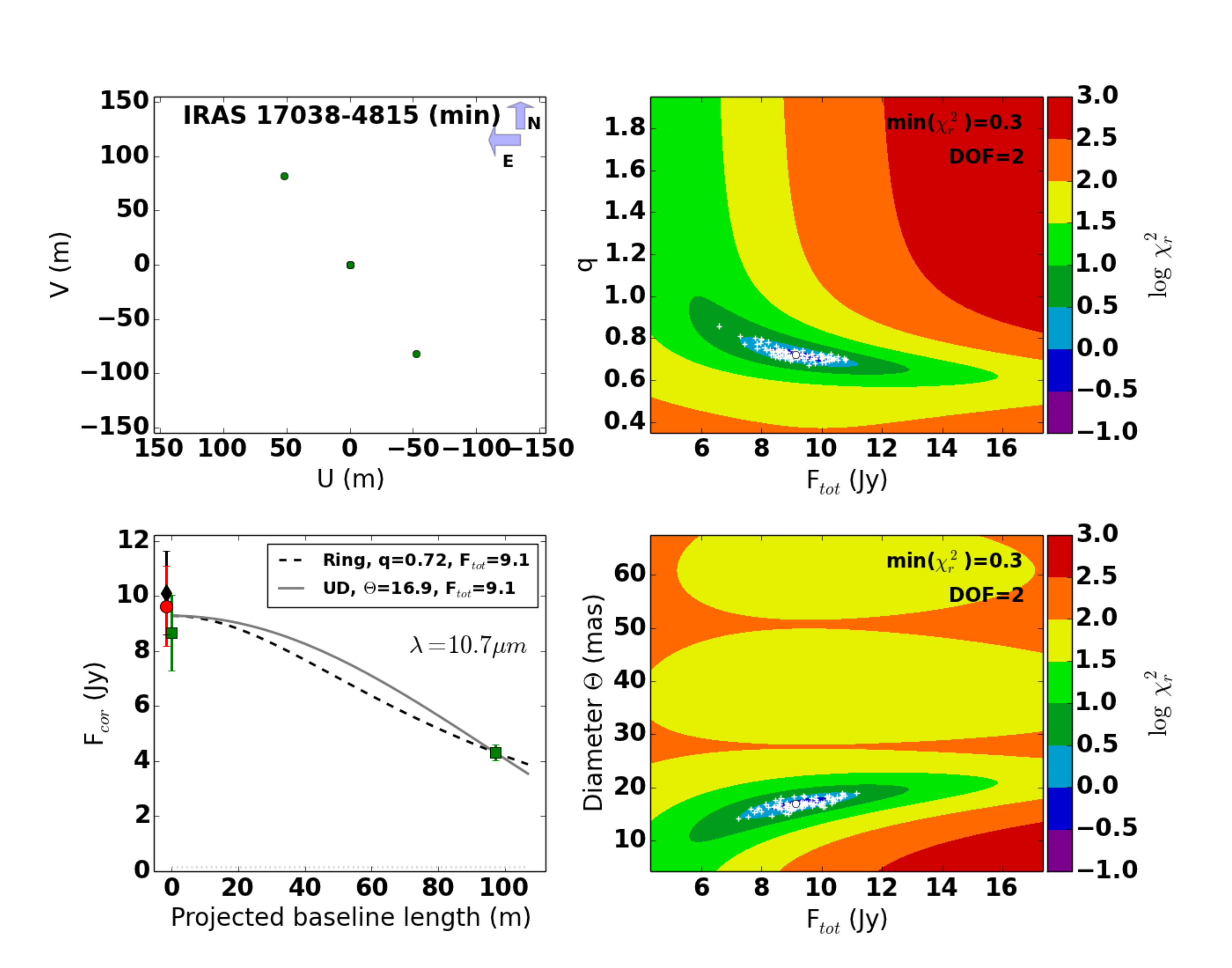}
   \includegraphics[width=9cm]{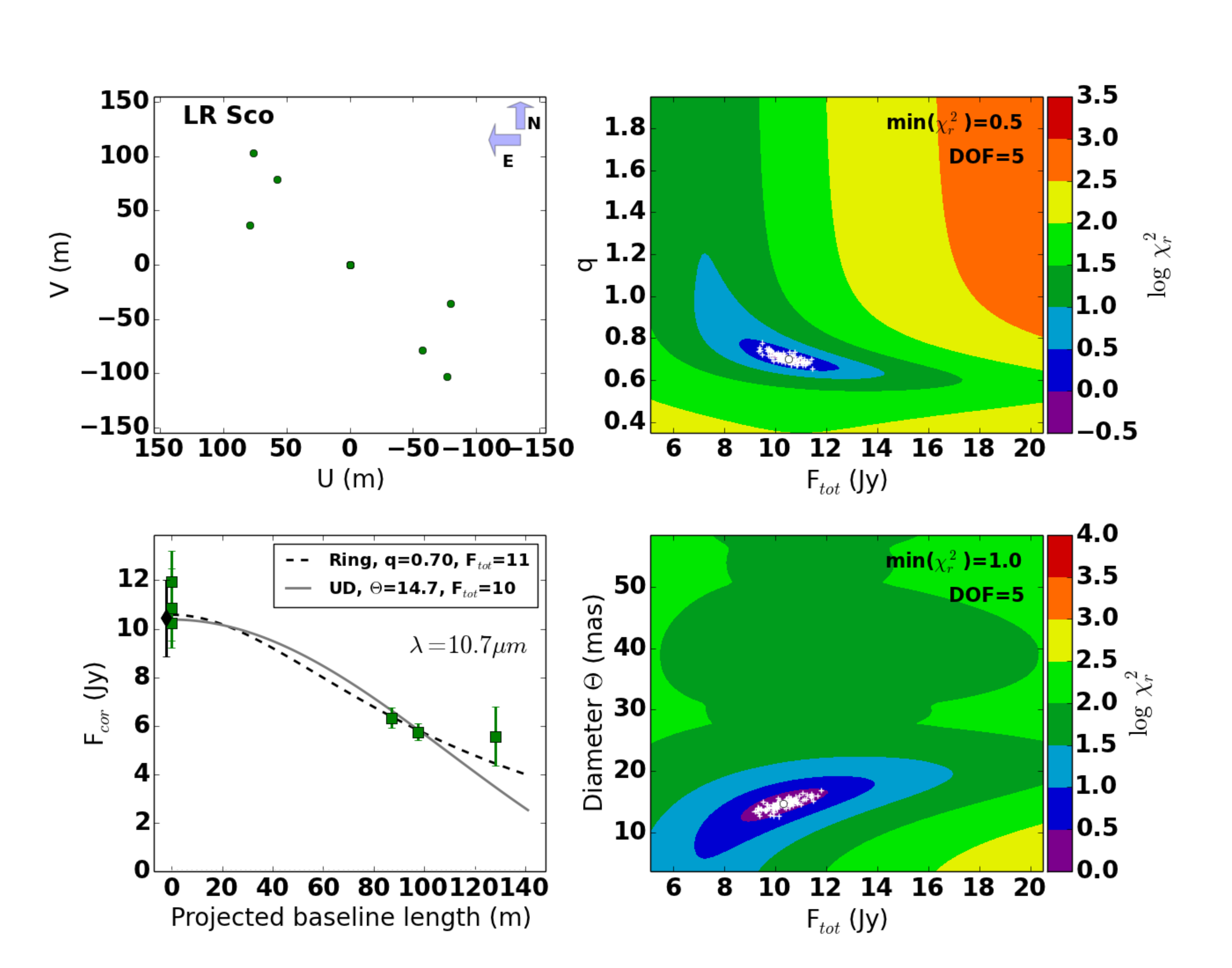}
   \includegraphics[width=9cm]{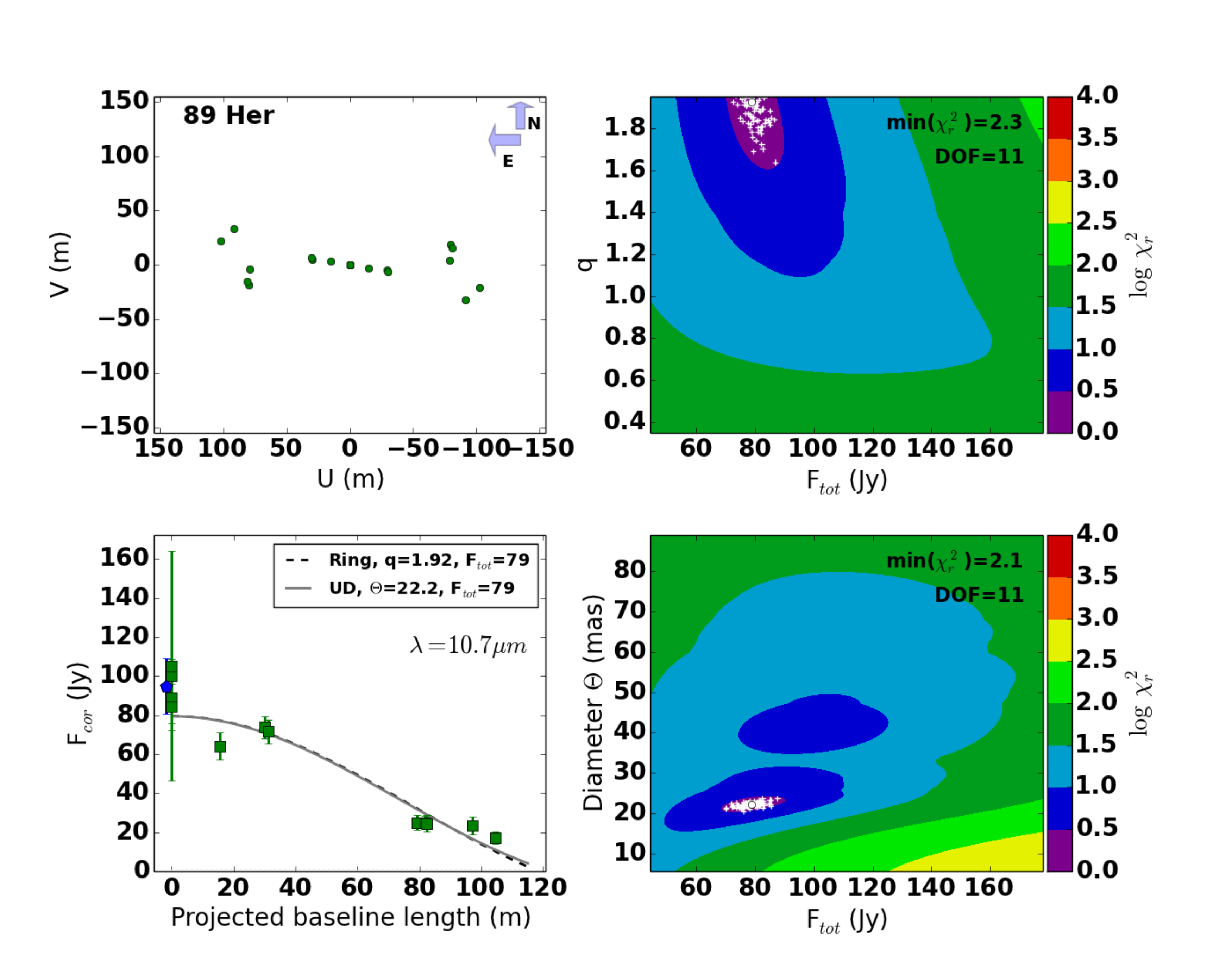}
   \includegraphics[width=9cm]{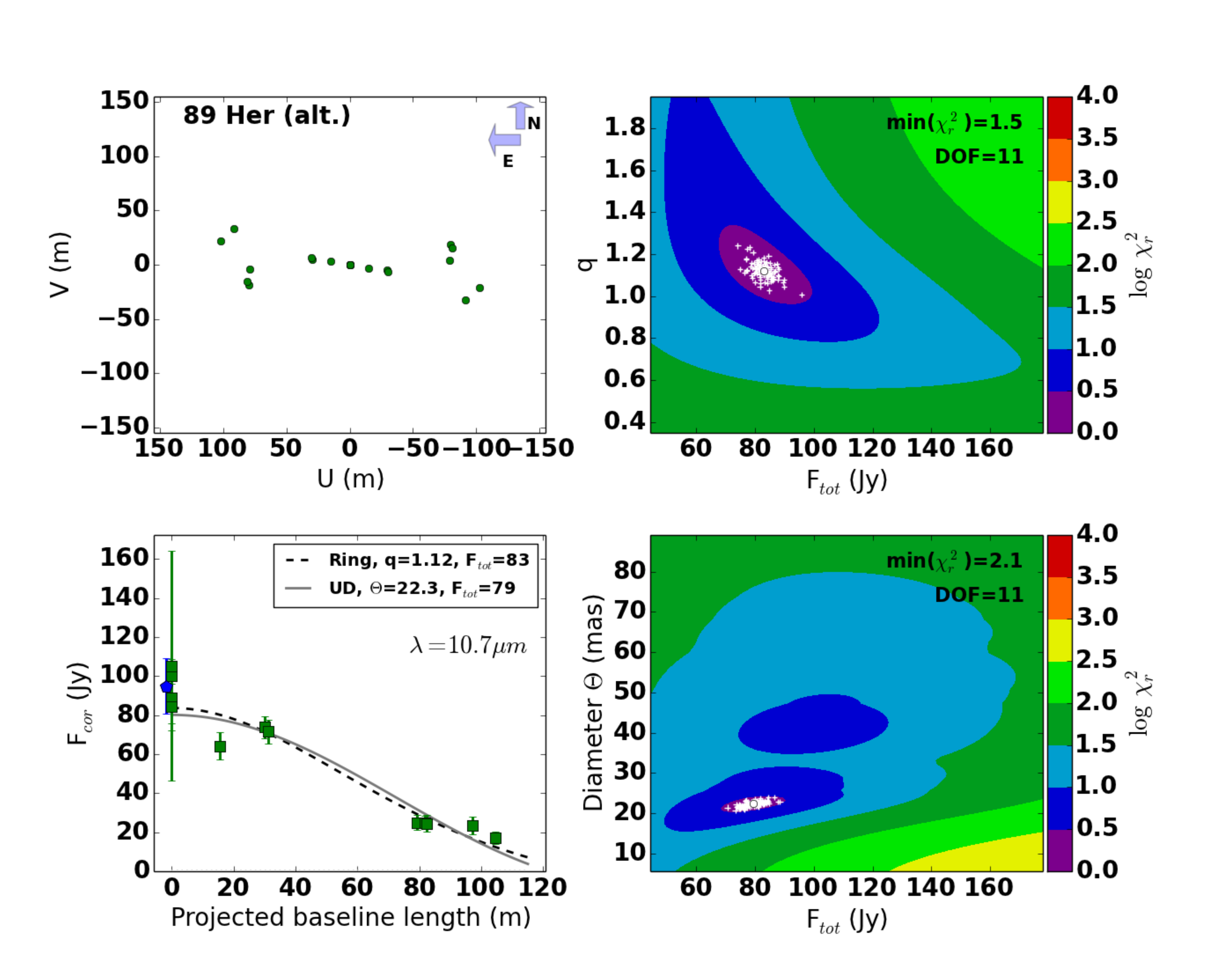}
   \caption{Continued. The lower right panel shows the result for the 89 Her data set with alternative stellar parameters that take
   the large detected fraction of scattered light into account \citep{2013AAHillen}.}
      \label{figure:visfit4}
\end{figure*} 
}

\onlfig{14}{
\begin{figure*}
   \centering
   \includegraphics[width=9cm]{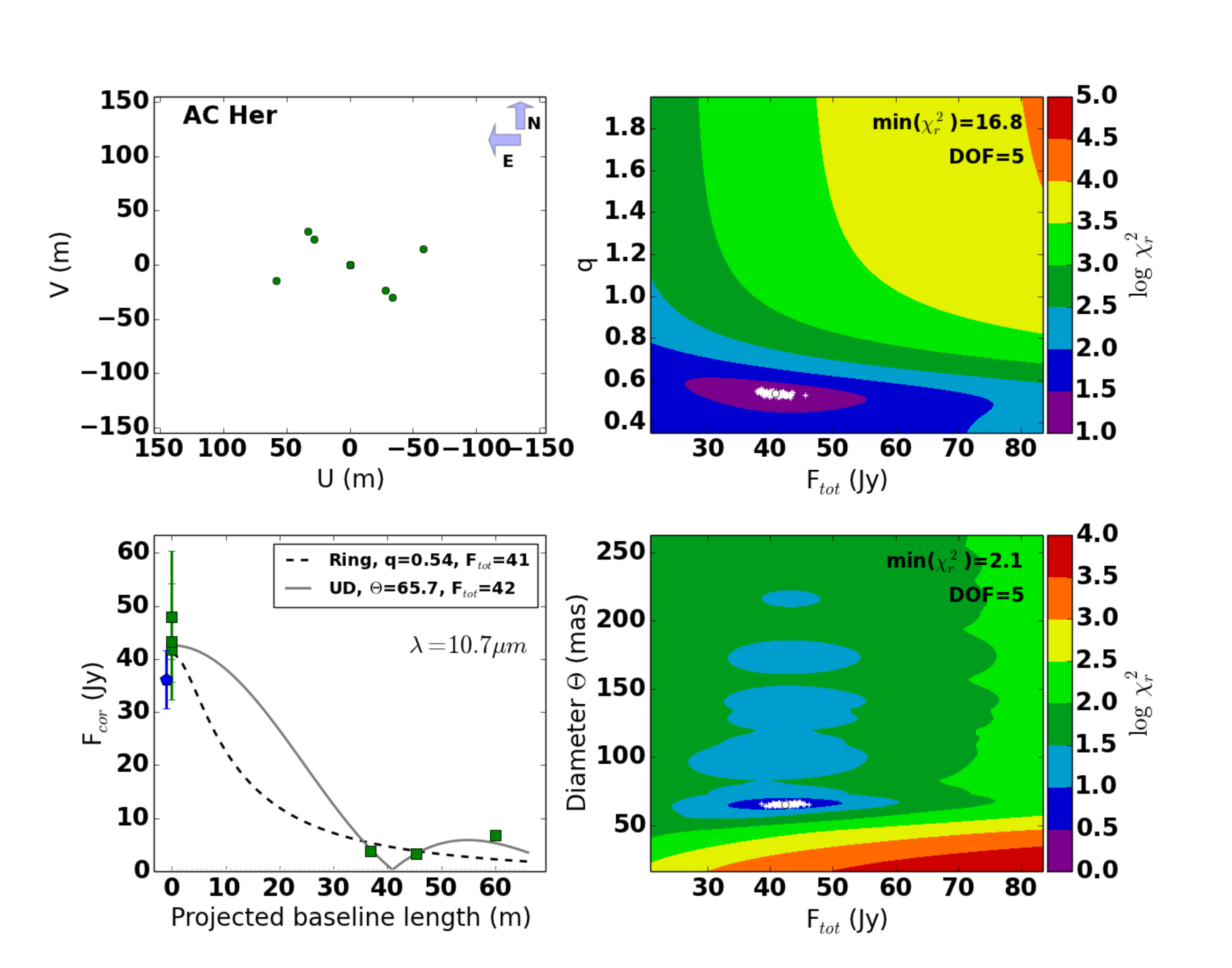}
   \includegraphics[width=9cm]{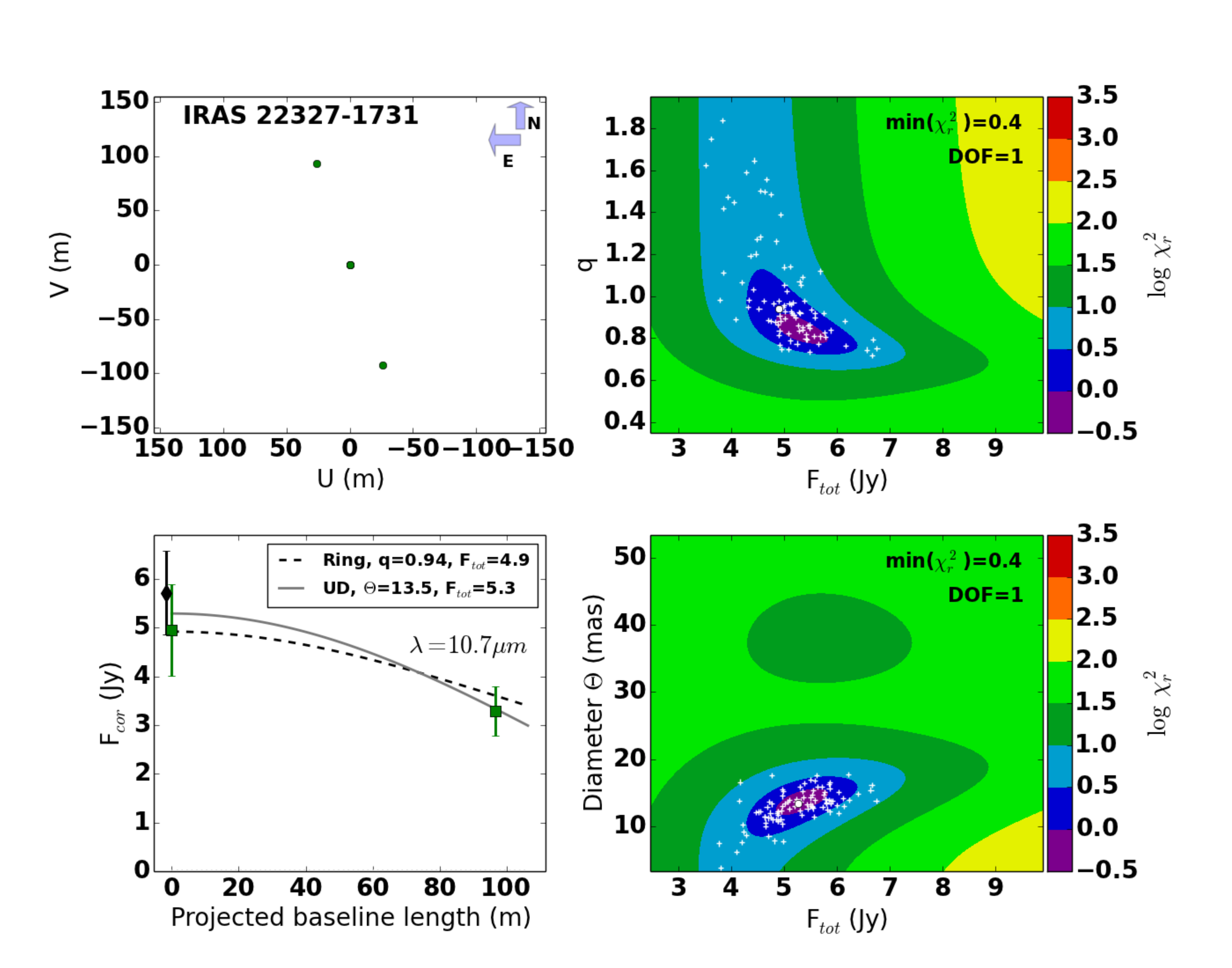}
   \caption{Continued.}
      \label{figure:visfit5}
\end{figure*} 
}

\begin{table*}   
  \caption{Result of the interferometric model fitting. The mid-IR variable sources identified 
  in Sect.~\ref{section:variability} have two entries, one for a minimum and 
  a maximum pulsation phase. For unresolved sources we give 3$\sigma$ upper or lower limits on the determined size parameters,
  otherwise the uncertainties are at the 1$\sigma$ level. We define $\tau{\rm eff}$ as the scale factor $(\tau{\rm eff} = (1-\exp{-\tau})$. For small $\tau$, the effective scale factor reduces to $\tau$.}\label{tab:interferometryPars}
\begin{footnotesize}
\begin{tabular}{clllllllllc}\hline
\multicolumn{1}{c}{Index} &
\multicolumn{1}{l}{IRAS} &
\multicolumn{1}{l}{$\Theta$ $^{+1\sigma}_{-1\sigma}$} &
\multicolumn{1}{l}{F$_{\rm{tot}}$ $^{+1\sigma}_{-1\sigma}$} &
\multicolumn{1}{l}{$\rho_{\rm{sub}}$ $^{+1\sigma}_{-1\sigma}$} &
\multicolumn{1}{l}{q $^{+1\sigma}_{-1\sigma}$} &
\multicolumn{1}{l}{F$_{\rm{tot}}$ $^{+1\sigma}_{-1\sigma}$} &
\multicolumn{1}{l}{$\tau_{\rm eff}$ $^{+1\sigma}_{-1\sigma}$} &
\multicolumn{1}{l}{hlr $^{+1\sigma}_{-1\sigma}$} &
\multicolumn{1}{c}{Bayes Factor B$_{10}$} &
\multicolumn{1}{c}{resolved?} \\
\multicolumn{1}{l}{} &
\multicolumn{1}{l}{} &
\multicolumn{1}{l}{(mas)} &
\multicolumn{1}{l}{(Jy)} &
\multicolumn{1}{l}{(mas)} &
\multicolumn{1}{l}{} &
\multicolumn{1}{l}{(Jy)} &
\multicolumn{1}{l}{} &
\multicolumn{1}{l}{(mas)} &
\multicolumn{1}{l}{} &
\multicolumn{1}{l}{} \\
\hline
1  & 04440+2605 & 13.9$^{+0.6}_{-0.6}$ & 11.8$^{+0.7}_{-0.8}$  & 2.3$^{+0.2}_{-0.2}$     & 1.11$^{+0.06}_{-0.05}$    & 12$^{+1}_{-1}$        & 0.27$^{+0.02}_{-0.02}$     & 4.3$^{+0.2}_{-0.2}$   & 10$^{14}$  & y \\
2  & 07008+1050 & 37$^{+6}_{-7}$       & 3.9$^{+0.5}_{-0.5}$   & 2.2$^{+0.1}_{-0.2}$     & 0.69$^{+0.13}_{-0.06}$    & 3.7$^{+0.6}_{-0.7}$   & 0.02$^{+0.01}_{-0.01}$     & 10$^{+3}_{-3}$        & $\sim$16   & y \\
3  & 07284-0940 (max) & 50.0$^{+0.5}_{-0.5}$ & 125$^{+3}_{-3}$ & 7.0$^{+0.9}_{-0.9}$     & 0.87$^{+0.02}_{-0.02}$    & 157$^{+7}_{-7}$       & 0.196$^{+0.005}_{-0.005}$  & 18.4$^{+0.6}_{-0.6}$  & 10$^{233}$ & y \\
   & 07284-0940 (min) & 41.2$^{+0.7}_{-0.6}$ & 75$^{+3}_{-3}$  & 7.0$^{+0.9}_{-0.9}$     & 0.97$^{+0.04}_{-0.04}$    & 90$^{+5}_{-5}$        & 0.15$^{+0.01}_{-0.01}$     & 15.6$^{+0.9}_{-0.9}$  & 10$^{99}$  & y \\
4  & 08011-3627 & 63$^{+1}_{-1}$       & 121$^{+6}_{-5}$       & 1.7$^{+0.3}_{-0.4}$     & 0.507$^{+0.005}_{-0.005}$ & 121$^{+5}_{-5}$       & 0.25$^{+0.01}_{-0.01}$     & 25$^{+1}_{-1}$        & 10$^{52}$  & y \\              
5  & 08544-4431 & 46$^{+2}_{-2}$       & 153$^{+4}_{-6}$       & 6.2$^{+0.8}_{-0.8}$     & 0.90$^{+0.02}_{-0.02}$    & 171$^{+5}_{-5}$       & 0.32$^{+0.01}_{-0.01}$     & 14.8$^{+0.4}_{-0.4}$  & 10$^{271}$ & y \\        
6  & 09256-6324 & 38$^{+3}_{-3}$       & 102$^{+4}_{-3}$       & 3.8$^{+0.5}_{-0.5}$     & 0.86$^{+0.04}_{-0.03}$    & 103$^{+4}_{-4}$       & 0.42$^{+0.05}_{-0.04}$     & 10.3$^{+0.8}_{-0.8}$  & 10$^{8}$   & y \\      
7  & 10158-2844 & 42$^{+2}_{-2}$       & 23.4$^{+0.5}_{-0.4}$  & 6.6$^{+0.9}_{-0.7}$     & 1.09$^{+0.05}_{-0.04}$    & 23.9$^{+0.6}_{-0.6}$  & 0.062$^{+0.006}_{-0.005}$  & 12.5$^{+0.6}_{-0.6}$  & 10$^{20}$  & y \\  
8  & 10174-5704 & 140$^{+1}_{-2}$      & 53$^{+2}_{-2}$        & 3.0$^{+0.1}_{-0.2}$     & 0.457$^{+0.004}_{-0.004}$ & 54$^{+2}_{-2}$        & 0.018$^{+0.001}_{-0.001}$  & 75$^{+4}_{-4}$        & 10$^{160}$ & y \\
9  & 10456-5712 & 58$^{+3}_{-3}$       & 163$^{+7}_{-7}$       & 7.4$^{+1.2}_{-1.2}$     & 0.92$^{+0.04}_{-0.04}$    & 170$^{+10}_{-10}$     & 0.23$^{+0.02}_{-0.02}$     & 18$^{+1}_{-1}$        & 10$^{19}$  & y \\  
10 & 11385-5517 & 27.9$^{+0.6}_{-0.6}$ & 80$^{+2}_{-2}$        & 4.0$^{+0.4}_{-0.6}$     & 1.00$^{+0.02}_{-0.02}$    & 81$^{+2}_{-2}$        & 0.45$^{+0.02}_{-0.02}$     & 8.5$^{+0.2}_{-0.2}$   & 10$^{101}$ & y \\
11 & 12185-4856 & $< 27$               & 5.3$^{+0.3}_{-0.4}$   & 1.4$^{+0.2}_{-0.3}$     & $> 0.75$                  & 5.2$^{+0.4}_{-0.3}$   & $> 0.1$                    & $< 5.2$               & $\sim$0.1  & n \\
12 & 12222-4652 & 28$^{+1}_{-1}$       & 31.9$^{+0.8}_{-0.8}$  & 1.9$^{+0.2}_{-0.2}$     & 0.72$^{+0.01}_{-0.01}$    & 33$^{+1}_{-1}$        & 0.29$^{+0.01}_{-0.01}$     & 7.7$^{+0.34}_{-0.4}$  & 10$^{33}$  & y \\            
13 & 15469-5311 & 36$^{+3}_{-4}$       & 40$^{+2}_{-2}$        & 2.9$^{+0.3}_{-0.4}$     & 0.77$^{+0.05}_{-0.05}$    & 40$^{+2}_{-2}$        & 0.20$^{+0.05}_{-0.04}$     & 10$^{+2}_{-1}$        & 10$^{11}$  & y \\      
14 & 17038-4815 (max) & 19.8$^{+0.8}_{-0.9}$ & 14.5$^{+0.7}_{-0.7}$ & 1.2$^{+0.2}_{-0.2}$ & 0.68$^{+0.02}_{-0.02}$   & 14.6$^{+0.8}_{-0.8}$  & 0.25$^{+0.02}_{-0.02}$     & 5.8$^{+0.5}_{-0.4}$   & 10$^{18}$  & y \\
   & 17038-4815 (min) & 17$^{+1}_{-1}$ & 9.1$^{+0.9}_{-0.8}$   & 1.2$^{+0.2}_{-0.2}$     & 0.72$^{+0.03}_{-0.03}$    & 9.1$^{+0.9}_{-0.8}$   & 0.20$^{+0.02}_{-0.02}$     & 5.0$^{+0.5}_{-0.5}$   & 10$^{5}$   & y \\
15 & 17243-4348 & 14.7$^{+0.8}_{-0.7}$ & 10.3$^{+0.5}_{-0.6}$  & 0.9$^{+0.2}_{-0.2}$     & 0.70$^{+0.02}_{-0.02}$    & 10.5$^{+0.5}_{-0.5}$  & 0.35$^{+0.03}_{-0.03}$     & 4.0$^{+0.3}_{-0.3}$   & 10$^{10}$  & y \\
16 & 17534+2603 & 22.2$^{+0.6}_{-0.5}$ & 79$^{+3}_{-3}$        & 5.4$^{+0.3}_{-0.3}$     & 1.9$^{+0.1}_{-0.1}$       & 79$^{+3}_{-3}$        & 0.90$^{+0.08}_{-0.07}$     & 7.0$^{+0.2}_{-0.2}$   & 10$^{51}$  & y \\
   & 17534+2603 (alt)\tablefootmark{a} &   &                   & 3.9$^{+0.3}_{-0.3}$     & 1.12$^{+0.04}_{-0.04}$    & 83$^{+4}_{-4}$        & 0.65$^{+0.05}_{-0.05}$     & 7.2$^{+0.3}_{-0.3}$   &   &  \\
17 & 18281+2149 & 65.7$^{+0.7}_{-0.7}$ & 42$^{+2}_{-2}$        & 3.3$^{+0.5}_{-0.4}$     & 0.537$^{+0.008}_{-0.009}$ & 41$^{+2}_{-2}$        & 0.030$^{+0.002}_{-0.003}$  & 38$^{+3}_{-2}$        & 10$^{101}$ & y \\
18 & 19125+0343 & 20.0$^{+0.5}_{-0.6}$ & 25$^{+1}_{-1}$        & 1.6$^{+0.1}_{-0.3}$     & 0.74$^{+0.01}_{-0.01}$    & 25$^{+1}_{-1}$        & 0.33$^{+0.02}_{-0.02}$     & 6.1$^{+0.3}_{-0.3}$   & 10$^{42}$  & y \\
19 & 22327-1731 & $< 20$               & 5.3$^{+0.6}_{-0.7}$   & 1.3$^{+0.2}_{-0.2}$     & $> 0.67$                  & 4.9$^{+0.8}_{-0.7}$   & $> 0.08$                   & $< 6.5$               & $\sim$1       & n \\
\hline
\end{tabular}
\tablefoot{\tablefoottext{a}{The results of the fit when assuming the stellar parameters of \citet{2013AAHillen}, i.e., T$_{\mathrm{eff}} = 6250$~K and
$\theta_\star = 0.45$~mas.}}
\end{footnotesize}
\end{table*}

\section{Notes on some individual targets}\label{section:individualtargets}

\subsection{U Mon (nr. 3) }
The best-fit value of F$_{\textrm{tot}}$ is the same for the uniform disk and the ring model for almost all objects in the sample. 
For one source, U\,Mon, the fitted F$_{\textrm{tot}}$ are discrepant both for the data in minimum as in maximum light.
It has the most extensive uv-coverage among the whole sample, in combination with very precise correlated 
flux measurements, but few total flux constraints. The data for this source were taken on several epochs, either corresponding to a maximum phase
in the pulsation cycle or to a secondary minimum phase. At the maximum phase, the only total flux measurement 
is based on the VISIR fluxes that are extrapolated with the shape of the MIDI spectrum observed for the minimum phase 
(the red cross in Fig.~\ref{figure:visfit2}; see also Fig.~\ref{figure:lightcurves}). 
Formally, we assume a large error for this measurement since it is based on an extrapolation. We note, however, that the VISIR fluxes 
are reliable and significantly larger than the minimum phase MIDI flux at the same wavelengths (Sect.~\ref{section:variability}). 
A visual inspection of the fit to these data shows that the ring model fits the fluxes at all baselines, in contrast 
to the uniform disk model which underestimates the zero-baseline flux. The $\chi^2$-map shows 
that a uniform disk model with the same F$_{\rm{tot}}$ as the ring model, fits the data significantly worse. This is due to the 
excellent sampling in baseline length at a single position angle, which puts precise constraints on the slope of the visibility curve. 

At the minimum phase, this source has a large coverage in baseline position angle, compared to any 
other star in our sample. The distinct fluxes at the various $\sim$70~m baselines, which have different position angles, suggests 
a clear variation in extension of the object as a function of position angle. This cannot be accommodated by our 1D models, as is 
reflected in the atypically high $\chi^2$ values for this data set. Finally, we note that the same uv point corresponding to the 
30~m baseline is probed at the maximum pulsation phase as well. The corresponding correlated fluxes are very different, thereby 
confirming our interpretation of the total flux variations.
The pulsation-induced flux variations and the high-quality correlated fluxes, justify a more detailed analysis in a future paper. 

\subsection{AR Pup (nr. 4)}

AR\,Pup is an RV\,Tauri pulsator and the only star in our sample for which the total infrared luminosity dominates over 
the dereddened optical fluxes (Fig.~\ref{figure:SEDexample5}). This is indicative that we see the disk close 
to edge-on \citep{2006AAdeRuyter}. As the disk is optically thick in the radial direction, the visible component of 
the SED is likely caused by scattering only, with no direct light being observed. This RV\,Tauri pulsator has a 
variable mean magnitude with a period of 1250$\pm$300 days \citep{2007MNRASKiss} which is likely the orbital period \citep{gezer15}.
The angular diameter of the post-AGB photosphere, obtained via the SED modelling, is therefore not very well constrained. 
Our angle coverage in the UV plane is too small to detect deviations from spherical symmetry. AR\,Pup is very well 
resolved despite the short baselines used (Fig.~\ref{figure:visfit2}). 

\subsection{HR4049 (nr. 7)}\label{subsection:HR4049}

The infrared properties of this source are well described in the literature \citep[e.g.]{dominik03,menut09,2014ApJMalek}.
Apart from a compact disk with a very small temperature range \citep{dominik03}, a large-scale outflow has been resolved
in this source by spatially extended emission of the mid-IR Polycyclic Aromatic Hydrocarbon (PAH) features \citep[e.g.]{2013AAAcke}. 

A similar over-resolved flux component is detected with MIDI, since 
the fitted zero-baseline flux of this source is significantly below the actual total-flux measurements (Fig.\ref{figure:visfit2}). 
This is significant as the correlated fluxes of this object have a very high S/N. The fitted zero-baseline flux can 
be considered a good proxy for the total flux of the marginally resolved disk in this object.

\subsection{IRAS10174-5704 (nr. 8)}

The infrared spectrum of this source stands out in our sample. The amorphous silicates dominate with no evidence 
for crystalline processing nor for the presence of large grains. \citet{2011AAGielen} suggest that this source might 
not be a post-AGB star but a massive luminous supergiant in the process of losing mass.
Also in our MIDI modelling, this object stands out (see Figs.\ref{figure:visfit1} and~\ref{figure:visfit2}). It is very well resolved and 
the correlated fluxes reach to the third lobe of the best ring model. This corroborates the suggestion 
of \citet{2011AAGielen} that this object is likely an outflow source with an optically thin wind.

\subsection{89 Her (nr. 16)}
This object was already subject to a very extensive multi-wavelength interferometric study \citep{2013AAHillen,2014AAHillen}.
One of the main conclusions was that in the optical, a resolved component is present which accounts for 35-40\% of 
the visible photons. This component is associated with optical scattering by the circumbinary disk and an outflow or jet emanating from 
the central system. For 89 Her, we present two fits (see Fig.~\ref{figure:visfit2}): one with and one without 
taking the scattered light into account, taking the Teff and angular diameter from here or as determined in \citep{2013AAHillen}. 
The $\chi^2$ improves when the scattered light is taken into account and the resulting $q$ value decreases to a less extreme value.

\subsection{AC Her (nr. 17)}

The results on AC\,Her are depicted in Fig.~\ref{figure:visfit2}. The source is very well resolved and the ring model allows to 
account for the higher correlated flux point of the 60m baseline coming from the second lobe in the visibility profile.
This source has been extensively modelled by \citet{2015AAHillen}. The radiative transfer modelling combined with the interferometric observables allowed to conclude that the disk in the AC Her system is in a very evolved state, 
as shown by its small gas/dust ratio and large inner hole. Indeed the inner rim is at significantly longer distance from the central source
then the sublimation radius.

In our parametric modelling of the sample, which assumes that the dust disk starts at the sublimation radius, these physical properties of the disk around AC\,Her translate in a low $q$ value (so flat temperature gradient) and a large half-light radius. 
The fitted uniform disk radius is significantly smaller than the half-light radius.

\section{Sample analysis}\label{section:analysis}
\subsection{The half-light radius} \label{subsection:halflightradius}

\begin{figure}
   \centering
   \includegraphics[width=9cm]{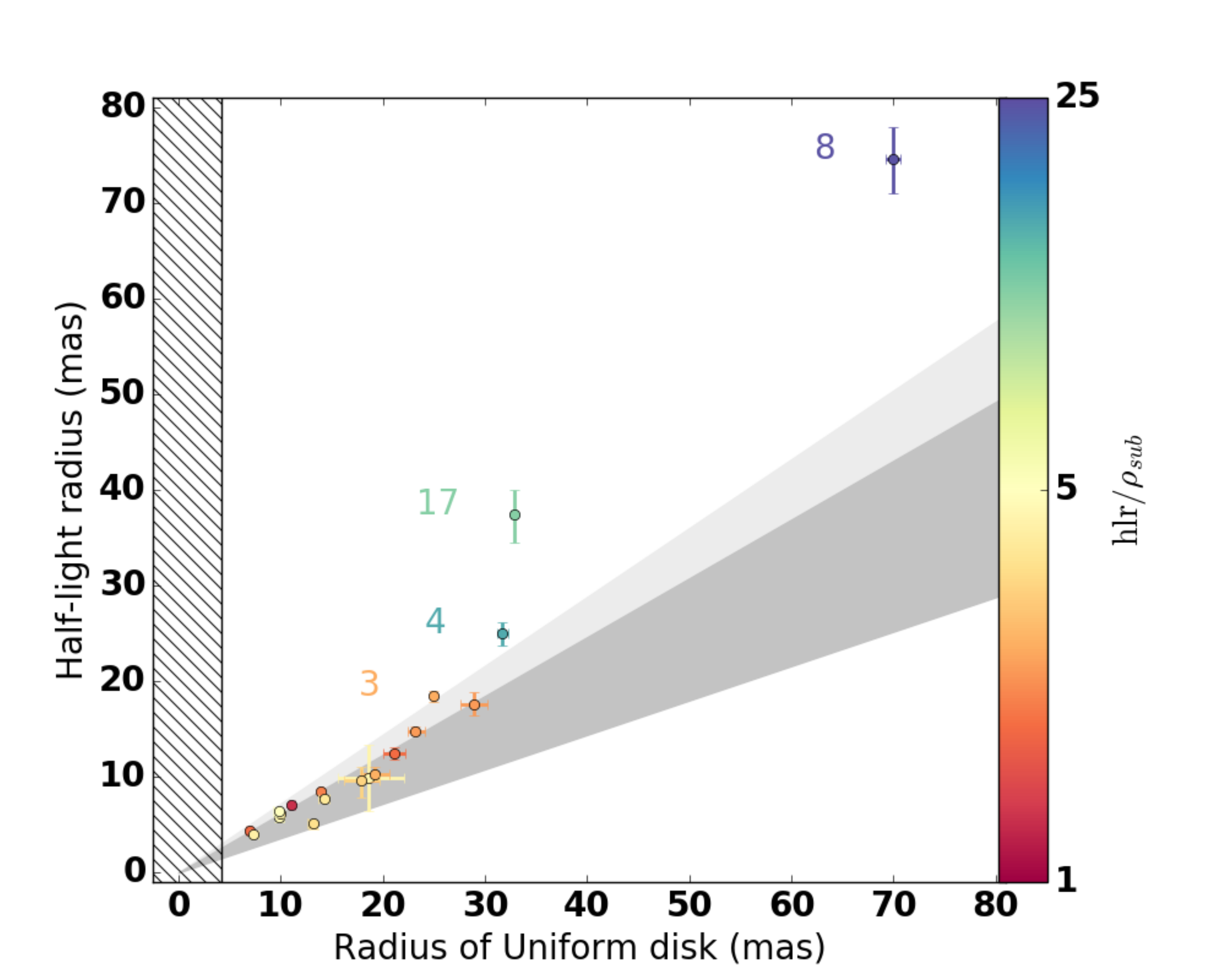}
   \caption{Radius of the fitted uniform disk versus the half-light radius. The different regions and colours are described in the text (Sect.~\ref{subsection:halflightradius}). 
   }
   \label{figure:UD-hlr}
\end{figure} 

In Fig.~\ref{figure:UD-hlr} we compare the radius of the uniform disk with the half-light radius.
The dark grey area denotes the region theoretically 
occupied by ring models for which the visibility $\mathrm{V}>0.4$. It is computed by fitting the uniform disk model to an ensemble of 
ring models. The grid of ring models covers the full range of $\rho_{\rm{sub}}$ of our sample  and with $q \in \{0.4;2.5\}$. The light grey area 
is an extension of this grid up to $\mathrm{V}>0.3$. The hatched area denotes the region inaccessible by the interferometer. 
The colour of the observed sources (see colour bar on the right of the Fig.\ref{figure:UD-hlr}) denotes the ratio of the half-light radius over the sublimation radius (see the colour bar on the right). 


Most objects occupy the dark grey region and are thus only moderately resolved. Their half-light radii 
are a bit larger than the sublimation radius (typically a factor of $\sim$3). This is expected for a
disk with a fairly optically thick inner rim located at the sublimation radius and an outer disk surface that is 
not flaring very strongly. The region near the inner rim of the disk dominates the flux at 10.7 $\mu$m.

Four sources which stand out are U~Mon (nr.3), AR~Pup (nr.4), IRAS~10174-5704 (nr.8), and AC~Her (nr.17) (see previous section for the details). 
These are the sources with the smallest $q$ values (see Table~\ref{tab:interferometryPars}) and the largest ratio of half-light radius over
sublimation radius. Since these objects have the largest appearance on the sky, they are also the most resolved -- hence have 
visibilities below 0.4 -- so that the details of the source morphology become important. As discussed previously, 
for two of these sources, IRAS~10174-5704 and AC~Her, the 10.7 $\mu$m flux seems to emanate predominantly from radii well beyond 
the sublimation radius. The other two objects most likely only deviate because they are nearby, thus
strongly resolved, and well inclined. 

\subsection{Brightness temperatures} \label{subsection:BT}

\begin{figure}
   \centering
   \includegraphics[width=9cm]{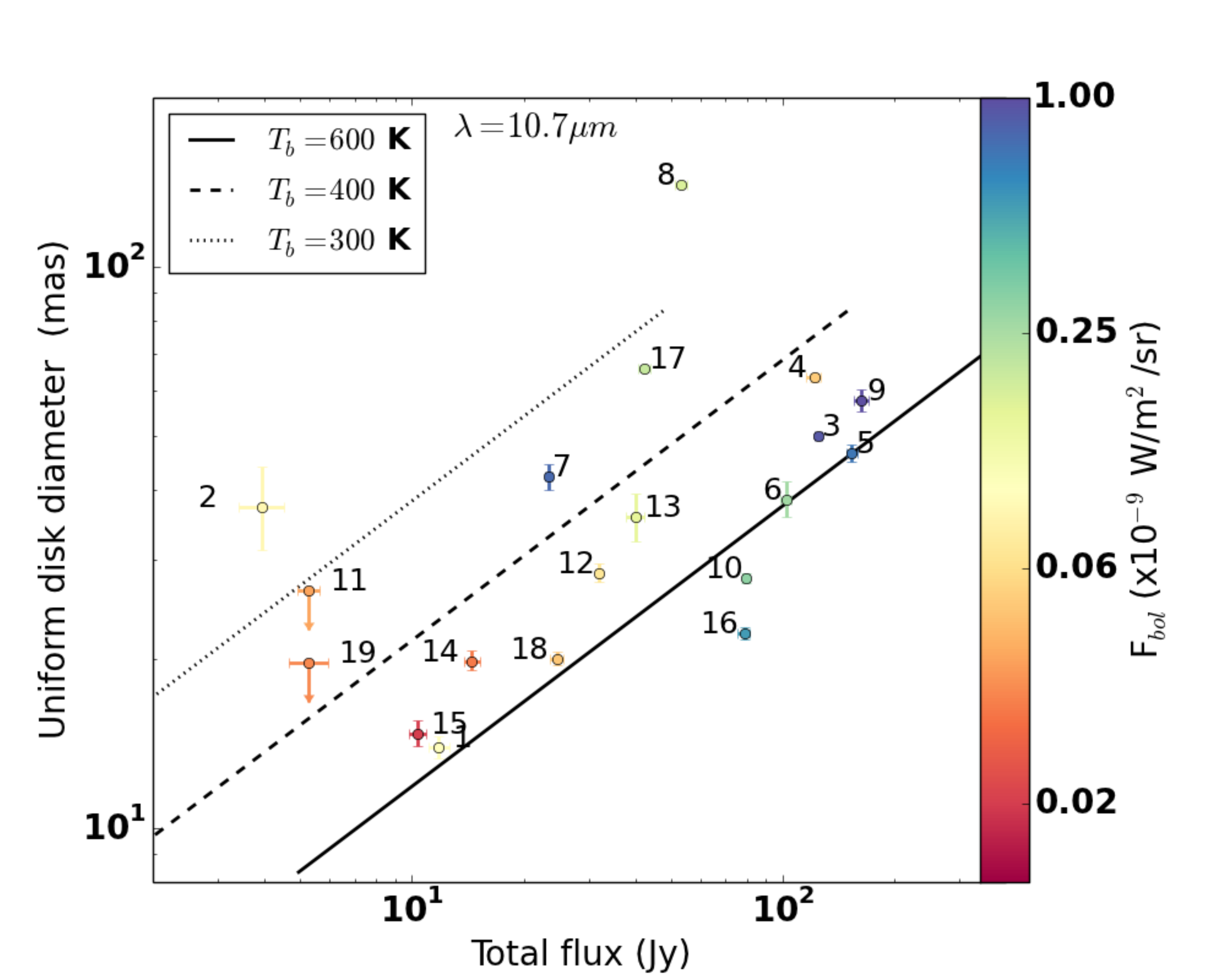}
   \caption{Uniform disk diameter versus total flux at 10.7~$\mu$m. The numbers indicate the index of the source in Table~\ref{tab:interferometryPars}. Symbols are explained in the text (Sect.~\ref{subsection:BT}).
   }
   \label{figure:Ftot-UD}
\end{figure}

We can use the uniform disk diameter to quantify the brightness temperature using the total flux at 10.7$\mu$m (see Fig.~\ref{figure:Ftot-UD}).
The numbers in the figures indicate the index of the source in Table~\ref{tab:interferometryPars} 
and are plotted to the left and right of the corresponding location for even and uneven indexes, 
respectively. The sources are coloured according to the stellar bolometric flux (see the colour bar on the right). 
The black dotted, dashed, and full lines represent uniform disks with brightness temperatures of 300, 400 and 600~K, respectively.
The dynamic range in 10.7~$\mu$m flux is similar to that in bolometric flux ($\sim$x50).

Overall the sources show a limited range in brightness temperature ($\geq$50\% of the sample falls between 400 and 600~K) 
and an increase in flux is mainly due to an increase in the extent of the source. The correlation between the 10.7~$\mu$m and 
stellar bolometric flux indicates that the observed range in angular extent mainly reflects variations in distance and 
stellar luminosity of the source, and that the disks are physically very similar within the sample. 

Noticeable outlier is again IRAS~10174-5704 (nr.8), but also HD\,52961 (nr.2) which is rather large for its flux. 
We suspect this is due to the contribution of PAH emission \citep{2009AAGielen} which likely has a different spatial 
scale than the thermal emission of the dust disk, similarly that HR\,4049 (nr.7, Sect.~\ref{subsection:HR4049}).

\subsection{A size-colour relation} \label{subsection:size-colour}

\begin{figure}
   \centering
   \includegraphics[width=9cm]{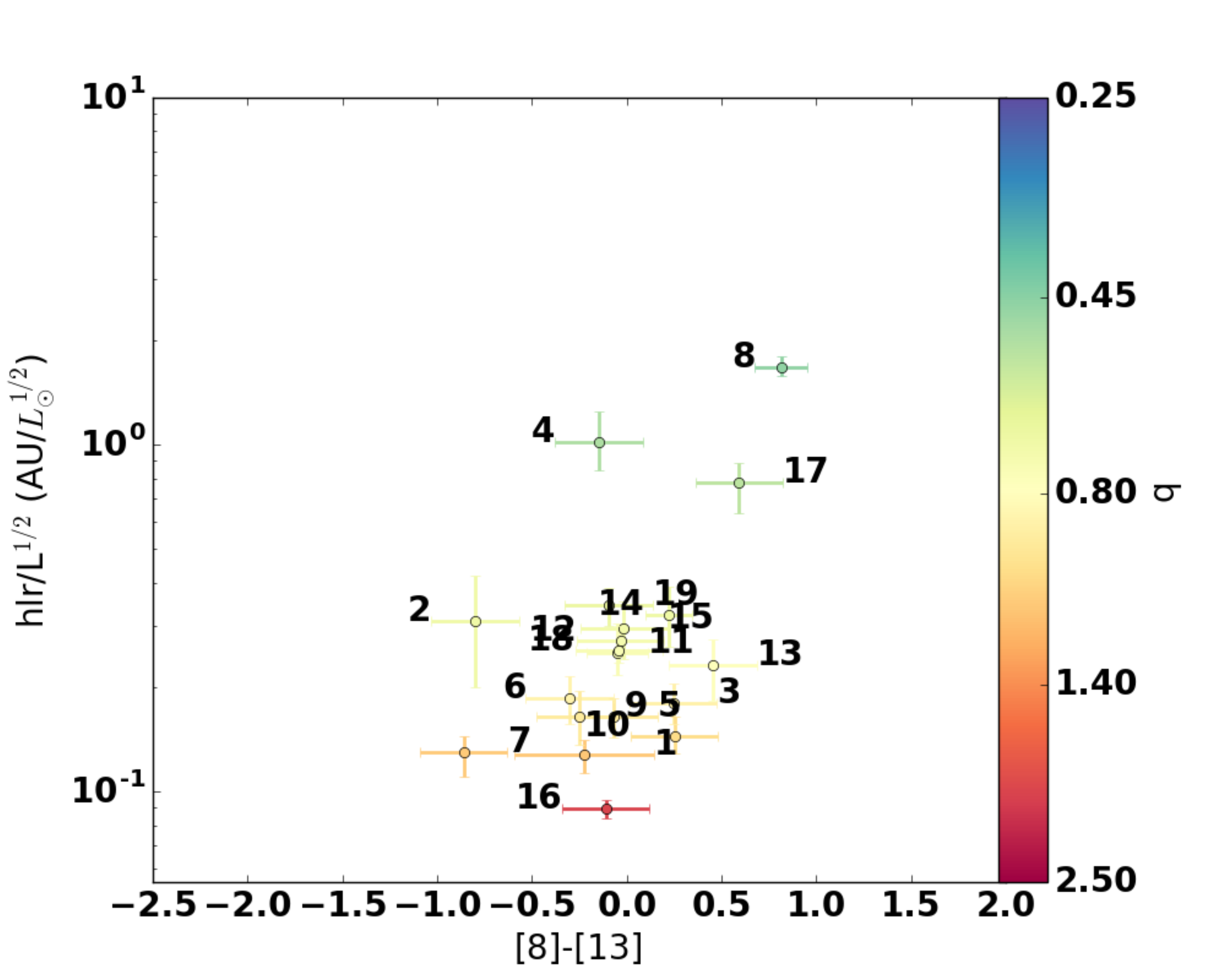}
   \caption{Size-colour diagram of our sample of post-AGB stars. We plot the half-light radius (hlr), normalized with the square root of the luminosity, versus 
   the 8-13~$\mu m$ colour (-2.5$\log (F_{\nu,8}/F_{_\nu,13})$). See Sect.~\ref{subsection:size-colour}.}
   \label{figure:size-colour}
\end{figure}

In Fig.~\ref{figure:size-colour} we compare the half-light radius to the 8-13~$\mu m$ colour ($-2.5\log (F_{\nu,8}/F_{_\nu,13})$). 
The horizontal axis represents the spectral slope in the N-band. 
To physically compare sources on the size-colour diagram, the effect of the highly uncertain distance to the source should be removed. 
Hence, the distance is divided away by normalizing the physical half-light radius with the square root of the luminosity. Arrows indicate 
3$\sigma$ upper limits. The numbers correspond to the index of the source in Table~\ref{tab:interferometryPars} and are plotted 
to the left and right of the corresponding source for even and uneven indexes, respectively. 
The colours indicate the q-value in the interferometric fit.
   
The homogeneity of the sample is again striking: all objects cluster in the same region, with very similar temperature 
power indices of the ring model. The largest sources (IRAS~10174-5704 (nr.8) and AC\,Her (nr.17) are also the reddest. 
AR\,Pup (nr.4), on the other hand, appears large but has a similar IR-colour as the others. As this source is 
seen close to edge-on, the luminosity determined by integrating under the dereddened photosphere 
(see Sect.~\ref{section:SEDfitting}) is a strong underestimate of the total luminosity of the central object.


\section{Comparison with a RT model grid}\label{section:RTgrid}

The most important conclusion from our global analysis is that the sample of objects is very homogeneous, both in SED characteristics and 
in N-band interferometric behaviour. Our previous detailed radiative transfer modelling \citep{2014AAHillen,2015AAHillen} 
of 89\,Her and AC\,Her showed that a model of a passively irradiated disk in hydrostatic equilibrium can provide a good match with
both the energetics and the multi-colour interferometric observables of both objects. 

Given the limited spatial constraints in our MIDI survey, we will not provide detailed models for all objects here,
but we do want to test our assertion that the model of a passively irradiated disk gives a good match for the whole sample.

We therefore compute a grid of physical models. 
Similarly to our previous work, we used the radiative transfer code MCMax \citep{2009AAMin} to compute disk structures. 
MCMax is based on the Monte Carlo method \citep{1999AALucy,2001ApJBjorkman}. Photon
packages are randomly emitted by a source at the origin of the coordinate system, which are then
absorbed or scattered by dust that is distributed in an axi-symmetric geometry.  The thermal structure is
hence determined from the interaction of the dust with the stellar radiation (i.e. a passively heated disk
in which the gas is in thermal equilibrium with the dust). The radial distribution of the gas and dust is a
basic input of the model, but the vertical structure is obtained by solving the equation of hydrostatic
equilibrium (i.e. the vertical component of the local gravitational force is balanced by the local gas
pressure gradient). The temperature and density profiles in the disk are iteratively determined; 
taking a grain size distribution with size-dependent settling into account in a self-consistent way \citep{2012AAMulders}.

Overall the same strategy is adopted as in \citep{2015AAHillen} and we summarize the main properties 
of our radiative transfer models:

\begin{itemize}
\item the disk is vertically in
hydrostatic equilibrium,
\item we assume isotropic photon scattering by the dust,
\item the composition of the dust is assumed to be an ISM-like mixture of silicates in the DHS (Distribution
of Hollow Spheres) approximation \citep{2007AAMin} (the alternative mixture has the same composition but with an additional 15\% of 
amorphous Carbon),
\item the size of the dust grains follow a power-law distribution of which the exponent is varied,
\item a grain-size-dependent settling of dust, counteracted by turbulence (with a variable turbulence strength parameter), 
is included self-consistently \citep{2012AAMulders} ,
\item  we adopt a double-power-law formalism to parametrize the
surface density distribution.
\end{itemize}

\subsection{The RT grid}

In Table~\ref{table:radiativetransfer} we list the grid parameters and their assumed values.
The grid of models contains both geometric dimensions, in the form of a range of inclinations 
and baselines (angles and lengths) and a series of physical dimensions, 
related to the structure and composition of the disk model. 

The dominant contributor to the N-band flux comes from the warm dust near the inner rim. 
Our model grid therefore focuses on the structure of the inner part of the disk, and on the dust properties.
Hence we fix the outer radius as well as the outer surface density power law.
We do apply different values for the inner surface density power law and for the turnover radius. 
Further dimensions in our physical grid comprise the grainsize distribution, the turbulence strength parameter $\alpha_{\textrm{turb}}$ 
and the dust mass. We assume that the disk inner radius is at the sublimation radius, according to Eq.~\ref{eq:subradius}.

We fix the central star properties, i.e., the central mass (total gravitational mass of the binary: 0.6 + 0.6 M$_\odot$), 
the luminosity, and the effective temperature of the post-AGB component. The assumed temperature is equal to the median over the observed sample. 
The luminosity is taken to be a typical post-AGB value of 6000~L$\odot$. The distance is then fixed by matching 
the median angular diameter in our sample with the assumed luminosity.
We compute the infrared [8]-[13]-colour for all our models. We fit the ring model to each synthetic visibility.
Despite assuming values for certain parameters, we still have a total of 
4320 models per dust composition type (times 98 baselines).

\begin{table}
 \caption{Input parameters of the radiative transfer model grid.}             
 \label{table:radiativetransfer}      
 \centering                          
 \begin{tabular}{l l r @{ -- } l r @{ $\pm$ } l}        
 \hline\hline\noalign{\smallskip}
 Parameter & Value \\
 \noalign{\smallskip}\hline\noalign{\smallskip}
 Star &  \\
 Mass M$_{\star}$ & 1.2 M$_{\odot}$  \\
 Luminosity L & 6000~L$_{\odot}$ \\
 Effective temperature T$_\mathrm{eff}$& 6250~K  \\
 Distance d & 2~kpc  \\
 & \\
 Disk &   \\
 outer surface density power p$_\mathrm{out}$  & 1.0 \\
 inner surface density power p$_\mathrm{in}$   & 1.0, 0.0, -1.0, -2.0  \\
 dust mass M$_\mathrm{dust}$  & 10$^{-4}$, 10$^{-3}$~M$_{\odot}$ \\
 inner radius R$_\mathrm{in}$  & R$_\mathrm{sub}$ \\
 outer radius R$_\mathrm{out}$ & 300~AU \\
 turnover radius R$_\mathrm{mid}$/R$_\mathrm{in}$  & 2.0, 2.5, 3.0 \\
 minimal grain size a$_\mathrm{min}$  & 0.01, 0.10, 1.00~$\mu$m  \\
 maximal grain size a$_\mathrm{max}$  & 1~mm  \\
 grain size distribution $q_\mathrm{g}$ & -3.50, -3.25, -3.00, -2.75 \\
 dust-to-gas ratio & 0.01 \\
 turbulence parameter $\alpha$ & 10$^{-4}$, 10$^{-3}$, 10$^{-2}$ \\
 inclination & 10$^\circ$, 30$^\circ$, 45$^\circ$, 55$^\circ$, 65$^\circ$, 75$^\circ$ \\
 dust composition & (1) \citet{2007AAMin} \\ 
                  & (2) \citet{2007AAMin} $+$ \\ 
                  & 15\% amorphous Carbon \\ 
 & \\
 Interferometric model observations & \\
 Baseline lengths & 10, 20, 30, ..., 130, 140~m \\
 Baseline angles & 0, 15, 30, 45, 60, 75, 90$^\circ$ \\
\noalign{\smallskip}\hline
\end{tabular}
\end{table}

\subsection{Model comparison}

We now compare the diversity in the models to the sample of observed post-AGB stars. To do so, we fit the ring model to each synthetic visibility.

\begin{figure*}[!htb]
   \centering
   \includegraphics[width=9cm]{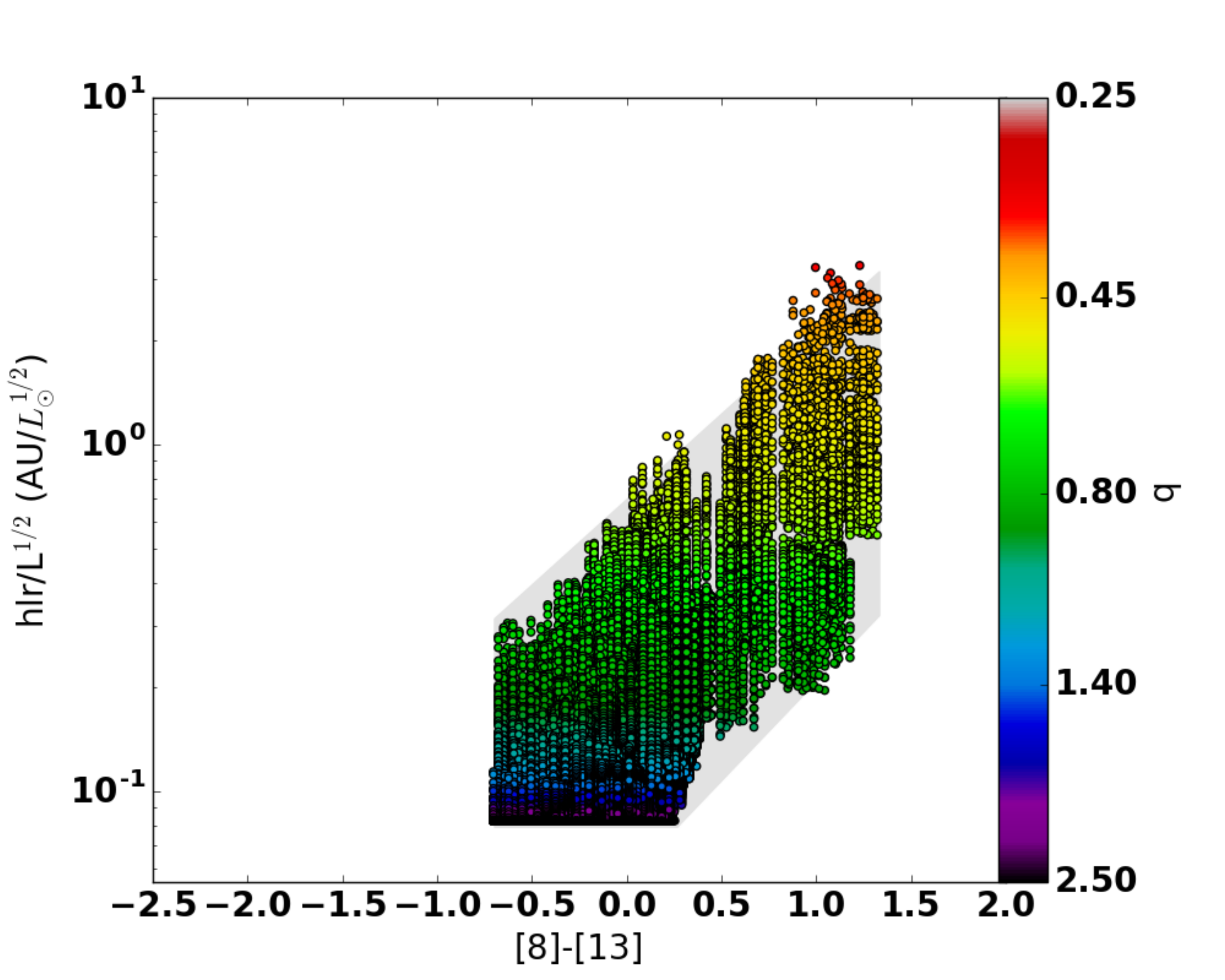}
   \includegraphics[width=9cm]{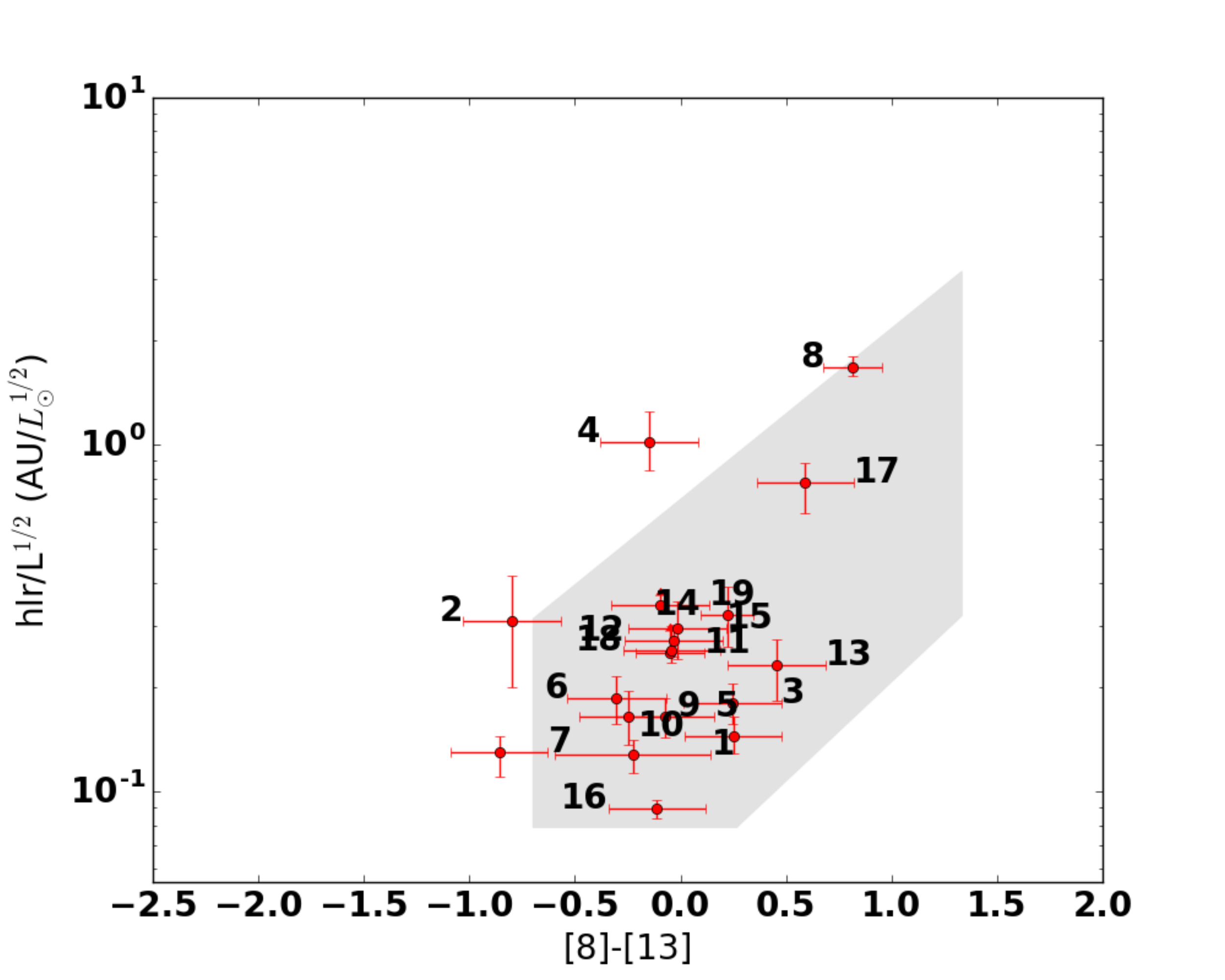}
   \caption{Left: Size-colour diagram with the distribution of observables computed for all models in the radiative transfer grid
   (i.e., one point corresponds to one combination of MCMax disk parameters and one baseline length+orientation).
   The colour coding is according to the q-value of the fit. Right: The location of the post-AGB disk stars 
   within the range of models, shown as grey-shaded region  \label{figure:modelsize-colour}}
\end{figure*}

The left panel of Fig.~\ref{figure:modelsize-colour} shows the position of each synthetic measurement in the size-luminosity diagram, again
colour coded according the $q$ value. The grey region embraces the range covered by the model grid.  
The figure only shows the models computed with the first dust mixture (Table~\ref{table:radiativetransfer}). The models 
computed with the second dust mixture shift the overall coverage only slightly upwards, so we limit our analysis to 
the models given in the figure. 

Models that give rise to an apparently small disk (i.e., a high $q$ value and steep temperature gradient) are blue and vice versa. 


In the right panel of Fig.~\ref{figure:modelsize-colour} we illustrate that the grid of passive disks in hydrostatic equilibrium 
cover very well our stars in this representation of the interferometric observables.

\begin{figure}
   \centering
   \includegraphics[width=9cm]{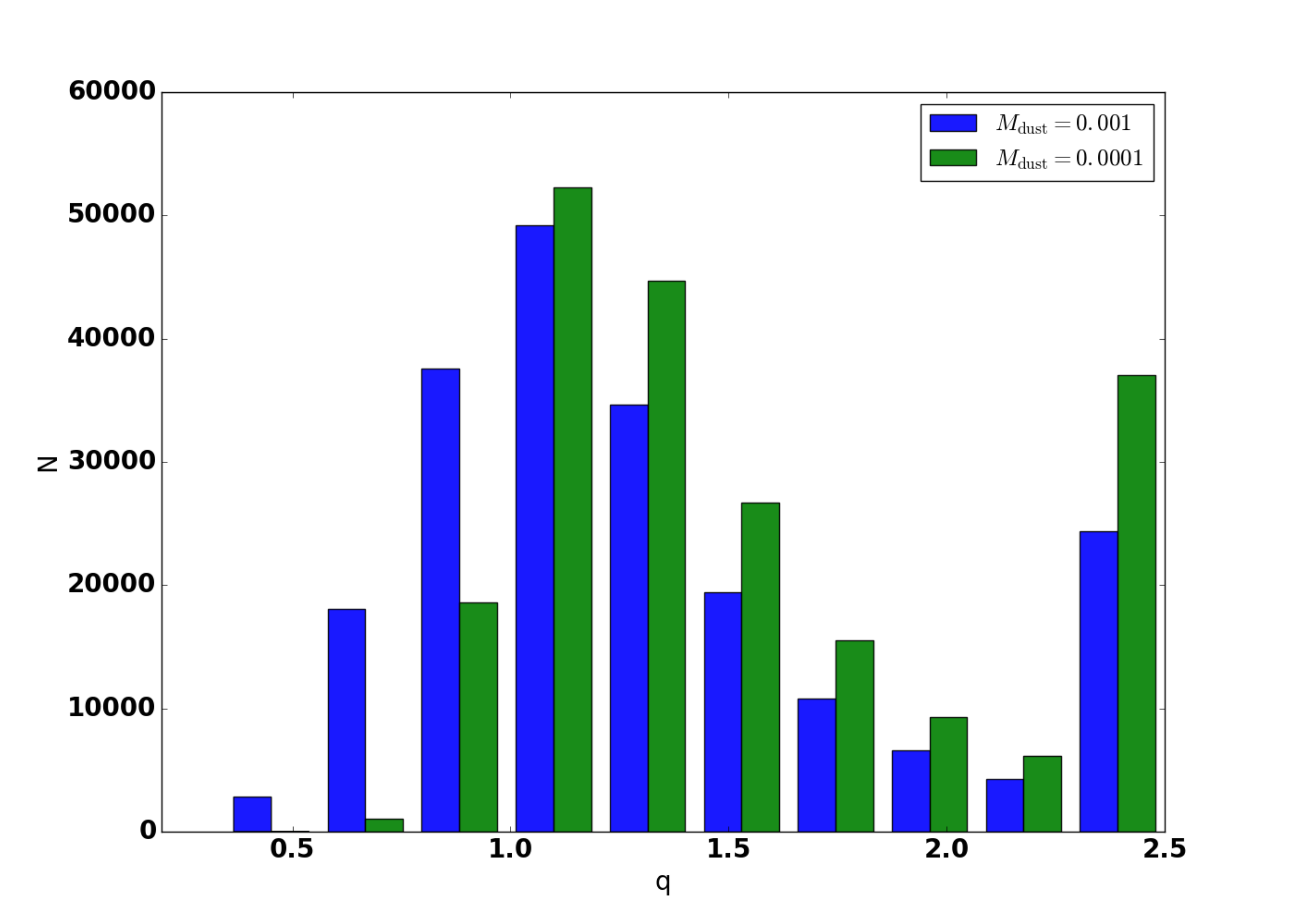}
   \caption{Distribution of q-values obtained by our models. The colours differentiate the models with 
   different dust mass.}  \label{figure:Q-dustmass}
\end{figure}

A full parameter study is beyond the scope of this paper, but we shortly discuss some general characteristics of the model grid.
Although our model disks are all continous, and start at the sublimation radius, they do cover a significant 
range in size, colour and temperature slope (q-value). To illustrate the dependence of the 
fitted q value on the RT model parameters, we show the distribution of q as function of dust mass in Fig.~\ref{figure:Q-dustmass}. 
Only the more massive models (our gas-to-dust ratio is constant) lead to very low $q$ values (hence large disk appearances).
Additionally, low q values only occur for highly-inclined disks, with baselines oriented along the major axis of the disk. 
The peak at the higher end of $q$ is artificial and represents the tail of the distribution, 
as we assume $q<2.5$ in our model fitting. Such q values predominantly come from highly-inclined, low-mass, single-power-law models with a small
value for the $\alpha$ parameter, and with baselines oriented along the minor axis of the disk. Hence, models with an inner rim 
that approach a vertical wall in appearance.


We conclude that our model comparison shows that the physical structures computed by assuming a passively 
irradiated disk give a very good representation of post-AGB disks in terms of N-band interferometric observables and SED.
This corroborates that the processes determining the physical structure of these disks are very similar to those operating in young YSOs,  
and that the models developed to represent proto-planetary disks around YSOs can be used very well for constraining 
the structure of disks at the opposite end of stellar evolution.

\subsection{Comparison with YSOs} \label{subsection:YSOcomparison}

The main difference between YSOs of the Herbig Ae class analysed in the MIDI survey of \citet{2015AAMenu}
and the post-AGB stars analysed here, is that the latter display significantly larger central luminosities, 
while the central gravitational mass is not too different.
This means that the inner rims of post-AGB disks, which are at the sublimation radius, lie further away from the illuminating star, 
where the local gravity field is smaller than for YSOs. The hot inner rim has a 
significant scale-height. This is also evident from the large infrared excess, compared to the 
energy that is available from the photosphere. 

\begin{figure}
   \centering
   \includegraphics[width=9cm]{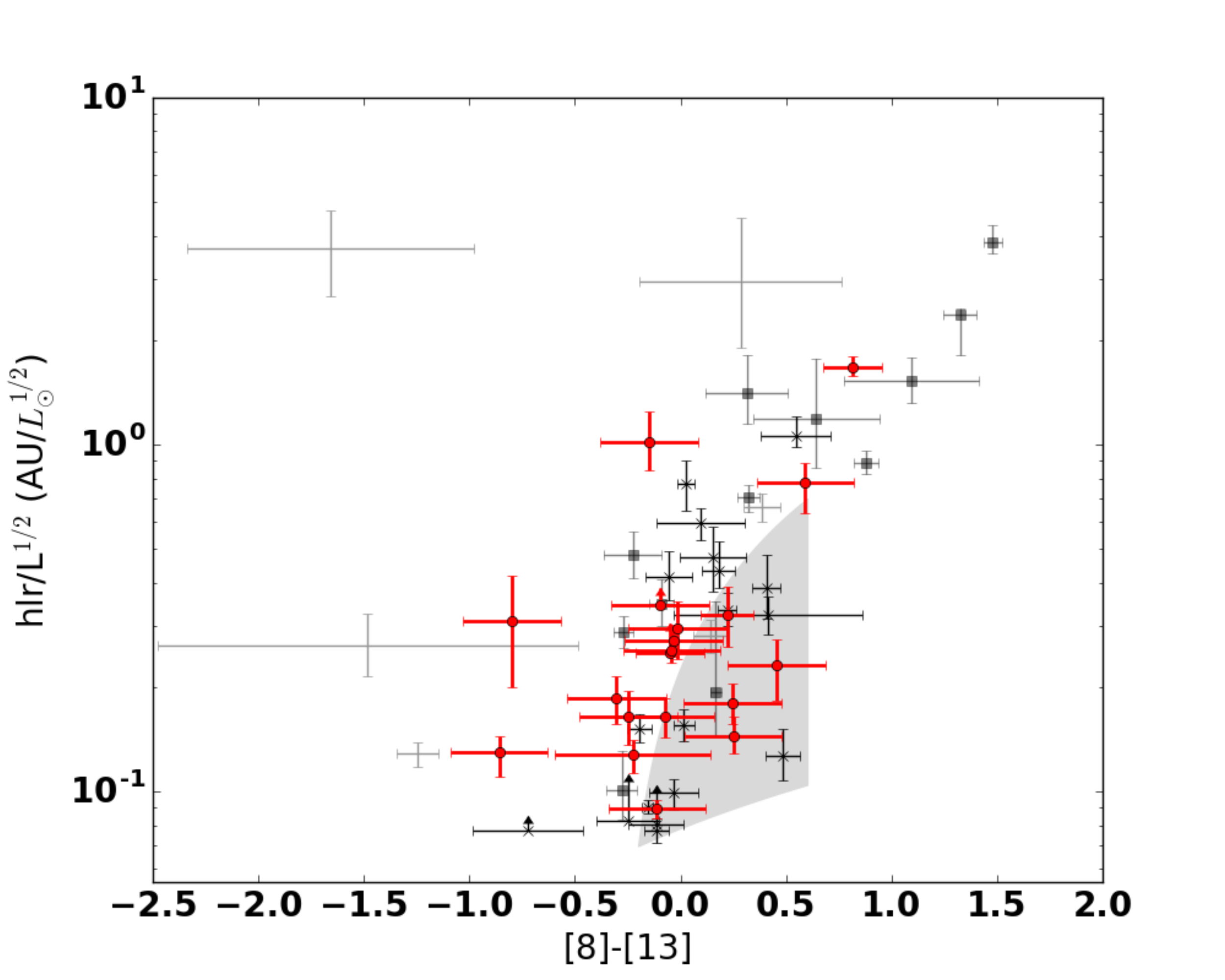}
   \caption{Size-colour diagram of our sample of post-AGB stars compared to the sample of 
   Herbig Ae disks that was studied by \citet{2015AAMenu}. Arrows indicate upper limits. The symbols are 
   explained in Sect.~\ref{subsection:YSOcomparison}. The grey area delimits the region of continuous HAEBE disk models
   computed by \citet{2015AAMenu}.
   \label{figure:size-colour2}}
\end{figure}

A direct comparison of the disks around YSO and post-AGB stars in the size-colour diagram is given in Fig.~\ref{figure:size-colour2}.
We used the sample of Herbig Ae stars of \citet{2015AAMenu}. 
A differentiation is made between three groups of Herbig Ae sources \citep[the classification of][]{2001AAMeeus}: group Ia (black squares), 
Ib (grey plusses) and II (black crosses).  This classification of YSOs was made on the basis of the far-IR SED \citep{2001AAMeeus},
and has been interpreted in terms of flaring (group I) vs. self-shadowed (group II) disks \citep{2001ApJDullemond,2004AADullemonda,2004AADullemondb}. More recent studies of transition and pre-transition disks
suggest that these group I SEDs show a large gap in the disk and display an inner disk and a large outer disk \citep[e.g.][and references therein]{Kraus12,2015AAMenu,vandermarel16}. 

Obviously, the 10~$\mu$m size-colour diagram is insufficient to demonstrate the full differences between the variety of 
YSO and post-AGB disks, but it is striking that these objects cluster very similarly, despite the different evolutionary stage 
of the host star(s). More precisely, and similar to their SED behaviour \citep{2006AAdeRuyter}, 
the post-AGB sources cover very much the same region as the group II Herbig Ae stars 
(with a few outliers that were discussed in Sects.~\ref{section:individualtargets} and~\ref{section:analysis}). 
It thus seems that post-AGB disks are significantly puffed-up at their inner rim, but do not have a strongly flaring outer surface
or evidence for a large outer disk.

The dark grey zone in Fig~\ref{figure:size-colour2} represents the region of continuous disks as computed in \citet{2015AAMenu}. 
They used the same code for the disk modelling, but with distinct assumptions. The main differences are 
the central star properties, our choice for a double-power-law formalism, the model parameter 
coverage (e.g. the dust mass) and the assumed presence of a halo at the sublimation radius. 
The combination of these different choices explains why our continuous models cover a larger zone in the 
size-colour diagram. Additionally, \citet{2015AAMenu} computed a range of models with a gap (i.e. the inner 
radius of the disk lies beyond the sublimation radius), to explain the largest and reddest objects. 
Our analysis shows, however, that being in 
the upper right corner of the size-colour diagram is not a sufficient condition to conclude about the presence of 
a gap in the disk. Sufficiently massive and highly inclined disks (see Fig.~\ref{figure:Q-dustmass}), 
observed with baselines aligned along the major axis, also appear big and red in the size-colour diagram.
On the other hand, there are several Herbig Ae sources and one post-AGB object \citep[nr.17 AC Her,][]{2015AAHillen},
for which a more detailed analysis of the wavelength-dependent MIDI visibilities and differential phases have shown
the presence of a gap. In general, however, a good uv-coverage (or an a priori constraint on the disk inclination and/or mass)
is required to firmly establish the presence of gaps in individual sources. This will be possible with the 
2nd-generation VLTI instrument MATISSE.


\section{Discussion} \label{section:discussion}

\begin{figure}[h]
   \centering
   \includegraphics[width=9cm]{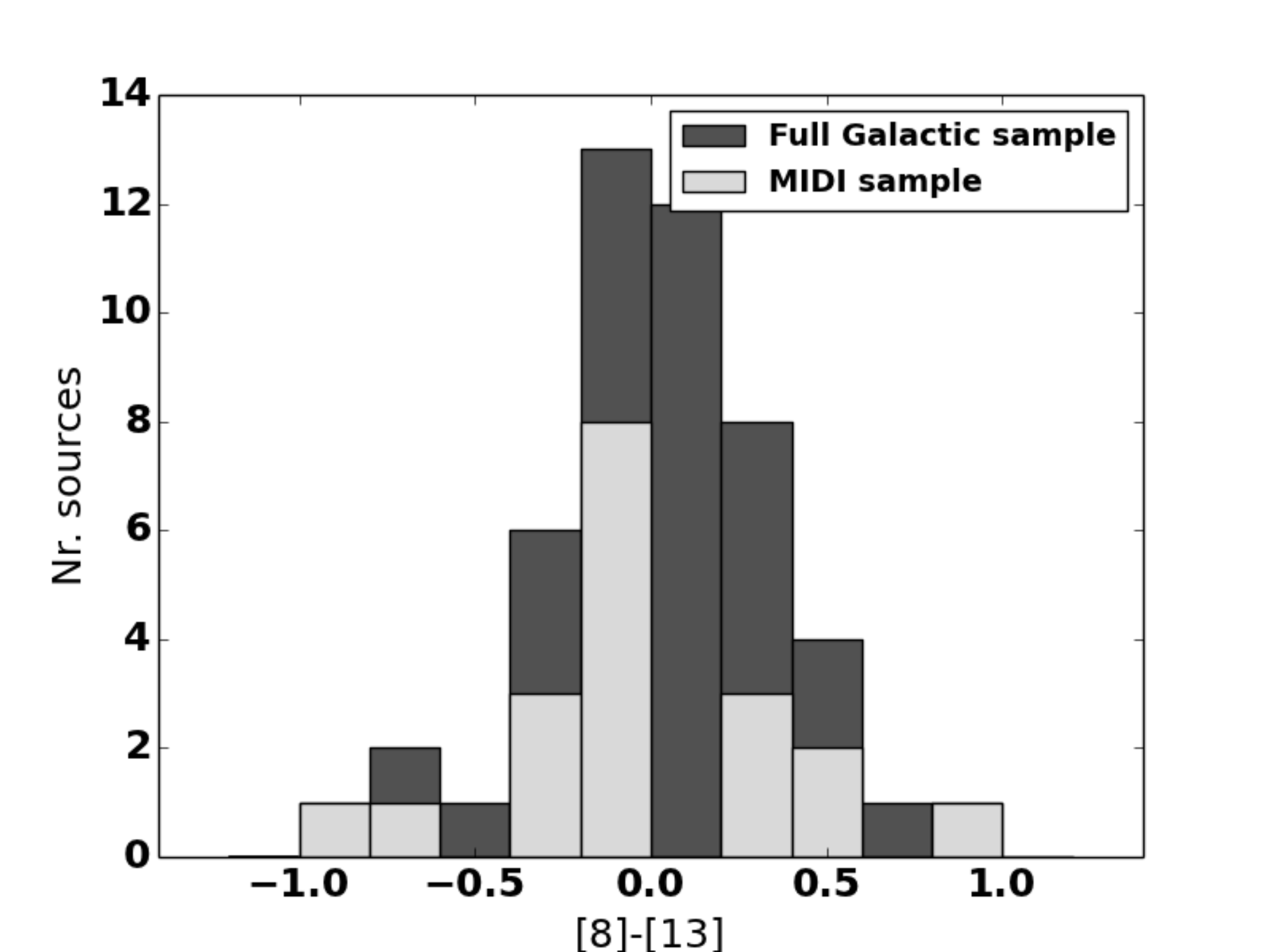}
   \caption{Histogram of the [8]-[13] colour, comparing our MIDI sample to the known Galactic population of which spectra were presented by \citet{2011AAGielen}.
   \label{figure:colour-hist}}
\end{figure}

We present a large interferometric N-band survey of 19 post-AGB objects, which are classified as proven or suspected binary 
stars with a circumbinary disk. This sample is representative of the whole Galactic population of similar objects 
(Fig.~\ref{figure:colour-hist}). For the interpretation of the correlated fluxes, we also constructed the full SEDs in a 
homogeneous way. We differentiated in the N-band between the faint and bright state in the pulsation cycle 
for two high-amplitude variables (U\,Mon (nr.3) and IRAS17038 (nr.14)).

A large fraction (17/19) of the objects is resolved with our MIDI interferometric baselines, showing a very compact 
emission region, corroborating the compact-disk interpretation of the SED.
Our model fitting and sample-wise comparison to a grid of continuous radiative transfer models shows that the inner rim in these disks 
is at the sublimation radius and dominates the N-band for most of the sources. Although diverse 
in N-band total flux, the objects show a very homogeneous behaviour in the interferometric observables. 
This shows that the disk structures are very similar within the sample. Noticeable exceptions are 
IRAS10174 (nr.8) and objects for which PAH emission, with a proven or suspected different spatial distribution, 
affects the result (HR4049 (nr.2) and HD52961 (nr.4)).
There is also one rather evolved disk in our sample, AC\,Her (nr. 17, \citet{2015AAHillen}), which has an inner rim 
that is significantly larger than the sublimation radius. The homogeneity in the sample is striking, given 
the covered range in photospheric temperature and orbital period. Although we do not constrain the post-AGB age, 
the sample homogeneity also indicates that the structure of these disks is very stable. 

\subsection{Aspect angle}
The inclination angle of the disk will of course impact on the observational properties.
Unfortunately, our angular coverage of the uv-plane is too limited to determine the inclinations systematically.
For objects like AR\,Pup (nr.4) an edge-on viewing angle can be anticipated, but it is striking
that actually many of the systems in Table~\ref{tab:sample} (10/19) show long-term photometric
behaviour that can be linked to orbital motion. A natural explanation of this so-called 'RVb phenomenon' 
is that a grazing aspect angle over the disk will affect the line-of-sight extinction
during the orbital motion. This will then result in light curves in which the orbital period is visible 
and in objects that appear redder when fainter.
A complicating factor is, however, that the optical scattering, which can represent up to 35-40\% 
of the optical flux \citep{2014AAHillen}, is known to be strongly angle-dependent. The large fraction of RVb objects must mean 
that orbital effects can show up in the lightcurves for a wide range of inclinations.

Our RT model grid of passively irradiated disks in hydrostatic equilibrium covers very well the observed properties. 
The bulk of the objects lie in the region of the more compact models, in which the inner rim dominates 
the N-band flux (Fig.~\ref{figure:modelsize-colour}). We note, however, that the most compact models (those with q$\sim$2.5) 
are not compatible with the observations, which indicates that there is some smoothing of the inner rim (i.e.
the rim does not show up as a vertical wall).

\subsection{Comparison with YSO}
To quantify the similarity between our evolved systems and YSOs, we compared them directly 
in the distance-independent N-band size-colour diagram in Fig.~\ref{figure:size-colour2}. 
Both in their integrated SED properties and in the N-band intensity distribution on the sky, the post-AGB stars 
resemble very well the settled group II disks around Herbig Ae stars. Disks of group Ia/Ib with a large
far infrared excess, are  not observed around post-AGB stars.
If the group Ia/Ib YSO disks can be linked to active planet formation, as suggested in \citet[e.g.]{2015AAMenu}, 
this means that planet formation does not occur in the same way around post-AGB stars, if it happens at all. 
Indeed, the fast evolution of the central evolved stars may prevent the formation of grains
beyond the mm- and cm-sized particles that are commonly detected
\citep[whenever sub-mm fluxes are available, see e.g.][]{2015AAHillen,2014AAHillen,2011ApJSahai}.

\subsection{Binary connection}
Second-generation protoplanetary disks are only found around post-AGB stars that are part of a binary system with an orbital period 
roughly in-between a hundred and a few thousand days \citep{2003ARAAVanWinckel,2009AAVanWinckel}. Such orbits 
are too small to accommodate an AGB star. All these objects must have gone through a phase of strong binary 
interaction. No spiralled-in system nor merged system has yet been detected among the objects with 
a disk-like SED that we are monitoring in a large radial velocity programme \citep{2010VanWinckel}. Spiralled-in 
systems in PNe typically display periods of a few days only, hence they have large radial-velocity amplitudes and the lack 
of such systems among the known post-AGB binaries is not an observational bias. Even if the formation of these 
second-generation disks remains elusive, the disparity between 
these populations of binary systems suggests that the formation of a disk (and the angular momentum contained in it), 
may prevent an evolution towards dramatic spiral-in.

\subsection{Formation and evolution}

Although the focus of this paper is on the current structure of post-AGB circumbinary disks, the impact of 
these disks on the further evolution of their host stars should not be underestimated. 
Many post-AGB stars display a chemical anomaly in their photosphere \citep{2003ARAAVanWinckel}.
The abundance trends in such atmospheres resemble the gas phase of the interstellar medium: refractory elements 
are underabundant, while volatiles retain their original abundances. Such depletion patterns are intimately linked
to the presence of a disk around the post-AGB star, but the latter is not a sufficient condition to observe depletion \citep[e.g.][]{gezer15}.
The anomaly can last long, as is illustrated by its presence in BD+33$^{\circ}$2642, the central star of a PN. 
It is long known that the chemical abundances show depletion in this object \citep{napiwotzki94}, but it was only recently 
proven to be a spectroscopic binary with a period of 1105 $\pm$ 24 days  \citep{vanwinckel14}. 

There is also increasing evidence that the presence of a circumbinary disk has a lasting effect on 
the dynamical evolution of the binary orbit, as resonant coupling between the binary and the disk can 
pump-up the eccentricity \citep[e.g][]{artymowicz91, lubow10, dermine13, vos15}. This provides
a way to explain the commonly observed large eccentricities \citep[e.g.][and references 
therein for the orbits]{2009AAVanWinckel,gezer15}. The integrated effect depends very 
much on the mass and longevity of the disk, as well as on the radial mass distribution within the disk. 

The MIDI spectral coverage and sensitivity makes that only objects with significant N-band excesses can be resolved.
But disks also evolve, as is likely illustrated by the BD+39$^{\circ}$4926 system. 
This is a highly depleted post-AGB binary in an orbit of $\sim$870 days 
for which only recently some circumstellar material was detected via a small excess at 22~$\mu$m \citep{kodaira70,venn14,gezer15}. 
The lack of an excess at shorter wavelengths, for a depleted star, indicates the presence of a highly evolved 
disk in which the innermost material has been removed. The remaining material is probably present in some form of 
debris disk.

It is by now well established that there are also white dwarfs (WD) with circumstellar material. 
The first discovery was made through the detection of an IR-excess \citep{zuckerman87}, and 
evaluated to be a disk created by the destruction of asteroids or even planets \citep{graham90}. 
These debris disks are now found around more than 30 WD \citep[e.g.][and references therein]{manser16}, 
but they are typically very close to the WD. When the WD accretes some of the solid material, 
the opposite happens to what is detected in the post-AGB binaries discussed in this paper: the refractory 
elements become more abundant. This is in fact an observable to trace circumstellar material 
around WD \citep[e.g.][]{clayton14}. The tidal disruption can even be detected as multiple transits 
caused by clouds of debris material \citep[see the lightcurve of WD1145+017,][]{gansicke16}. There is no
link to binary evolution for these systems. 

This is different for post-common envelope WD plus M-dwarf systems. Interesting here is that 
the transit timing variations of NN Ser can be explained very well with a two-planet model \citep{marsh14}. 
The recent detection of circumbinary dust \citep{hardy16} leads to the conclusion that these planets 
may well be of secondary origin and are formed from the material ejected in the common envelope event. 
This system has spiralled-in, so there is no evolutionary connection between post-AGB wide binaries, 
which avoided spiral-in, and NN\,Ser. However, the potential formation of secondary planets is 
also in the context of long-lived disks around post-AGB binaries an important item for focussed future research.

The dust in old WD discussed above, is located near the disruption radius, so close to the WD itself. 
Larger disks have been discovered as well, and mainly around WDs which are in the centre of a
planetary nebula \citep[e.g.][and references therein]{bilikova12}. The disk in 
the Helix nebula is likely the most famous example \citep{su07}. The systematic study 
by \citet{clayton14} shows that the observational evidences are not conclusive, but 
these wider disks are linked to a formation scenario of secondary disks on the AGB 
(and not via destruction of asteroids) as well as to binary evolution (as most CSPNe 
with larger disks are binaries). The orbital properties of these CSPNe are, however, diverse and not complete. 

\subsection{Conclusion}
We conclude that our N-band interferometric survey confirms the presence of long-lived disks around 
post-AGB stars. These post-AGB disks have an inner structure that resembles very well that of proto-planetary disks around YSO. 
As the luminous central star is in a fast evolutionary phase, these objects offer an alternative
environment to study proto-planetary disk structure, evolution and evaporation.  
Moreover, with multi-beam combiners, these objects provide an ideal opportunity to obtain, via 
interferometric imaging, a direct probe for disk-binary interactions and their possible 
dependence of orbital phase \citep{2016AAHillen}. Given the results of our interferometric and radial velocity surveys, we show 
that the radiative and dynamical interaction between these binary systems and their dusty environments 
can now be resolved for a significant fraction of the known Galactic population of these objects. These scaled-up
versions of proto-planetary disks can be used to study poorly understood
disk evolution and dispersal processes under a much wider range of physical
conditions. The disks must have an important role in the future evolution of these stars and the 
potential evolutionary connection to the circumstellar material around WD needs to be further explored.


 
\begin{acknowledgements}
MH, RM, DK and HvW acknowledge support from the Research Council of the KU Leuven under grant number GOA/2013/012.
DK acknowledges support of the Research Foundation - Flanders under contract G.0B86.13.
The authors acknowledge Christoffel Waelkens, Bram Acke, Pieter Deroo,
Cees Dullemond and Tom Lloyd Evans (RIP) for enlightening
discussions during the whole project. We acknowledge with thanks the variable star observations 
from the AAVSO International Database contributed by observers worldwide and used in this research.
This research has made use of the SIMBAD database,
operated at CDS, Strasbourg, France. This research has made use of NASA's Astrophysics Data System Bibliographic Services.
This research has made use of the Jean-Marie Mariotti Center \texttt{Aspro}
service \footnote{Available at http://www.jmmc.fr/aspro}.
\end{acknowledgements}

\bibliographystyle{aa}

\Online

\begin{appendix}
\section{Origin of literature photometry} \label{section:photorigin}
In this appendix we give full reference to the sources of the photometry used in this work. Our Python-based SED fitting tools 
allow to automatically retrieve the available photometry in a predefined list of Simbad/Vizier catalogues. This is the case 
for the following photometry: (UV) from the TD1 \citep{1978csufThompson} and 
the ANS satellite \citep{1982AASWesselius}, (optical and near-IR) from the JOHNSON/COUSINS 11-band, 
GENEVA \citep{1972VAGolay}, Str{\"o}mgren \citep{1956VAStromgren}, 
Tycho-2 \citep{2000AAHog}, 2MASS \citep{2003yCatCutri} and SDSS \citep{1996AJFukugita} systems,  
(far-IR) from the Infra-Red Astronomy Satellite \citep[IRAS,][]{1984ApJNeugebauer}, the 
AKARI satellite \citep{2007PASJMurakami}, the DIRBE instrument on the COBE satellite \citep{2004ApJSSmith}, 
the WISE satellite \citep{2012yCatCutri} and the Midcourse Space Experiment \citep[MSX,][]{1995SSRvPrice}.
In addition, we included some newly acquired near-IR photometry from the South African Astronomical Observatory (SAAO).

For sources with small 
variability amplitudes, the photometry retrieved in this way is generally sufficient for our purposes. In a few cases, however, 
some additional photometry was added as well (see below). For the sources with large-amplitude flux variations, more care was taken 
to only select photometry at specific epochs. To this end, we also made an extensive search through the literature to find 
multicolour photometry for which a phase attribution could be made with the light curves from the ASAS and AAVSO databases.  

Finally, we note that for some stars, measurements in specific bands of certain systems were not retained, e.g. due to saturation 
issues. This is particularly relevant for WISE data (mainly W1 and W2), for which we employed the following limits: $W_1 > 2.5$, $W_2 > 2.0$,
$W_3 > -2.5$ and $W_4 > -3.5$. Upper limits were not retained either (e.g. for some IRAS.F100 measurements).

\begin{footnotesize}
\begin{longtable}{lllll}
 \caption{Full reference list to the photometric data used to construct the SEDs \label{tab:photsource}} \\
 \hline
 \hline
Star & Photometric system & Retained bands & Vizier catalogue & Note \\
     &                    &                & or bibl. reference     &      \\
\hline
\endfirsthead
\caption{continued.}\\
\hline\hline
Star & Photometric system & Retained bands & Vizier catalogue & Note \\
     &                    &                & or reference     &      \\
\hline
\endhead
\hline
\endfoot
\hline
\endlastfoot
\multicolumn{5}{l}{\normalsize{Stars with negligible variability}} \\
IRAS07008+1050 & 2MASS       & J, H, Ks                 & II/246/out  & \\
               & IRAS        & F12, F25, F60            & II/275/fsr  & \\
               & TD1         & 1965, 2365, 2740         & II/59B/catalogue  & \\
               & TYCHO-2     & BT, VT                   & I/259/tyc2   & \\
               & JOHNSON     & B, V                     & V/137C/XHIP  & \\
               & COUSINS     & R, I                     & V/137C/XHIP  & \\
               & JOHNSON     & B, V                     & I/280B/ascc  & \\
               & JOHNSON     & V                        & II/271A/patch2  & \\
               & COUSINS     & I                        & II/271A/patch2  & \\
               & AKARI       & WIDES                    & II/298/fis  & \\
               & AKARI       & S9W, L18W                & II/297/irc & \\
               & WISE        & W1, W2, W3, W4           & II/311/wise  & \\
               & GENEVA      & U, B, V, B1, B2, V1, G   & GCPD\tablefootmark{a}  &  \\
IRAS08544-4431 & 2MASS       & J, H, Ks                 & II/246/out  & \\   
               & IRAS        & F12, F25, F60            & II/275/fsr  & \\  
               & TYCHO-2     & BT, VT                   & I/259/tyc2  & \\
               & JOHNSON     & U, B, V                  & \citet{2006AAdeRuyter} & \\
               & COUSINS     & R, I                     & \citet{2006AAdeRuyter} & \\
               & AKARI       & N60, WIDES, WIDEL        & II/298/fis  & \\   
               & AKARI       & S9W, L18W                & II/297/irc  & \\   
               & MSX         & B1, B2, A, C, D, E       & V/114/msx6\_gp  & \\
               & WISE        & W3, W4                   & II/311/wise  & \\
               & SAAO        & J, H, K, L               & \citet{2006AAdeRuyter} & \\
IRAS09256-6324 & 2MASS       & J, H, Ks                 & II/246/out  & \\          
               & IRAS        & F12, F25, F60, F100      & II/275/fsr  & \\          
               & TYCHO-2     & BT, VT                   & I/259/tyc2  & \\          
               & JOHNSON     & J, H, K, L, M            & II/237/colours  & \\       
               & AKARI       & N60, WIDES, WIDEL        & II/298/fis  & \\          
               & AKARI       & S9W, L18W                & II/297/irc  & \\          
               & STROMGREN   & Y, B, V, U               & II/215/catalogue   & \\     
               & DIRBE       & F4\_9, F12, F25          & J/ApJS/154/673/DIRBE  & \\
               & WISE        & W1, W3, W4               & II/311/wise  & \\         
               & JOHNSON     & B, V                     & \citet{1996MNRASPollard} & max. RVb phase \\        
               & COUSINS     & R, I                     & \citet{1996MNRASPollard} & max. RVb phase \\    
IRAS10158-2844 & 2MASS       & J, H, Ks                 & II/246/out   & \\         
               & IRAS        & F12, F25, F60            & II/125/main   & \\        
               & TYCHO2      & BT, VT                   & I/259/tyc2   & \\         
               & JOHNSON     & B, V                     & V/137C/XHIP   & \\        
               & COUSINS     & R, I                     & V/137C/XHIP   & \\        
               & JOHNSON     & B, V                     & I/280B/ascc   & \\        
               & JOHNSON     & B, V                     & II/168/ubvmeans   & \\    
               & AKARI       & WIDES                    & II/298/fis   & \\         
               & AKARI       & S9W, L18W                & II/297/irc   & \\         
               & STROMGREN   & Y, B, V, U               & J/A+A/373/625/table2  & \\
               & DIRBE       & F1\_25, F2\_2, F3\_5, F4\_9 & J/ApJS/154/673/DIRBE  & \\
               & WISE        & W3, W4                   & II/311/wise   & \\        
               & GENEVA      & U, B, V, B1, B2, V1, G   & GCPD\tablefootmark{a}  &   \\               
               & STROMGREN   & Y, B, V, U               & GCPD\tablefootmark{a}  &   \\               
IRAS10174-5704 & 2MASS       & J, H, Ks                 & II/246/out & \\                 
               & IRAS        & F12, F25                 & II/275/fsr & \\ 
               & TYCHO2      & BT, VT                   & I/259/tyc2 & \\ 
               & AKARI       & N60, WIDES               & II/298/fis & \\ 
               & AKARI       & S9W, L18W                & II/297/irc & \\ 
               & MSX         & B1, B2, A, C, D, E       & V/114/msx6\_gp & \\
               & WISE        & W1, W2, W3, W4           & II/311/wise  & \\
               & SAAO        & J, H, K, L               & - & P. Whitelock \\
IRAS10456-5712 & 2MASS       & J, H, Ks                 & II/246/out  & \\     
               & IRAS        & F12, F25, F60, F100      & II/125/main  & \\   
               & TYCHO-2     & BT, VT                   & I/259/tyc2  & \\    
               & JOHNSON     & B, V                     & V/137C/XHIP  & \\   
               & COUSINS     & I                        & V/137C/XHIP  & \\   
               & JOHNSON     & B, V                     & I/280B/ascc  & \\   
               & JOHNSON     & U, B, V                  & II/168/ubvmeans  & \\
               & JOHNSON     & B, V                     & II/5A/data  & \\    
               & AKARI       & N60, WIDES               & II/298/fis  & \\    
               & AKARI       & S9W, L18W                & II/297/irc  & \\    
               & MSX         & B1, B2, A, C, D, E       & V/114/msx6\_gp  & \\ 
               & WISE        & W3, W4                   & II/311/wise  & \\   
               & SAAO        & J, H, K, L               & - & P. Whitelock \\     
IRAS11385-5517 & 2MASS       & J, H, Ks                 & II/246/out  & \\   
               & IRAS        & F12, F25, F60, F100      & II/275/fsr  & \\
               & TYCHO-2     & BT, VT                   & I/259/tyc2  & \\
               & JOHNSON     & U, B, V                  & II/168/ubvmeans & \\
               & JOHNSON     & U, B, V, J, H, K, L, M   & II/237/colours & \\
               & COUSINS     & R, I                     & V/137/XHIP & \\
               & AKARI       & N60, WIDES, WIDEL, N160  & II/298/fis  & \\
               & AKARI       & S9W, L18W                & II/297/irc & \\
               & STROMGREN   & Y, B, V, U               & II/215/catalogue & \\
               & ALMA        & Band6                    & \citet{2015AAOlofsson} & \\
IRAS12222-4652 & 2MASS       & J, H, Ks                 & II/246/out  & \\    
               & IRAS        & F12, F25, F60, F100      & II/125/main  & \\   
               & TYCHO-2     & BT, VT                   & I/259/tyc2  & \\    
               & JOHNSON     & B, V                     & I/280B/ascc  & \\   
               & AKARI       & N60, WIDES               & II/298/fis  & \\    
               & AKARI       & S9W, L18W                & II/297/irc  & \\    
               & STROMGREN   & Y, B, V, U               & II/215/catalogue  & \\
               & WISE        & W1, W2, W3, W4           & II/311/wise  & \\   
               & GENEVA      & U,B,V,B1,B2,V1,G         & GCPD\tablefootmark{a}  &  \\          
               & STROMGREN   & Y, B, V, U               & GCPD\tablefootmark{a}  &  \\          
               & SAAO        & J, H, K, L               & - & P. Whitelock \\
IRAS15469-5311 & 2MASS       & J, H, Ks                 & II/246/out  & \\                  
               & IRAS        & F12, F25, F60            & II/125/main  & \\  
               & TYCHO-2     & BT, VT                   & I/259/tyc2  & \\   
               & JOHNSON     & B, V                     & I/280B/ascc  & \\  
               & AKARI       & N60, WIDES               & II/298/fis  & \\   
               & AKARI       & S9W, L18W                & II/297/irc  & \\   
               & MSX         & B1, B2, A, C, D, E       & V/114/msx6\_gp  & \\
               & WISE        & W1, W3, W4               & II/311/wise  & \\  
               & SAAO        & J, H, K, L               & -  & P. Whitelock \\   
IRAS17535+2603 & 2MASS       & J,H,Ks                   & II/246/out  &  \\          
               & IRAS        & F12,F25,F60,F100         & II/125/main  &  \\         
               & TD1         & 1965, 2365, 2740         & II/59B/catalogue  &  \\      
               & ANS         & 18, 25, 33               & II/97/ans  &  \\           
               & TYCHO2      & BT, VT                   & I/259/tyc2  &  \\          
               & JOHNSON     & B, V                     & I/280B/ascc  &  \\         
               & JOHNSON     & U, B, V, R, I            & II/7A/catalogue   &  \\      
               & JOHNSON     & B, V, R, I, H, K, L, M, N & II/237/colours  &  \\       
               & JOHNSON     & U, B, V                  & II/168/ubvmeans  &  \\     
               & AKARI       & N60, WIDES, WIDEL, N160  & II/298/fis  &  \\          
               & AKARI       & S9W, L18W                & II/297/irc  &  \\
               & DIRBE       & F3\_5, F4\_9, F12        & J/ApJS/154/673/DIRBE  &  \\
               & WISE        & W4                       & II/311/wise  &  \\         
               & GENEVA      & U, B, V, B1, B2, V1, G   & GCPD\tablefootmark{a}  &  \\                
               & STROMGREN   & Y, B, V, U               & GCPD\tablefootmark{a}  &  \\                
               & VILNIUS     & V, Z, S, Y, X, P, U      & GCPD\tablefootmark{a}  &  \\
IRAS19125+0343 & 2MASS       & J, H, Ks                 & II/246/out  & \\                    
               & IRAS        & F12, F25, F60            & II/125/main  & \\    
               & TYCHO-2     & BT, VT                   & I/259/tyc2  & \\     
               & JOHNSON     & B, V                     & I/280B/ascc  & \\    
               & JOHNSON     & V                        & II/271A/patch2  & \\ 
               & COUSINS     & I                        & II/271A/patch2  & \\ 
               & SDSS        & RP                       & I/304/out  & \\      
               & AKARI       & S9W, L18W                & II/297/irc  & \\     
               & MSX         & B2, A, C, D, E           & V/114/msx6\_gp  & \\  
               & WISE        & W1, W3, W4               & II/311/wise  & \\    
               & GENEVA      & U, B, V, B1, B2, V1, G   & \citet{2006AAdeRuyter} & \\    
IRAS22327-1731 & 2MASS       & J, H, Ks                 & II/246/out  & \\                    
               & IRAS        & F12, F25, F60            & II/125/main  & \\    
               & TYCHO-2     & BT, VT                   & I/259/tyc2  & \\     
               & JOHNSON     & B, V                     & V/137C/XHIP  & \\    
               & COUSINS     & R, I                     & V/137C/XHIP  & \\    
               & JOHNSON     & B, V                     & I/280B/ascc  & \\    
               & JOHNSON     & U, B, V, J, H, K, L      & II/237/colours  & \\  
               & AKARI       & WIDES                    & II/298/fis  & \\     
               & AKARI       & S9W, L18W                & II/297/irc  & \\     
               & WISE        & W1, W2, W3, W4           & II/311/wise  & \\    
               & STROMGREN   & Y, B, V, U               & GCPD\tablefootmark{a} & \\          
               & GENEVA      & U, B, V, B1, B2, V1, G   & \citet{2006AAdeRuyter} & max. RVb phase \\    
\multicolumn{5}{l}{\normalsize{Stars with significant variability}} \\
IRAS04440+2605 & JOHNSON     & B, V                     & \citet{1992AJWahlgren} & max. puls. and RVb phase \\ 
               & JOHNSON     & B, V                     & \citet{1992AJWahlgren} & min. puls. and max. RVb phase \\ 
               & JOHNSON     & J, H, K, L, M            & \citet{2009AstLTaranova} & max. puls. and RVb phase \\ 
               & GENEVA      & U, B, V, B1, B2, V1, G   & \citet{2006AAdeRuyter} & max. puls. and RVb phase \\    
               & GENEVA      & U, B, V, B1, B2, V1, G   & -\tablefootmark{b} &  min. puls. and max RVb. phase \\    
               & AKARI       & N60, WIDES               & II/298/fis & \\
               & AKARI       & S9W, L18W                & II/297/irc & \\
               & WISE        & W1, W2, W3, W4           & II/311/wise & \\
               & 2MASS       & J, H, K                  & II/246/out & \\
IRAS07284-0940 & JOHNSON     & B, V                     & \citet{1996MNRASPollard} & max. puls. and RVb phase \\ 
               & COUSINS     & R, I                     & \citet{1996MNRASPollard} & max. puls. and RVb phase \\
               & GENEVA      & U, B, V, B1, B2, V1, G   & \citet{2006AAdeRuyter} &  max. puls. and RVb phase \\    
               & JOHNSON     & U, B, V                  & \citet{1963ApJPreston} &  max. puls. and RVb phase \\  
               & JOHNSON     & U, B, V                  & \citet{1963ApJPreston} &  max. puls. and RVb phase \\  
               & SAAO        & J, H, K, L               & \citet{1985MNRASEvans} & max. puls. and RVb phase \\
               & JOHNSON     & B, V                     & \citet{1996MNRASPollard} & min. puls. and max. RVb phase \\ 
               & COUSINS     & R, I                     & \citet{1996MNRASPollard} & min. puls. and max. RVb phase \\
               & GENEVA      & U, B, V, B1, B2, V1, G   & \citet{2006AAdeRuyter} & min. puls. and max. RVb phase \\   
               & JOHNSON     & U, B, V                  & \citet{1963ApJPreston} & min. puls. and max. RVb phase \\
               & SAAO        & J, H, K, L               & \citet{1985MNRASEvans} & min. puls. and max. RVb phase \\
               & 2MASS       & J, H, K                  & II/246/out &   \\
               & IRAS        & F12, F25, F60, F100      & II/125/main &  \\
               & AKARI       & N60, WIDES               & II/298/fis &   \\
               & AKARI       & S9W, L18W                & II/297/irc &   \\
               & MSX         & B2, A, C, D, E           & V/114/msx6\_gp & \\
               & WISE        & W1, W3, W4               & II/311/wise &  \\
IRAS08011-3627 & JOHNSON     & B, V                     & \citet{1996MNRASPollard} & max. puls. and RVb phase \\   
               & GENEVA      & U, B, V, B1, B2, V1, G   & \citet{2006AAdeRuyter} & max. puls. and RVb phase \\  
               & SAAO        & J, H, K, L               & \citet{1985MNRASEvans} & max. puls. and RVb phase \\
               & JOHNSON     & B, V                     & \citet{1996MNRASPollard} & min. puls. and max. RVb phase \\   
               & SAAO        & J, H, K, L               & \citet{1985MNRASEvans} & min. puls. and max. RVb phase \\
               & 2MASS       & J, H, Ks                 & II/246/out  & \\         
               & IRAS        & F12, F25, F60            & II/125/main  & \\        
               & AKARI       & WIDES                    & II/298/fis  & \\         
               & AKARI       & S9W, L18W                & II/297/irc  & \\         
               & DIRBE       & F1\_25, F2\_2, F3\_5, F4\_9  & J/ApJS/154/673/DIRBE & \\
               & WISE        & W3, W4                   & II/311/wise  & \\        
IRAS12185-4856 & JOHNSON     & U, B, V                  & \citet{1987MNRASGoldsmith}  & max. puls. and RVb phase \\  
               & COUSINS     & R, I                     & \citet{1987MNRASGoldsmith}  & max. puls. and RVb phase \\ 
               & SAAO        & J, H, K, L               & \citet{1987MNRASGoldsmith}  & max. puls. and RVb phase \\ 
               & JOHNSON     & U, B, V                  & \citet{1987MNRASGoldsmith}  & min. puls. and max. RVb phase \\ 
               & COUSINS     & R, I                     & \citet{1987MNRASGoldsmith}  & min. puls. and max. RVb phase \\ 
               & SAAO        & J, H, K, L               & \citet{1987MNRASGoldsmith}  & min. puls. and max. RVb phase \\ 
               & 2MASS       & J, H, Ks                 & II/246/out  & \\   
               & IRAS        & F12, F25, F60            & II/125/main  & \\
               & AKARI       & WIDES                    & II/298/fis  & \\ 
               & AKARI       & S9W, L18W                & II/297/irc   & \\
               & WISE        & W1, W2, W3, W4           & II/311/wise  & \\  
IRAS17038-4815 & JOHNSON     & U, B, V                  & \citet{2006AAdeRuyter} & max. puls. and RVb phase \\                  
               & COUSINS     & R, I                     & \citet{2006AAdeRuyter} & max. puls. and RVb phase \\  
               & JOHNSON     & U, B, V                  & \citet{2006AAdeRuyter} & min. puls. and max. RVb phase \\  
               & COUSINS     & R, I                     & \citet{2006AAdeRuyter} & min. puls. and max. RVb phase \\  
               & 2MASS       & J, H, Ks                 & II/246/out & \\    
               & IRAS        & F12, F25, F60            & II/125/main & \\   
               & AKARI       & WIDES, WIDEL             & II/298/fis & \\  
               & AKARI       & S9W, L18W                & II/297/irc & \\  
               & MSX         & A, C, D, E               & V/114/msx6\_gp & \\
               & WISE        & W1, W2, W3, W4           & II/311/wise & \\ 
IRAS17243-4348 & JOHNSON     & U, B, V                  & \citet{2006AAdeRuyter} & max. puls. phase \\    
               & COUSINS     & R, I                     & \citet{2006AAdeRuyter} & max. puls. phase \\    
               & SAAO        & J, H, K, L               & \citet{2006AAdeRuyter} & T. Lloyd Evans (mean)\tablefootmark{c} \\ 
               & JOHNSON     & U, B, V                  & \citet{2006AAdeRuyter} & min. puls. phase \\   
               & COUSINS     & R, I                     & \citet{2006AAdeRuyter} & min. puls. phase \\    
               & 2MASS       & J, H, Ks                 & II/246/out & \\   
               & IRAS        & F12, F25, F60            & II/275/fsr & \\   
               & AKARI       & WIDES                    & II/298/fis & \\   
               & AKARI       & S9W, L18W                & II/297/irc & \\   
               & MSX         & A, C, D, E               & V/114/msx6\_gp & \\
               & WISE        & W1, W2, W3, W4           & II/311/wise & \\
IRAS18281+2149 & GENEVA      & U, B, V, B1, B2, V1, G   & \citet{2006AAdeRuyter} & max. puls. phase \\               
               & JOHNSON     & U, B, V                  & \citet{1993AAZsoldos} & max. puls. phase \\
               & JOHNSON     & J, H, K, L               & \citet{2010yCatTaranova} & max. puls. phase \\
               & GENEVA      & U, B, V, B1, B2, V1, G   & \citet{2006AAdeRuyter} & min. puls. phase \\
               & JOHNSON     & U, B, V                  & \citet{1993AAZsoldos} & min. puls. phase \\ 
               & JOHNSON     & J, H, K, L               & \citet{2010yCatTaranova} & min. puls. phase \\
               & 2MASS       & J, H, Ks                 & II/246/out & \\
               & IRAS        & F12, F25, F60, F100      & II/125/main & \\
               & AKARI       & N60, WIDES, WIDEL        & II/298/fis & \\ 
               & AKARI       & S9W, L18W                & II/297/irc & \\ 
               & WISE        & W1, W2, W3, W4           & II/311/wise & \\             
\end{longtable}
\end{footnotesize}
\tablefoot{\tablefoottext{a}{\citet{1997AASMermilliod}},
           \tablefoottext{b}{Data taken from the archive of the Mercator telescope. We did not use the values from \citet{2006AAdeRuyter}
           in this case because they correspond to a minimum RVb phase.},
           \tablefoottext{c}{For this star the mean JHKL values were used in both pulsation phases.}}
\end{appendix}
\end{document}